\documentclass[traditabstract]{aa} 
\usepackage[english,greek]{babel}
\usepackage{graphicx}
\usepackage{txfonts}
\usepackage{wasysym}

\newcommand{\ptolemaios}{{\em Ptolemaios}}
\newcommand{\ulughbeg}{{\em UlughBeg}}

\newcommand{\gtap}{\mathrel{\hbox{\rlap{\lower.55ex \hbox {$\sim$}}
                   \kern-.3em \raise.4ex \hbox{$>$}}}}
\newcommand{\ltap}{\mathrel{\hbox{\rlap{\lower.55ex \hbox {$\sim$}}
                   \kern-.3em \raise.4ex \hbox{$<$}}}}

\begin{document}
\selectlanguage{english}
  \title{The star catalogues of Ptolemaios and Ulugh Beg}
  \subtitle{Machine-readable versions and comparison with the modern
    Hipparcos Catalogue\thanks{The full Tables
       are available in electronic form only at the CDS via anonymous ftp
      to cdsarc.u-strasbg.fr (130.79.128.5) or via
      http://cdsweb.u-strasbg.fr/cgi-bin/qcat?J/A+A/}}

  \author{Frank Verbunt\inst{1,2} \and Robert H. van Gent\inst{3,4}}

  \institute{Department of Astrophysics/IMAPP, Radboud University Nijmegen, PO Box 9010,
    6500 GL Nijmegen, The Netherlands; \email{F.Verbunt@astro.ru.nl}
  \and SRON Netherlands Institute for Space Research, Utrecht
  \and Until Jan.2010: URU-Explokart, Faculty of Geosciences, Utrecht University, PO Box 80\,115,
    3508 TC Utrecht, The Netherlands 
  \and Institute for  History and Foundations of Science,  PO Box 80\,010,
    3508 TA Utrecht, The Netherlands; \email:{r.h.vangent@uu.nl}}

  \date{Received 14 May 2012 / Accepted 30 May 2012}

  \abstract{In late antiquity and throughout the middle ages, the
    positions of stars on the celestial sphere were obtained from the
    star catalogue of Ptolemaios.  A catalogue based on new
    measurements appeared in 1437, with positions by Ulugh Beg,
    and magnitudes from the 10th-century astronomer al-Sufi.  We provide
    machine-readable versions of these two star catalogues, based on
    the editions by Toomer (1998)\nocite{toomer} and Knobel (1917),
    \nocite{knobel} and determine their accuracies by comparison with
    the modern Hipparcos Catalogue. The magnitudes in the catalogues
    correlate well with modern visual magnitudes; the indication
    `faint' by Ptolemaios is found to correspond to his magnitudes 5 and
    6. Gaussian fits to the error distributions in longitude /
    latitude give widths $\sigma\simeq27\arcmin$ / 23\arcmin\ in the range
    $|\Delta\lambda,\Delta\beta|<50\arcmin$ for Ptolemaios and
    $\sigma\simeq22\arcmin$ /18\arcmin\  in Ulugh Beg. Fits to the range
    $|\Delta\lambda,\Delta\beta|<100\arcmin$ gives 10-15\%\ larger widths,
    showing that the error distributions are broader than
    gaussians. The fraction of stars with positions wrong by more than
    150\arcmin\ is about 2\%\ for Ptolemaios and 0.1\%\ in Ulugh Beg;
    the numbers of unidentified stars are 1 in Ptolemaios and 3 in
    Ulugh Beg. These numbers testify to the excellent quality of both
    star catalogues (as edited by Toomer and Knobel). }


  \maketitle

\section{Introduction}

Ancient Greek astronomy culminated in the work of Ptolemaios, and in
particular in his Mathematike Syntaxis, i.e.\ Mathematical
Composition, of astronomy. By its pre-eminence this book, known in
later ages by its Arabic name Almagest, eclipsed much of the earlier
work, and thus its star catalogue with epoch 137 AD is the oldest extant
major star catalogue that we have.  Knowledge of the Almagest has
reached the modern world in Arabic via the Arabic/Islamic culture,
in particular through Spain, and in Greek via Byzantium. In both cases, the
oldest manuscripts we have are copies of copies of copies\ldots, and
as a result one and the same star may have rather different positions 
in different manuscripts.

The Arabic manuscripts and the translation into Latin by Gerard of
Cremona based on them have been studied by Kunitzsch (1974a), who also
edited the star catalogue in these manuscripts (Kunitzsch 1986, 1990, 1991).  
\nocite{kunitzsch}\nocite{kunitzsch86}\nocite{kunitzsch90}\nocite{kunitzsch91}
A critical edition of the Greek text, based on the available Greek
manuscripts, was produced by Heiberg (1898, 1903), and translated
into German by Manitius (1913, reprinted
with corrections by Neugebauer 1963). 
\nocite{heiberg98}\nocite{heiberg03}\nocite{manitius}
More
recently, an English translation of the Almagest which takes into
account both the Arabic and the Greek manuscripts has been made by
Toomer (1984, 1998). \nocite{toomer84}\nocite{toomer}

The medieval Arabic and Latin editions of star catalogues in general
did not involve new observations. The star positions of Ptolemaios
were merely updated by the addition of a constant to the ecliptic
longitudes, to correct for precession. In the 10th century the Persian
astronomer al-Sufi estimated the magnitudes anew, but Ulugh Beg was
the first astronomer to produce a star catalogue with positions based
on new, independent measurements. He took the magnitudes from al-Sufi
(Knobel 1917). Thus his catalogue, published in manuscript in 1437,
was new with respect to Ptolemaios both in positions and in
magnitudes. A modern translation of the catalogue, based on the
Persian manuscripts of the catalogue present in Great Britain, was
published by Knobel (1917).\nocite{knobel} The results of an analysis
of the positional accuracy of the catalogue of Ulugh Beg, based on
comparison with positions from the modern Hipparcos Catalogue, have
been published by Heiner Schwan (2002),\nocite{schwan} in a paper that
acted as a stimulus for our work on old star catalogues. Shevchenko
(1990)  separates systematic and random errors in the analysis
of the accuracy of the star catalogues of Ptolemaios and Ulugh Beg, 
for zodiacal stars only. Krisciunas (1993) extends this analysis to all
stars in Ulugh Beg.\nocite{shevchenko}\nocite{krisciunas}

A machine-readable version of the star catalogue of Ptolemaios has been made
available by Jaschek (1987),\nocite{jaschek} based on Manitius
(1913).\nocite{manitius}
In the present paper we provide machine-readable versions of the
star catalogue of Ptolemaios, according to the
edition of Toomer, and of the star catalogue of
Ulugh Beg according to the edition by Knobel.
We analyse the accuracy of both catalogues by comparison
with the modern Hipparcos Catalogue.
We compare the Knobel and Toomer editions of the star
catalogue of Ptolemaios (Appendix\,\ref{a:knobtoom}).

In the following we refer to (our machine-readable versions of) the
catalogues of Ptolemaios and Ulugh Beg as \ptolemaios\ and \ulughbeg,
respectively. Individual entries are numbered in order of appearance,
i.e.\ P\,350 is the 350th entry in \ptolemaios, U\,250 the 250th entry
in \ulughbeg. The sequence number within a constellation is given
by a number following the abbreviated constellation name: Aql\,5
is the fifth star in Aquila.

\section{Description of the catalogues}

\subsection{The star catalogue of Ptolemaios}

The star catalogue of Ptolemaios is organized by constellation, and
begins with 21 northern constellations, followed by the 12 zodiacal
and 15 southern constellations.  To many constellations some stars are
added that lie outside the figure (`amorfotos') that
defines the constellation (Table\,\ref{t:ptol}).  The total number of
entries is 1028.  For each entry, a description is followed by the
longitude, latitude and magnitude. The longitude is expressed in
zodiacal sign, degrees and fractions of degrees, the latitude with
bo[reios] for northern or no[tios] for southern, degrees and fractions
of degrees. The fractions in the Greek manuscripts are sometimes
composite, e.g.  $\frac{5}{6}$ is expressed as $\frac{1}{2}$
\,$\frac{1}{3}$. In Toomer (1984,
1998)\nocite{toomer84}\nocite{toomer} this is converted to a single
fraction, e.g.\ $\frac{5}{6}$.  The fractions $F$ are always such that
they correspond to an integer number $M$ of minutes, $M=60F$.  Writing
the zodiacal sign as $Z$ and the degrees as $G$, we may write the
longitude as
\begin{equation}
\lambda = (Z-1)\times30 + G +  F \equiv  (Z-1)\times30 + G + {M\over60}
\label{e:long}
\end{equation}
Writing the degrees and minutes of the latitude as $G$ and
$M$, respectively, we may write the latitude as
\begin{equation}
\beta = \pm (G+F) \equiv \pm (G+{M\over60}); \qquad +/-
\mathrm{for}\quad S = \,\mathrm{bo/no}   \label{e:lat}
\end{equation}

The magnitudes range from 1 to 6, occasionally qualified with
m[eizoon] (greater, i.e.\ brighter) or el[assoon] (less, i.e.\
fainter).  The magnitude of some stars is indicated `faint'
(`amauros'), a nebulous star as nebel[oeides] .
 
At the end of each constellation the total number of stars, and the
numbers of stars for each magnitude are given. 

\begin{table}
\caption{Constellations in the catalogue of Ptolemaios \label{t:ptol}}
\begin{tabular}{rc@{\hspace{0.2cm}}c@{\hspace{0.2cm}}rr|rc@{\hspace{0.2cm}}c@{\hspace{0.2cm}}rr}
C & abbv & $N$ & $N_\mathrm{a}$ & P & C & abbv & $N$ & $N_\mathrm{a}$ & P \\
\hline
 1 & UMi &  7 &  1 &    1 & 26 & Leo & 27 &  8 &  462 \\
 2 & UMa & 27 &  8 &    9 & 27 & Vir & 26 &  6 &  497 \\
 3 & Dra & 31 &  0 &   44 & 28 & Lib &  8 &  9 &  529 \\
 4 & Cep & 11 &  2 &   75 & 29 & Sco & 21 &  3 &  546 \\
 5 & Boo & 22 &  1 &   88 & 30 & Sgr & 31 &  0 &  570 \\
 6 & CrB &  8 &  0 &  111 & 31 & Cap & 28 &  0 &  601 \\
 7 & Her & 29 &  1 &  119 & 32 & Aqr & 42 &  3 &  629 \\
 8 & Lyr & 10 &  0 &  149 & 33 & Psc & 34 &  4 &  674 \\
 9 & Cyg & 17 &  2 &  159 & 34 & Cet & 22 &  0 &  712 \\
10 & Cas & 13 &  0 &  178 & 35 & Ori & 38 &  0 &  734 \\
11 & Per & 26 &  3 &  191 & 36 & Eri & 34 &  0 &  772 \\
12 & Aur & 14 &  0 &  220 & 37 & Lep & 12 &  0 &  806 \\
13 & Oph & 24 &  5 &  234 & 38 & CMa & 18 & 11 &  818 \\
14 & Ser & 18 &  0 &  263 & 39 & CMi &  2 &  0 &  847 \\
15 & Sge &  5 &  0 &  281 & 40 & Arg & 45 &  0 &  849 \\
16 & Aql &  9 &  6 &  286 & 41 & Hya & 25 &  2 &  894 \\
17 & Del & 10 &  0 &  301 & 42 & Crt &  7 &  0 &  921 \\
18 & Equ &  4 &  0 &  311 & 43 & Crv &  7 &  0 &  928 \\
19 & Peg & 20 &  0 &  315 & 44 & Cen & 37 &  0 &  935 \\
20 & And & 23 &  0 &  335 & 45 & Lup & 19 &  0 &  972 \\
21 & Tri &  4 &  0 &  358 & 46 & Ara &  7 &  0 &  991 \\
22 & Ari & 13 &  5 &  362 & 47 & CrA & 13 &  0 &  998 \\
23 & Tau & 33 & 11 &  380 & 48 & PsA & 12 &  6 & 1011 \\
24 & Gem & 18 &  7 &  424 \\
25 & Cnc &  9 &  4 &  449 & \multicolumn{2}{c}{total}  & 920 & 108 & 1028  
\end{tabular}
\tablefoot{For each constellation the columns give the sequence number $C$,
  the abbreviation for it that we use, the number of stars in the constellation
  $N$, the number of associated stars outside the constellation figure $N_\mathrm{a}$,
  and the sequence number of the first star in the constellation $P$}
\end{table}
\begin{table}
\caption{Repeated entries in \ptolemaios\label{t:ptoldoub}}
\begin{tabular}{c|c|c}
$\nu^2$\,Boo P\,96 -- P\,147  & $\beta$\,Tau P\,230 -- P\,400 &
$\alpha$\,PsA P\,670-- P\,1011 
\end{tabular}
\end{table}

The catalogue contains three repeated entries
(Table\,\ref{t:ptoldoub}). From the descriptions of the stars it is
clear that Ptolemaios was aware of this. In the translation of Toomer
(1998):\nocite{toomer}
P\,147 `The star on the end of the right leg [of Hercules] is the same
as the one of the tip of the staff [of Bootes].'
P\,400 `The star on the tip of the northern horn [of Taurus], which is
the same as the one on the right foot of Auriga.'
P\,1011 `The star in the mouth [of Piscis Austrinus], which is the
same as the beginning of the water [i.e.\, in Aquarius].''

The epoch given by Ptolemaios for his catalogue is 1 Thoth 885
Nabonassar, which corresponds to 20 July 137 = JD\,1771298.  He notes
in (book VII.3 of) the Almagest that the change in longitude is {\em
  1\degr\ in about 100 years, or 2$\frac{2}{3}$\degr\ in the 265 years
  between Hipparchos' and our observations} (Toomer, 1998, p.333).
Subtraction of 265 Egyptian years of 365 days each then gives the
approximate epoch of the measurements by Hipparchos as 1 Thoth 620
Nabonassar, which corresponds to 24 September $-$128 (=129 B.C.) =
JD\,1674573.

 \begin{figure}
 \centerline{\includegraphics[angle=270,width=\columnwidth]{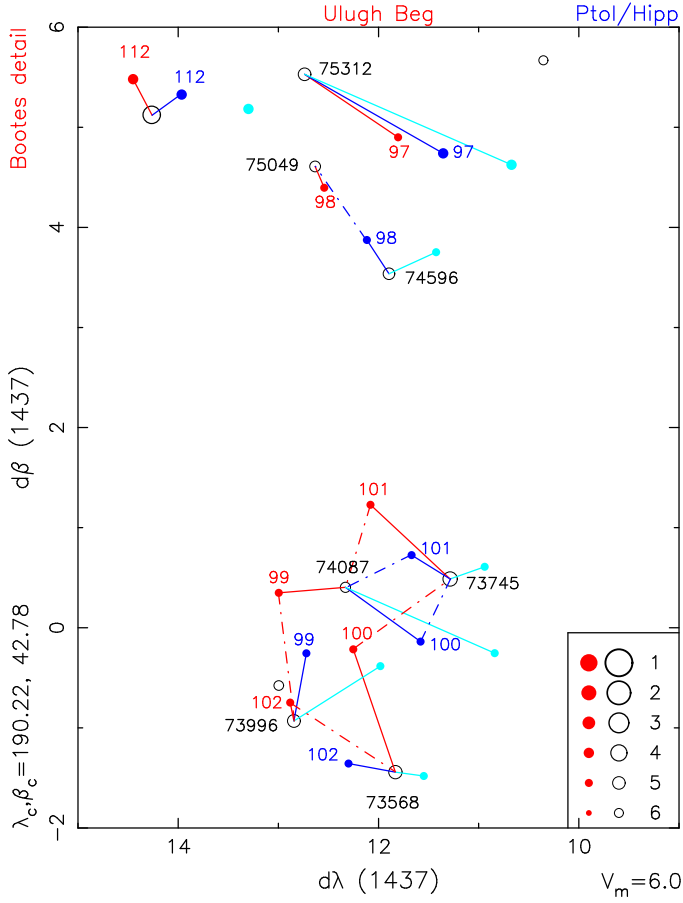}}
 \caption{Detail of the constellation Bootes (see
   Fig.\,\ref{f:bootes}), showing the relative
   positions of the stars in \ulughbeg\ (red), \ptolemaios\ (light
   blue, epoch 137), and \ptolemaios\ after correction to the epoch
  of Hipparcos ($-$128), by subtraction of 2\degr40\arcmin\ from the
  longitude  (dark blue).  Stars from the modern Hipparcos catalogue 
   are indicated with black open circles.
   For illustrative purposes, the positions of these stars and of the stars from both
   epochs of  Ptolemaios were converted with modern precession equations to the
   epoch of Ulugh Beg, 1437.  It is seen that the catalogue positions
   as given by Ptolemaios for 137 are systematically too low.
  Our identifications are indicated with solid lines; those by Knobel
  (1917) and Toomer (1998) with dash-dotted lines,  when different
  from our identifications.
  For the projection used see Sect.\,\ref{s:figures}.
   \label{f:bootesdet}}
 \end{figure}

The positions given in the star catalogue of Ptolemaios show a
systematic offset: its longitudes are on average about 1\degr\ too
small. An example is shown in Fig.\,\ref{f:bootesdet}.  
The difference in longitude due to precession between the
epochs of Ptolemaios and Hipparchos, according to modern computation,
is about 3\degr40\arcmin, about 1\degr\ more than the value given by
Ptolemaios.  This has led to the suggestion that Ptolemaios did not
make independent measurements, but merely copied the catalogue of
Hipparchos, applying a (wrong) correction for precession.  An
alternative explanation is that the zero point of the longitude scale,
i.e.\ the vernal equinox, as used by Ptolemaios is about 1\degr\ off,
due to errors in his solar theory.  Opinion as to which of these two
explanations is the correct one have oscillated ever since Tycho
Brahe, as reviewed by Grasshoff (1990).\nocite{grasshoff}  The ingeneous statistical
analysis by Duke (2003),\nocite{duke} who compares the positions in Ptolemaios to
the times for rising and setting of stars in the {\em Commentary to
  Aratus} by Hipparchos, indicates that Ptolemaios indeed copied most of
his positions from Hipparchos.

\begin{table}
\caption{Constellations in the catalogue of Ulugh Beg \label{t:ulug}}
\begin{tabular}{rc@{\hspace{0.2cm}}c@{\hspace{0.2cm}}rr|rc@{\hspace{0.2cm}}c@{\hspace{0.2cm}}rr}
C & abbv & $N$ & $N_\mathrm{a}$ & U & C & abbv & $N$ & $N_\mathrm{a}$ & U \\
\hline
 1 & UMi &  7 &  1 &    1 & 26 & Leo & 27 &  8 &  459 \\
 2 & UMa & 27 &  8 &    9 & 27 & Vir & 26 &  6 &  494 \\
 3 & Dra & 31 &  0 &   44 & 28 & Lib &  8 &  9 &  526 \\
 4 & Cep & 11 &  2 &   75 & 29 & Sco & 21 &  3 &  543 \\
 5 & Boo & 22 &  1 &   88 & 30 & Sgr & 31 &  0 &  567 \\
 6 & CrB &  8 &  0 &  111 & 31 & Cap & 28 &  0 &  598 \\
 7 & Her & 28 &  1 &  119 & 32 & Aqr & 42 &  3 &  626 \\
 8 & Lyr & 10 &  0 &  148 & 33 & Psc & 34 &  4 &  671 \\
 9 & Cyg & 17 &  2 &  158 & 34 & Cet & 22 &  0 &  709 \\
10 & Cas & 13 &  0 &  177 & 35 & Ori & 38 &  0 &  731 \\
11 & Per & 26 &  3 &  190 & 36 & Eri & 34 &  0 &  769 \\
12 & Aur & 13 &  0 &  219 & 37 & Lep & 12 &  0 &  803 \\
13 & Oph & 24 &  5 &  232 & 38 & CMa & 18 & 11 &  815 \\
14 & Ser & 18 &  0 &  261 & 39 & CMi &  2 &  0 &  844 \\
15 & Sge &  5 &  0 &  279 & 40 & Arg & 45 &  0 &  846 \\
16 & Aql &  9 &  6 &  284 & 41 & Hya & 25 &  2 &  891 \\
17 & Del & 10 &  0 &  299 & 42 & Crt &  7 &  0 &  918 \\
18 & Equ &  4 &  0 &  309 & 43 & Crv &  7 &  0 &  925 \\
19 & Peg & 20 &  0 &  313 & 44 & Cen & 37 &  0 &  932 \\
20 & And & 23 &  0 &  333 & 45 & Lup & 19 &  0 &  969 \\
21 & Tri &  4 &  0 &  356 & 46 & Ara &  7 &  0 &  988 \\
22 & Ari & 13 &  5 &  360 & 47 & CrA & 13 &  0 &  995 \\
23 & Tau & 32 & 11 &  378 & 48 & PsA & 11 &  0 & 1008 \\ 
24 & Gem & 18 &  7 &  421 & \\
25 & Cnc &  9 &  4 &  446 & \multicolumn{2}{c}{total}  & 916 & 102 & 1018  
\end{tabular}
\tablefoot{For each constellation the columns give the sequence number
  $C$, the abbreviation for it that we use, the number of stars in the constellation
  $N$, the number of associated stars outside the figure of the constellation $N_\mathrm{a}$,
  and the sequence number of the first star in the constellation $U$}
\end{table}

\subsection{The star catalogue of Ulugh Beg}

This star catalogue is organized as that of Ptolemaios, giving mostly
the same stars in the same 48 constellations in the same order
(Table\,\ref{t:ulug}). There are some differences, however
(Table\,\ref{t:ptolulug}).  Eleven stars from \ptolemaios\ are absent
in \ulughbeg, including the three repeated entries of \ptolemaios; on
the other hand, the entry P\,657 is split into two stars. As a result,
the total number of entries in \ulughbeg\ is 1018. Of these, one
(U\,961) has no coordinates, as its Ptolemaic original (P\,964) was
not seen by Ulugh Beg. 
27 entries were too far south for Ulugh Beg to measure, and for
these he took over the positions from al-Sufi, adding 6\degr59\arcmin\
to the longitudes to correct for precession; al-Sufi already had added
12\degr42\arcmin\ to the longitudes of Ptolemaios to correct for
precession (Knobel 1917); the correction to the longitudes
between Ulugh Beg and Ptolemaios thus is 19\degr41\arcmin.
These 27 entries include one star, U\,979 = P\,982, which Ulugh Beg
remarks he did not see. 

\begin{table}
\caption{Comparison of entries in \ulughbeg\ and \ptolemaios\label{t:ptolulug}}
\begin{tabular}{cc||cc}
P & U & P & U   \\
P\,147* & absent & P\,884-889,892,893 & copied \\
P\,233 & absent & P\,961-963,965-971 & copied \\ 
P\,400* & absent & P\,964 & empty  \\
P\,434,435 & U\,432,431 &  P\,981,982 & copied \\
P\,611,612 & U\,609,608 & P\,991-997 & copied \\
P\,651 & absent &  P\,1011* & absent\\
P\,657 & U\, 653,654 & P\,1023-1028 & absent \\
P\,665,666 & U\,663,662 \\
\end{tabular}
\tablefoot{P\,964 and P\,982 are annotated by Ulugh Beg with {\em not
    seen}. An asterisk indicates a repeat entry in \ptolemaios.}
\end{table}

The original star catalogue of Ulugh Beg was probably written in
Persian, and Knobel (1917)\nocite{knobel} based his edition on the Persian
manuscripts available to him. The Knobel edition gives for each entry
the sequence number (as in Baily 1843)\nocite{baily43} for the catalogue as a whole,
the sequence number within the constellation, a brief description of
the star, the modern identification, the longitude in zodiacal sign
$Z$ and integer degrees $G$ and minutes $M$, the latitude in sign $S$,
integer degrees $G$ and minutes $M$, and the magnitude.
The sign of the latitude is often omitted,
implicitly assumed to be the same as for the previous entry.
The ecliptic longitude is found from
\begin{equation}
\lambda = Z\times 30 + G + {M\over 60}  \label{e:ublong}
\end{equation}
and the latitude from
\begin{equation}
\beta = \pm(G+{M\over60})\qquad +/- \quad\mathrm{for} \quad S = +/-
\label{e:ublat}
\end{equation}
Note that the indication of the zodiacal sign in \ulughbeg\ differs
from that in \ptolemaios\ by one: for example, a star with longitude
10\degr\ has $Z$=1 in \ptolemaios, and $Z$=0 in \ulughbeg.

The magnitude in \ulughbeg\ is an integer, or occasionally indicated
as two integers bracketing the actual magnitude. For the modern
identifications according to Knobel (1917), see Sect.\,\ref{s:toomerid}.
\nocite{knobel}

The epoch given by Ulugh Beg is the beginning of the Islamic year 841,
i.e.\ 5 July 1437 = JD\,2246108.

\section{Identification procedure \label{s:identification}}

The procedure that we follow for the identification of each star from
the catalogues of Ptolemaios and Ulugh Beg is {\em mutatis mutandis}
identical to the procedure that we followed for the catalogue of
Brahe, and we refer to Verbunt \&\ van Gent (2010a) for details.
Briefly, we select all stars from the {\em Hipparcos Catalogue} with a
Johnson visual magnitude brighter than 6.0, we correct their
equatorial positions for proper motion between the Hipparcos epoch
1991.25 and the epoch of the old catalogue, then precess the resulting
equatorial coordinates from the Hipparcos equinox 2000.0 to the
equinox of the old catalogue, and finally convert the coordinates from
equatorial to ecliptic, using the obliquity appropriate for the old
equinox.

For each entry in the old catalogue we find the nearest -- in terms of
angular separation -- counterpart with $V\leq6.0$ in the Hipparcos
Catalogue. In general, this counterpart is selected by us as a secure
identification, and given an identification flag 1. If a significantly
brighter star is at a marginally larger angular distance, we select
that star as the secure counterpart, and give it flag 2.  Especially
for larger angular distances we may decide that the identification is
uncertain (flag 3); and occasionally several Hipparcos stars appear to
be comparably plausible as counterparts for the same entry (flag
4). An entry for which we do not find a plausible identification is
flagged 5; and an entry which is identified with an Hipparcos star
that already is the identification of another entry -- i.e.\ a repeat
entry -- is flagged 6. This notation is summarized in Table\,\ref{t:idflags}.

\begin{table}[!]
\caption{Meaning of flags I classifying our identifications \label{t:idflags}}
\begin{tabular}{ll}
1 & nearest star, secure identification \\
2 & not nearest star, secure identification \\
3 & probable identification, not secure \\
  & because too far or too faint \\
4 & possible identification \\
  & other identification(s) also possible \\
5 & not identified \\
6 & repeat entry 
\end{tabular}
\end{table}

For \ptolemaios, we decided to use the equinox of Hipparchos,
JD\,1674573 ($-$128), i.e.\ we convert both modern and old positions
to this equinox: the modern Hipparcos positions as indicated above,
and the positions in \ptolemaios\ by subtracting 2\degr40\arcmin\ from
the longitude. These converted positions are then used in the search for
counterparts. For \ulughbeg\ we use the equinox JD\,2246108
(1437).

 \begin{figure}
 \centerline{\includegraphics[angle=270,width=\columnwidth]{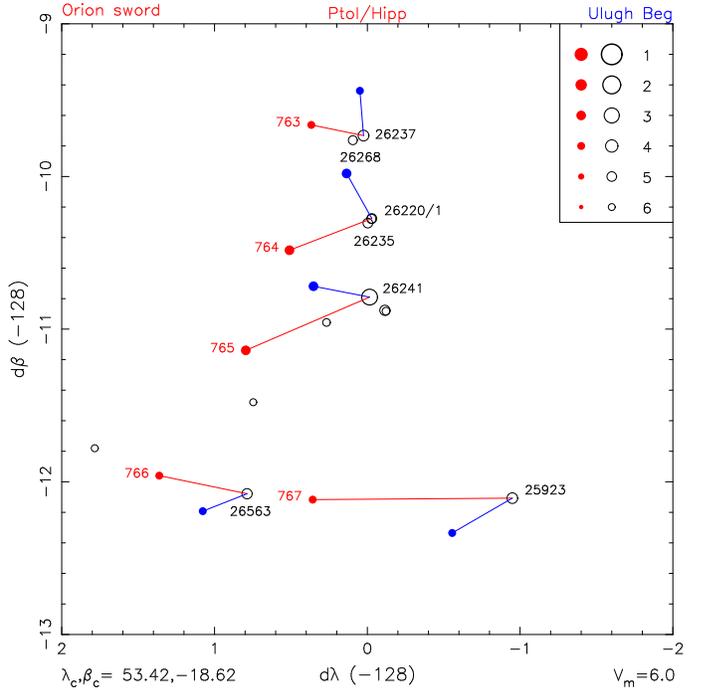}}
 \caption{Detail of the constellation Orion (see Fig.\,\ref{f:orion}) 
  in \ptolemaios\ (red) and \ulughbeg\ (blue). 
   Stars from the modern Hipparcos catalogue are indicated with black
   open circles.
    For comparison purposes the  positions of these stars and of stars from
   \ulughbeg\ have been converted to the epoch of
   Hipparchos/Ptolemaios, with modern precession equations.
   \label{f:orisword}}
 \end{figure}

In making our identifications we not only look at individual stars, but
also at star patterns. Two examples are shown in Fig.\,\ref{f:orisword}:
the middle of three stars in the sword of Orion in \ulughbeg\ is
closer to the modern most northern star, but we identify it with the
middle modern star; and P\,767 is closest to HIP\,26563, but since
this star is already identified with P\,766, we identify P\,767 with
HIP\,25923. Another example is furnished by two stars in Ara in
\ptolemaios, P\,996 and P\,997, near $-$1.5,$-$4.9 and $-$5.0,$-$5.9
in Fig.\,\ref{f:ara} (left), respectively.  Because all other stars in
Ara have identifications to the east (left in the Figure)
of the old catalogue position we identify both stars with a star to
the east as well, HIP\,85258 near 0.4,$-$3.5 and HIP\,83081 near
$-$3.3,$-$4.4; even though P\,996 is closer to HIP\,83081 and P\,997
closer to HIP\,82363 near $-$3.9,$-$7.6.  Figure\,\ref{f:ara} for Ara
in \ulughbeg\ illustrates also how we prefer brighter, further
counterparts to fainter nearby ones.
In very crowded constellations, identifications may be made more
easily if only brighter modern Hipparcos stars are considered, as
may be seen most spectacularly in Argo by comparing Fig.\,\ref{f:argo} 
(limiting magnitude $V$=6) with Fig.\,\ref{f:argbright}
(limiting magnitude $V$=5). Further examples are furnished by
the illustrations for \ptolemaios\ and \ulughbeg\ for  the
constellations Centaurus (Fig.\,\ref{f:centaurus}) and Ara
(Fig.\,\ref{f:ara}). This selection between only bright stars
makes us confident that we have often found the correct identification, 
even in crowded constellations.

Nonetheless, it must be noted that our identification flags are to
some extent subjective. For example, P\,41 near $-$14.1,$-$12.7 in
Fig.\,\ref{f:ursamaior} (left) is identified by us with HIP\,44248
with the flag `probable', but we might have chosen to leave it
unidentified, or even to identify it with the closer HIP\,47029, near
$-$10.7,$-$12.1, the counterpart of the corresponding entry in
\ulughbeg. 

 \begin{figure}
 \centerline{\includegraphics[angle=270,width=\columnwidth]{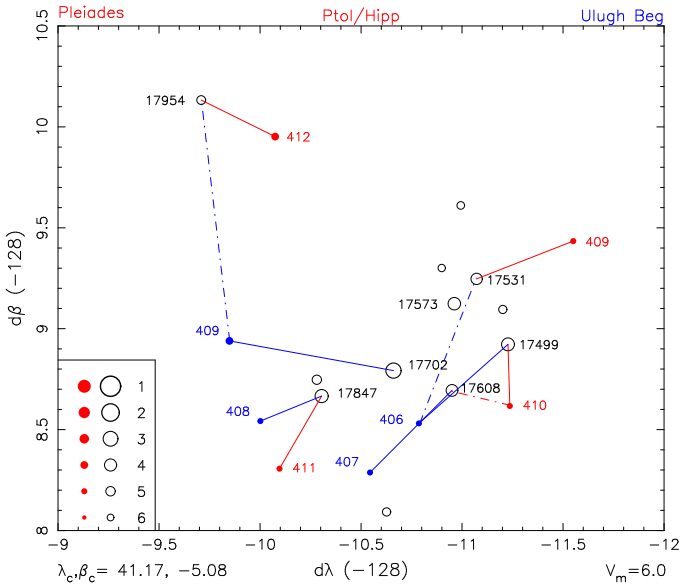}}
 \caption{Stars in the Pleiades in \ptolemaios\ (red) and \ulughbeg\
   (blue), together with stars according to the modern Hipparcos
   Catalogue (open circles). (See Figs.\,\ref{f:taurus} and \ref{f:taucent}.)
   For comparison purposes the positions of these stars and of
   stars from \ulughbeg\ have been
   converted to the epoch of Hipparcos/Ptolemaios, with modern
   precession equations. 
   Solid lines indicate our identifications,
   dash-dotted lines the identifications by Toomer (1998) and
   Knobel (1917), when these differ from our identifications.
   \label{f:pleiades}}
 \end{figure}

The identifications of the stars in the Pleiades gives a further
illustration of the ambiguities that occasionally occur (see
Fig.\,\ref{f:pleiades}. P\,409, P\,411 and P\,412 are identified 
with relative ease, but P\,410 is ambiguous: we choose the brighter
HIP\,17499 as the counterpart, in agreement with the
description `the sourthern end of the advance side', but Toomer (1998)
prefers the closer albeit fainter HIP\,17608. The descriptions of
the stars in the Pleiades by Ulugh Beg is virtually identical to those
by Ptolemaios, but the positions differ markedly, which leads us
to identify three stars in \ulughbeg\ differently from their
counterparts in \ptolemaios. Thus U\,409 is identified to HIP\,17702,
its counterpart P\,412 to HIP\,17954. Identification with HIP\,17702
(Alcyone) is in accordance with it being the brightest star in the  Pleiades,
in accordance with the magnitude 4 assigned to U\,409/P\,412
both in \ptolemaios\ and in \ulughbeg, which give the three other
Pleiades members magnitude 5. Identification with HIP\,17954
fits better with the description `the small star outside the Pleiades
towards the North', and Knobel identifies U\,409 accordingly.

\subsection{Identifications by Toomer and Knobel \label{s:toomerid}}

Toomer (1998) and Knobel (1917) give Bayer names and/or Flamsteed
numbers, and occasionally an HR (Bright Star Catalogue) number, as
modern identifications.  We convert these to HD numbers through the
{\em Bright Star Catalogue} (Hoffleit \&\ Warren 1991) and the HD
numbers to Hipparcos numbers with the modern Hipparcos catalogue (ESA,
1997).\nocite{toomer}\nocite{knobel}\nocite{hoffleit}\nocite{esa} For
those cases where the Flamsteed number is not given in the {\it Bright Star
Catalogue}, we convert stars from the Hipparcos catalogue to the epoch
of Flamsteed, 1690.0 = JD\,2338331 and identify the position in the
catalogue of Flamsteed (1725) with the nearest Hipparcos match, which 
is always within 2\arcmin.

In a number of cases Knobel gives an identification consisting of a
roman and an arabic number; in two cases a zero and a number. These
refer to the catalogue of Piazzi (1803; we use the 1814
reprint).\nocite{piazzi} To find the corresponding \textit{Hipparcos}
numbers we convert the positions of all \textit{Hipparcos} entries
brighter than $V=6.0$ to the equinox of the Piazzi catalogue 1800.0 =
JD\,2378497 and find the nearest \textit{Hipparcos} entry to each of
the Piazzi stars.  This leads to an unambiguous identification in all
cases with positional differences of about 0\farcm2 or less.

In a number of cases Knobel gives an identification from Lacaille
(Baily 1847).\nocite{baily}  We convert Hipparcos entries to the epoch of this
catalogue, JD\,2360235, to identify Lacaille numbers with Hipparcos
stars.  Knobel identifies U\,867 with Brisbane\,2249. This star is
identified in Baily (1847) with Lac.\,3580 and we identify it through
this with HIP\,43603.
 
In two cases Knobel gives an identification to a star in the edition
by Weisse (1846)\nocite{weisse} of observations by Bessel, viz.\ W.B.\,2h788 for
U\,785 and W.B.\,9h439 for U\,901, which by converting Hipparcos 
entries to the epoch of this catalogue, JD\,2387628, we identify with 
HIP\,13421 and HIP\,46404, respectively.

For one star the only identification given by Knobel is from Hevelius
(U\,217 = 14H Cam). We convert this to HIP\,19949 using our
edition of the Hevelius Catalogue (Verbunt \&\ Van Gent 2010b).
\nocite{verbunt10b}

\section{The machine-readable catalogues}

\begin{table*}
\caption{The machine-readable catalogue of Ptolemaios \label{t:ptolmach}}
\begin{tabular}{rr@{\hspace{0.1cm}}cr@{\,}lr@{\hspace{0.2cm}}r@{\hspace{0.2cm}}rr@{\hspace{0.2cm}}r@{\hspace{0.2cm}}lr@{\,}lr@{\hspace{0.2cm}}r@{\hspace{0.2cm}}rrrrrc}
  P  & C &   abb.  &  i\phantom{i}  & F & Z & G & M & G & M &S&  V & q & HIP & I & T &
  $V_\mathrm{H}$ & $\Delta\lambda$ & $\Delta\beta$ & $\Delta$(\arcmin)
  & a \\   
\hline
   1 & 1 & =UMi &  1& &  3& 00 &10& 66 &00&B & 3&& 11767& 1& 1& 2.0&   88.8& $-$9.4& 37.4\\
   2 & 1 & =UMi &  2& &  3& 02 &30& 70 &00&B & 4&& 85822& 1& 1& 4.3&  108.5&  $-$18.4& 41.6\\
   3 & 1 & =UMi &  3& &  3& 10 &10& 74 &20&B & 4&& 82080& 1& 1& 4.2 & 116.9 & $-$41.5& 52.6\\
   4 & 1 & =UMi &  4& &  3& 29 &40& 75 &40&B & 4&& 77055& 2& 1& 4.3& 21.9 & $-$48.4& 48.7\\
   5 & 1 & =UMi &  5& &  4& 03 &40& 77 &40&B & 4&& 79822& 1& 1& 4.9&  $-$28.7&  3.1&  6.8\\
   6 & 1 & =UMi &  6& &  4& 17 &30& 72 &50&B & 2&& 72607& 1& 1& 2.1& $-$101.7& $-$2.0& 30.1\\
   7 & 1 & =UMi &  7& &  4& 26 &10& 74 &50&B & 2&& 75097& 2& 1& 3.0& $-$136.1& 14.5& 38.2\\
   8 & 1 & =UMi &  8&a & 4& 13& 00& 71&10&B & 4&& 70692& 1& 1& 4.3& $-$119.5&  3.1& 38.7\\
\multicolumn{5}{l}{\ldots}\\
  18 & 2 & =UMa & 10 & &4 &11 &00& 44 &00&B & 4&f&  48402& 2& 1&  4.6&
  78.4 &$-$358.0  &362.8 & * \\
\end{tabular}
\tablefoot{For explanation of the columns see Sect.\,\ref{s:ptolmach}}
\end{table*}

\begin{table*}
\caption{The machine-readable catalogue of Ulugh Beg \label{t:ulugmach}}
\begin{tabular}{r@{\hspace{0.1cm}}crr@{\hspace{0.1cm}}cr@{\,}lr@{\hspace{0.2cm}}r@{\hspace{0.2cm}}rr@{\hspace{0.2cm}}r@{\hspace{0.2cm}}cr@{\,}lr@{\hspace{0.2cm}}r@{\hspace{0.2cm}}r@{\hspace{0.2cm}}rrrrrc}
  U & $u$ &P  & C & abb.  &  i  & F & Z & G & M & G & M & S & V &q & HIP & I & K
  & $I_P$ &  $V_\mathrm{H}$ & $\Delta\lambda$ & $\Delta\beta$ &
  $\Delta$(\arcmin) & a\\   
\hline
   1 &&   1  & 1 &=UMi & 1 &&   2 &20& 19  & 66& 27 &B& 3&&   11767& 1& 1& 1 & 2.0&   23.4 & $-$24.9 &  26.7\\
   2 & &  2  & 1 &=UMi & 2 & &  2 &22& 25  & 70& 00 &B& 4&&   85822& 1& 1& 1 & 4.3 &  55.8  & $-$7.2  & 20.4 \\
   3  & & 3  & 1 &=UMi & 3  & & 3 &00& 55 &  73& 45 &B& 4  && 82080& 1& 1& 1 & 4.2 &  19.6  &  6.0  &  8.1 \\
   4  & & 4 &  1 &=UMi&  4  & & 3 &17& 43&   75& 36 &B& 4&&   77055& 1& 1& 1&  4.3 & 102.1 & $-$32.6 &  41.6 \\
\multicolumn{5}{l}{\ldots}\\
   8  & & 8 &  1& =UMi  &8&a&   4& 00& 55 &  71& 45 &B& 4&&   70692& 1& 1&1&  4.3 & $-$30.4 & $-$21.6  & 23.7 \\
   9 &&   9 &  2& =UMa&  1&&    3 &14& 55&   40& 15& B& 4&&   41704& 1& 1& 1& 3.3  & 13.4 &  $-$3.0&10.6 \\
  10 & &10 &  2& =UMa&2&&    3& 15& 43&   43& 48& B&  5&&   42080& 3& 1& 1& 5.5& $-$117.7& 43.4& 94.9& *\\
\multicolumn{5}{l}{\ldots}\\
  21 & & 21 &  2& =UMa &13 &&   3& 25& 43 &  29& 00&B&  3&f&  44471& 1 &1& 1&  3.6&   21.0  & $-$4.1 &  18.9\\
  22  && 22&   2& =UMa &14 &&   3& 25& 16&   36& 00 &B& 5&b&  45493&  1& 1& 1&  4.8 &   9.6&    2.2   & 8.1\\
\multicolumn{5}{l}{\ldots}\\
 881&c&  884&  40& =Arg &36&&4& 28 &11&   69& 40 &A&  4&&   38827 &2& 1& 1&  3.5& $-$306.0 & -41.5&  112.5\\
\end{tabular}
\tablefoot{For explanation of the columns see Sect.\,\ref{s:ulugmach}}
\end{table*}

\subsection{The star catalogue of Ptolemaios\label{s:ptolmach}}

The machine-readable table \ptolemaios\ contains the following
information (see Table\,\ref{t:ptolmach}). The first column gives the
sequence number P. The second and third column give the sequence
number of the constellation $C$ and the abbreviation of the
constellation name.  The fourth column gives the sequence number $i$
within the constellation; a star outside the constellation figure is
flagged 'a' in column five.  Columns 6, 7 and 8 give the ecliptic
longitude in zodiacal sign $Z$, degrees ($G$) and minutes ($M$), and
columns 9, 10 and 11 the latitude in degrees ($G$), minutes ($M$), and
sign $S$. These may be converted to longitude and latitude with
Eqs.\,\ref{e:long},\ref{e:lat}, where sign B stands for bo[reios] and A
for no[tios].  Column 12 gives the magnitude $V$ according to
Ptolemaios, and column 13 the qualifier $q$, usually blank, but an f or b
for fainter or brighter, respectively. The stars indicated by Ptolemaios as
`faint' are written as magnitude 7 in the machine-readable catalogue,
and the nebulous stars as 9.

Columns 14-20 provide additional information from our analysis, viz.\
the Hipparcos number of our identification HIP, the flag I indicating
the quality of the identification, the flag T which compares our
identification with that by Toomer (see Table\,\ref{t:toomer}), the
visual (Johnson) magnitude $V_\mathrm{H}$ given in the {\em Hipparcos
  Catalogue} for our identification, the differences in longitude
$\Delta\lambda$ and latitude $\Delta\beta$ in minutes as tabulated, and the angle
$\Delta$ between the catalogue entry and our Hipparcos identification,
in arcminutes (\arcmin).  If the catalogue entry in degrees as
given by \ptolemaios\ is $\lambda$, $\beta$ and the value computed from the
position and proper motion in the {\em Hipparcos Catalogue}
$\lambda_\mathrm{HIP}$, $\beta_\mathrm{HIP}$, then columns 18 and 19 give
give $60(\lambda_\mathrm{HIP}-\lambda)$ and
$60(\beta_\mathrm{HIP}-\beta)$.

Column\,21 indicates with an asterisk those entries which are annotated
in Appendix \ref{s:idptol}.

\subsection{The star catalogue of Ulugh Beg\label{s:ulugmach}}

The machine-readable table \ulughbeg\ contains the following
information (see Table\,\ref{t:ulugmach}). The first column gives the
sequence number U, the second column a flag $u$ which is set to 'c' when
the position of the entry is stated by Ulugh Beg to be derived from the catalogue
of Ptolemaios via al-Sufi. Column 3 gives the P number of the corresponding entry
in \ptolemaios.  The fourth and fifth column give the sequence number
of the constellation $C$ and the abbreviation of the constellation
name.  The sixth column gives the sequence number $i$ within the
constellation; a star outside the constellation figure is flagged 'a'
in column seven.  Columns 8, 9 and 10 give the ecliptic longitude in
zodiacal sign $Z$, degrees ($G$) and minutes ($M$), and columns 11, 12
and 13 the latitude in degrees ($G$), minutes ($M$), and sign $S$. A
These may be converted to longitude and latitude with
Eqs.\,\ref{e:ublong},\ref{e:ublat}, where sign B stands for + and A
for $-$, respectively. In our machine-readable catalogue, we
follow strictly the convention for $Z$ of (the Knobel edition
of) the catalogue of Ulugh Beg, and thereby accept that it differs from the
convention in most other ancient catalogues, e.g.\ that of Ptolemaios,
as reflected in the difference between Eqs.\,\ref{e:long} and \ref{e:ublong}.
Column 14 gives the magnitude $V$ according to
Ulugh Beg / al-Sufi, and column 15 the qualifier $q$, usually blank, but an f or b
for fainter or brighter, respectively.  Note that Ulugh Beg (in the
Knobel 1917 edition) indicates intermediate magnitude values by giving
two integers, either $n$ -- $n$+1, which we indicate $n$ f in columns 14,
15; or $n$+1 -- $n$, which we indicate $n$+1 b in columns 14, 15.
For example, the magnitude of U\,5 is indicated by Knobel as 5--4,
which in the machine-readable table is given as 5 b; for U\,15
magnitude 4--5 is given as 4 f. 

Columns 16-23 provide additional information from our analysis, viz.\
the Hipparcos number of our identification HIP, the flag I indicating
the quality of the identification, the flags K which compares our
identification with that by Knobel and $I_P$ which indicates whether
the corresponding entries in \ulughbeg\ and \ptolemaios\ have the same
Hipparcos counterpart (see Table\,\ref{t:knobel}), the visual (Johnson)
  magnitude $V_\mathrm{H}$ given in the {\em Hipparcos Catalogue} for
  our identification, the differences in longitude $\Delta\lambda$ and
  latitude $\Delta\beta$ in minutes as tabulated, and the angle $\Delta$ between
  the catalogue entry and our Hipparcos identification, in arcminutes
  (\arcmin).  If the catalogue entry  in degrees as given by Ulugh
  Beg is $\lambda$, $\beta$ and the value computed from the position and
  proper motion in the {\em Hipparcos Catalogue}
  $\lambda_\mathrm{HIP}$, $\beta_\mathrm{HIP}$,
  then columns 21 and 22 give $60(\lambda_\mathrm{HIP}-\lambda)$
  and $60(\beta_\mathrm{HIP}-\beta)$.

Column\,24 indicates with an asterisk those entries which are annotated
in Appendix \ref{s:idulug}

\section{Results}

\subsection{\ptolemaios \label{s:results}}

It is indicative of the high quality of the (reconstructed) catalogue
of Ptolemaios that the number of identifications we consider secure
(flags 1-2) is 1009, not counting the 3 repeat entries. For 15 entries
we have a probable counterpart, and only one entry we leave
unidentified.  The large agreement between our identifications and
those of Toomer (1998)\nocite{toomer} for \ptolemaios\ as seen in
Table\,\ref{t:toomer} also suggests that most identifications may be
considered secure.  The cases where we disagree with the
identification by Toomer (1998) usually arise when he prefers a
fainter star (see for example the annotations with P\,98, P\,132 and
P\,152 in Sect.\,\ref{s:idptol}).  Occasionally the descriptions of
the stars in \ptolemaios\ do not match the
positions, and if followed may lead to a different identification,
as for P\,410 in the Pleiades (Fig.\,\ref{f:pleiades}), or to a
permutation of identifications, as for P\,100 and P\,101 in Bootes
(Fig.\,\ref{f:bootesdet}).

\begin{table}
\caption{Frequency of flags T of identifications by Toomer
(1998) as a function of our flags I. \label{t:toomer}}
\begin{tabular}{l|rrrrr}
I\verb+\+T & 0 & 1 & 2 & 3 & all \\
\hline 
1 &   0 & 807 &   2 &  23 & 832 \\
2 &   1 & 168 &   1 &   7 & 177 \\
3 &   0 &   2 &   1 &  11 &  14 \\
4 &   0 &   0 &   1 &   0 &   1 \\
5 &   0 &   0 &   0 &   1 &   1 \\
6 &   0 &   2 &   1 &   0 &   3 \\

all & 1 & 979 & 6 & 42 & 1028 
\end{tabular}
\tablefoot{The meanings of flags $I$ are explained in
  Sect.\,\ref{s:identification}, those of flags $T$ are as follows:
  0 unidentified in Toomer (1998)\nocite{toomer}, 1 Toomer gives same identification 
  as we do, 2 Toomer choses one of two plausible identifications
  and we the other, 3 we reject the identification given by
  Toomer.}
\end{table}

\begin{figure}
\includegraphics[angle=270,width=\columnwidth]{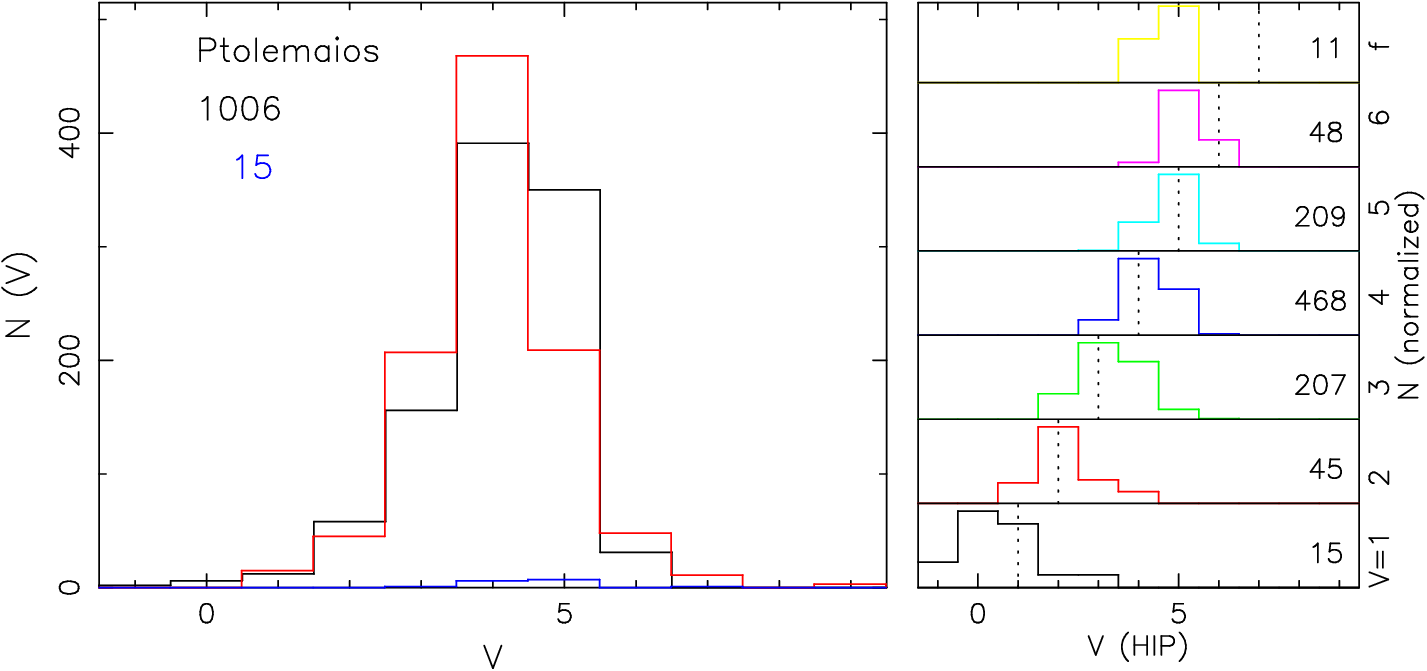} 

\centerline{
\includegraphics[angle=270,width=0.5\columnwidth]{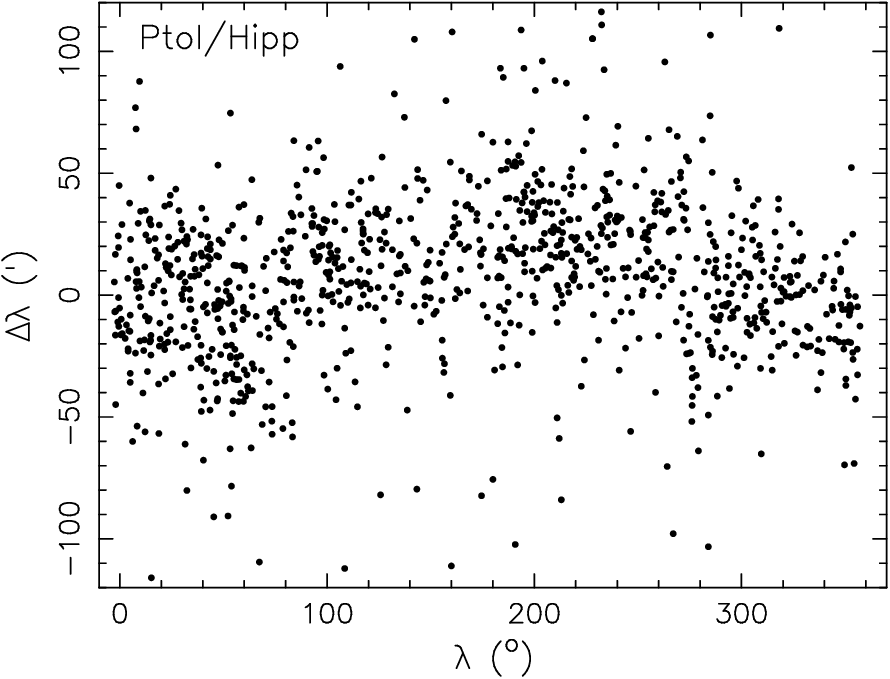} \hfill
\includegraphics[angle=270,width=0.5\columnwidth]{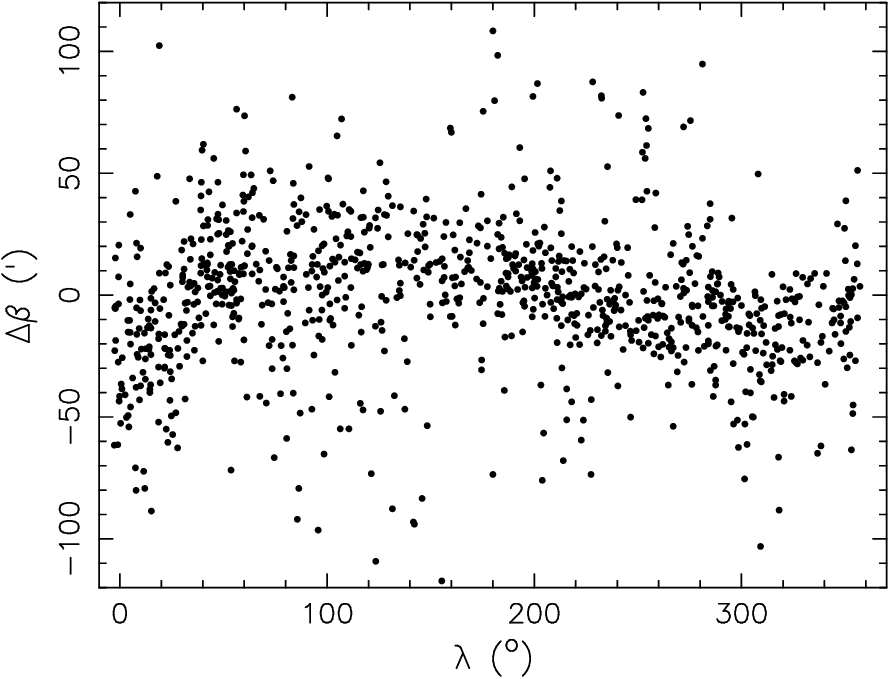} 
} 

\includegraphics[angle=270,width=\columnwidth]{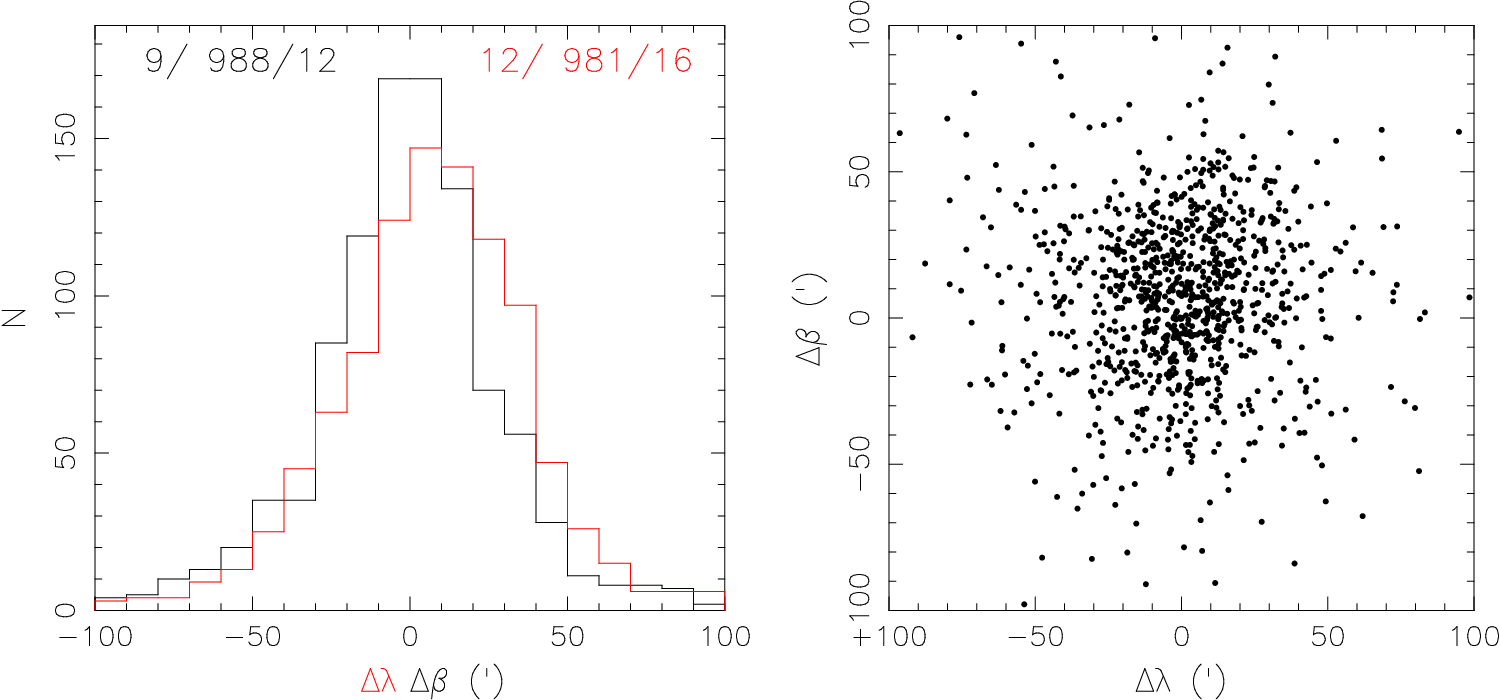} 

\includegraphics[angle=270,width=\columnwidth]{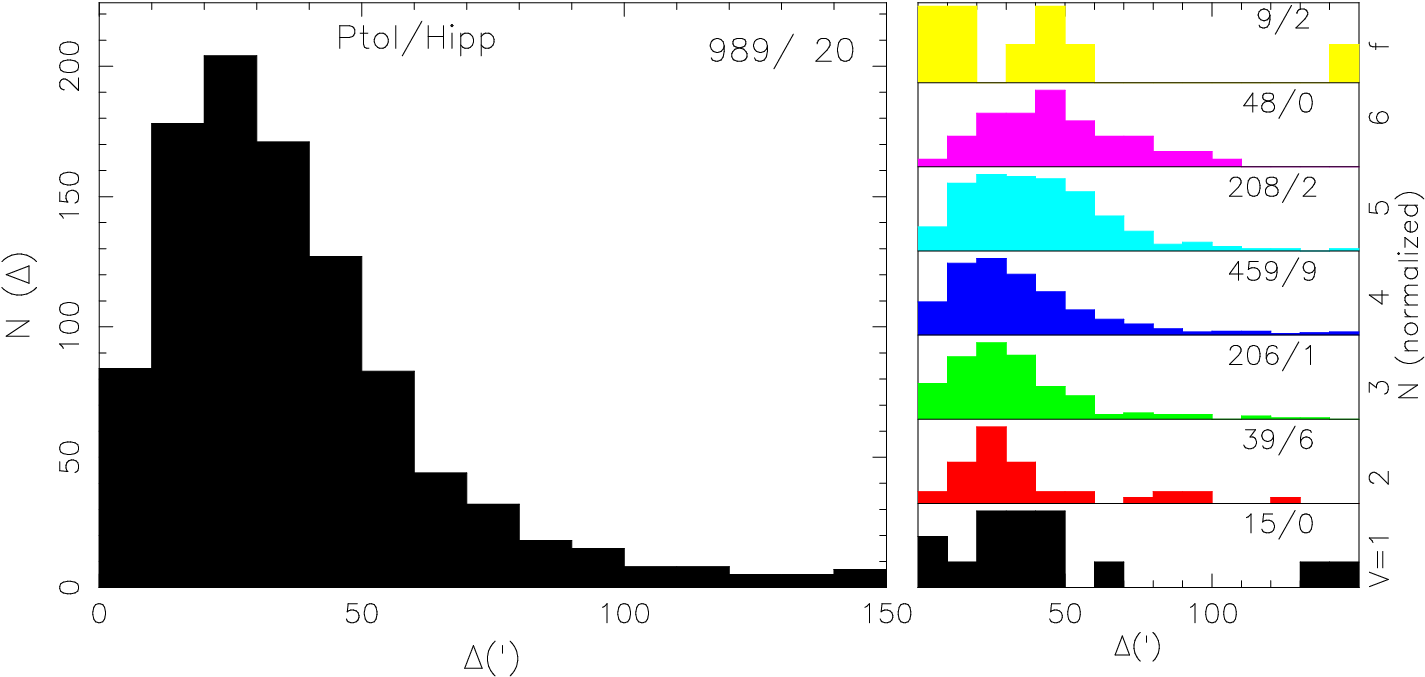} 

\caption{Magnitude and position errors and their correlations of
   \ptolemaios. The top graph shows the distributions of magnitudes
according to Ptolemaios (7=faint, 9=nebulous) for securely identified stars (identification
flags $I-1,2$, red), for not securely identified stars (flags 3,4, blue),
and of the magnitudes in the Hipparcos catalogue for all
secure identifications (black) in the large frame, and for each
magnitude according to Ptolemaios in the small frames. 
The middle
frames show the errors in longitude and latitude as a function of
longitude and of one another, and histograms of the position errors
of the securely identified stars separately for the longitude (red)
and latitude (black). The numbers indicate stars within the frame
(middle) and outside the frame to the left or right. The lower frame
shows the distribution of the position angles $\Delta$, for all
securely identified entries in the large frame, of for each magnitude
according to Ptolemaios separately in the small frames. The numbers
indicate the number of entries within/outside the frame.
\label{f:vdptol}}
\end{figure}

Figure\,\ref{f:vdptol} shows that the correlation between the
magnitudes as given by Ptolemaios and the magnitudes from modern
measurements is good. Note that in this figure we ignore the
qualifiers fainter and brighter made to some magnitudes (q in
Table\,\ref{t:ptolmach}). Remarkably, the 11 stars indicated by
Ptolemaios as `faint' are not fainter than those given magnitudes 5 or
6 by him. The errors $\Delta\lambda$ in longitude and $\Delta\beta$
latitude show systematic trends with longitude and latitude. The
errors $\Delta\lambda$ and $\Delta\beta$ are not correlated; the
spread in errors at each $\lambda$ is slightly larger for
$\Delta\lambda$ than for $\Delta\beta$.  

Using maximum likelihood (i.e.\ Poisson statistics), we fit gaussians
to the histograms of the errors for all errors in the range
$-$100\arcmin\ to +100\arcmin\ (and a more limited range $-$50\arcmin\
and +50\arcmin), and find an offset of about 7\arcmin\ (9\arcmin) and
a width $\sigma=$30\arcmin\ (27\arcmin) for $\Delta\lambda$; and an
offset of $-$0.7\arcmin\ (+0.3\arcmin) and a width $\sigma=$29\arcmin\
(23\arcmin) for $\Delta\beta$. These widths are slightly larger than
those found by Shevchenko (1990, for zodiacal stars only) and by Schwan
(2002),\nocite{schwan} who subtracted the systematic trends in the
errors before fitting gaussians to the remaining spread.  (Shevchenko
gives rms-errors, excluding outliers from the computations; the
article by Schwan, written for a semi-popular journal, gives no
details on his determination of $\sigma$.)  The total error $\Delta$ peaks
between 20\arcmin\ and 30\arcmin; the errors $\Delta$ increase only
slowly with magnitude.

 \begin{figure}
\includegraphics[angle=270,width=\columnwidth]{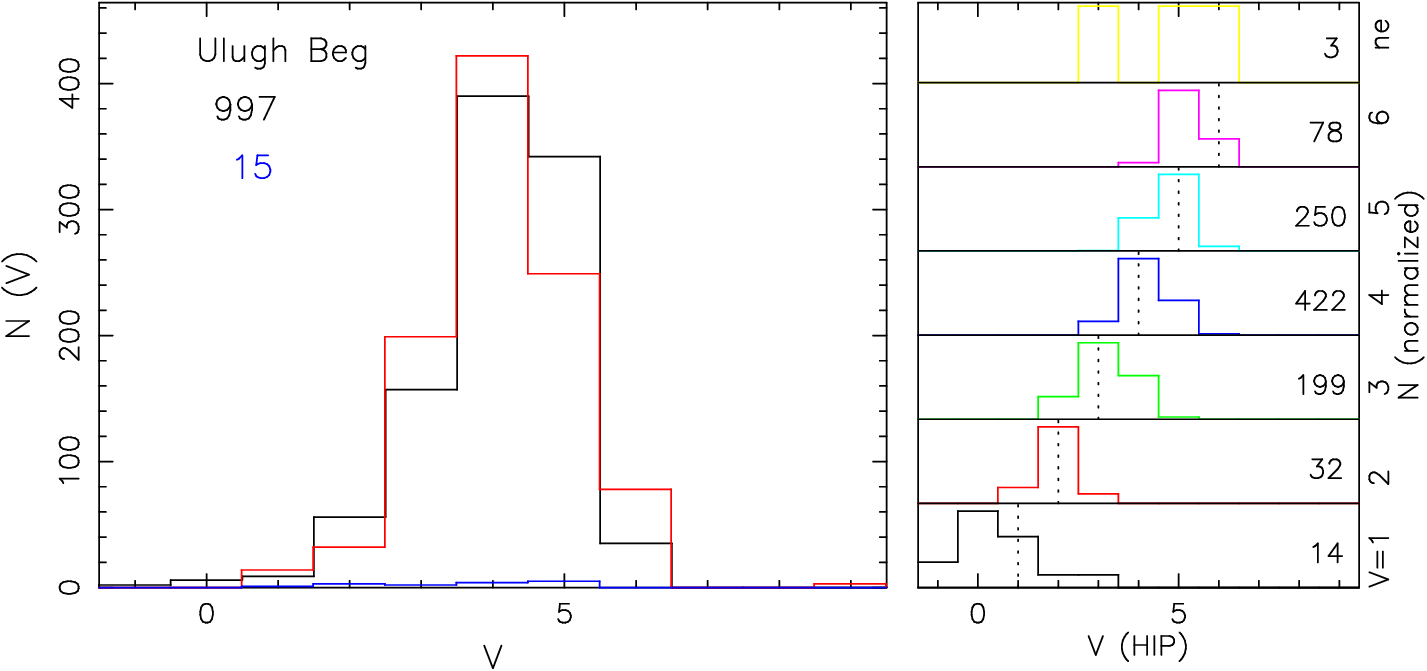}

\centerline{
\includegraphics[angle=270,width=0.5\columnwidth]{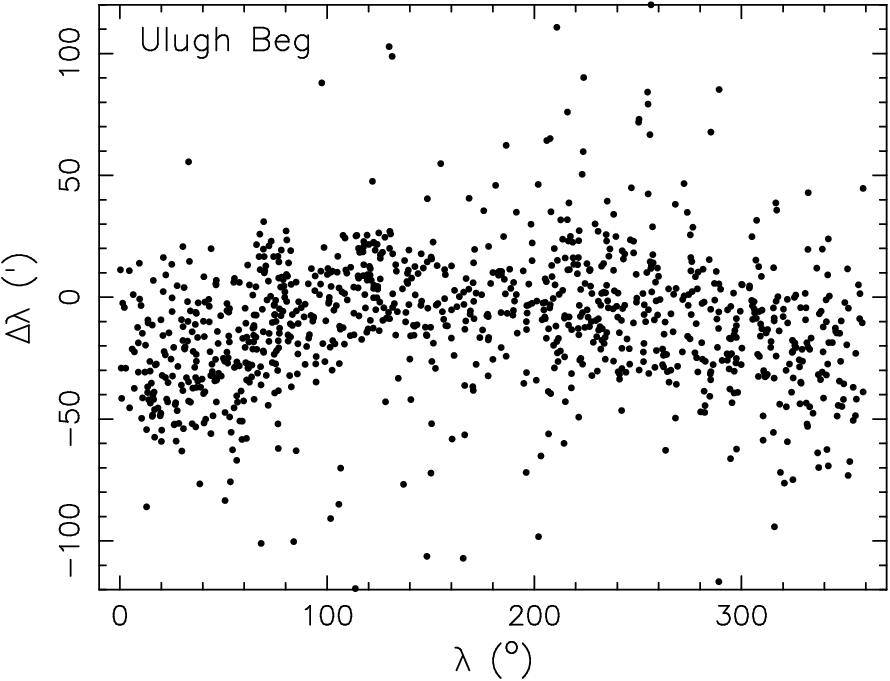} \hfill
\includegraphics[angle=270,width=0.5\columnwidth]{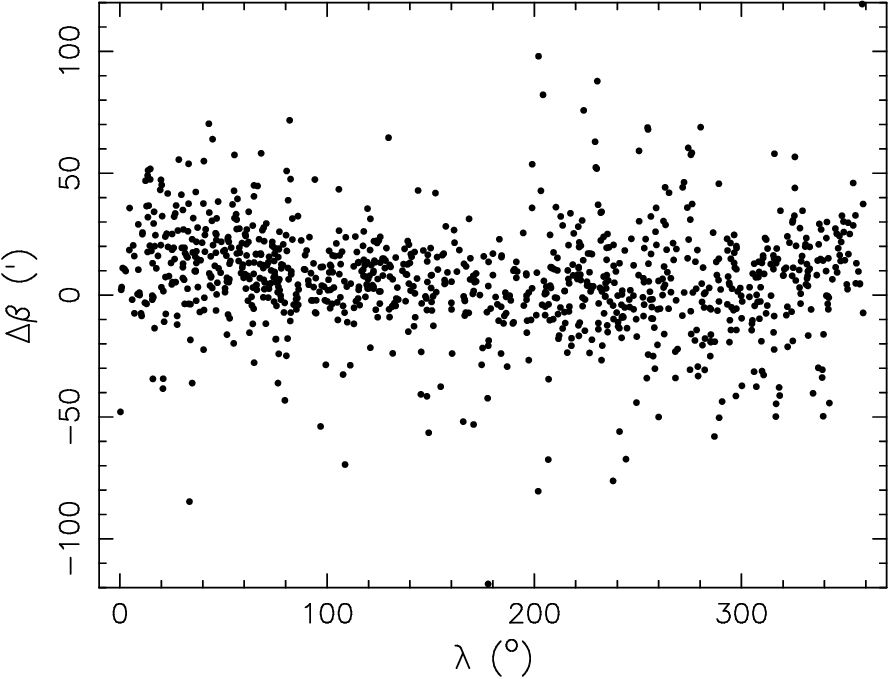} 
} 

\includegraphics[angle=270,width=\columnwidth]{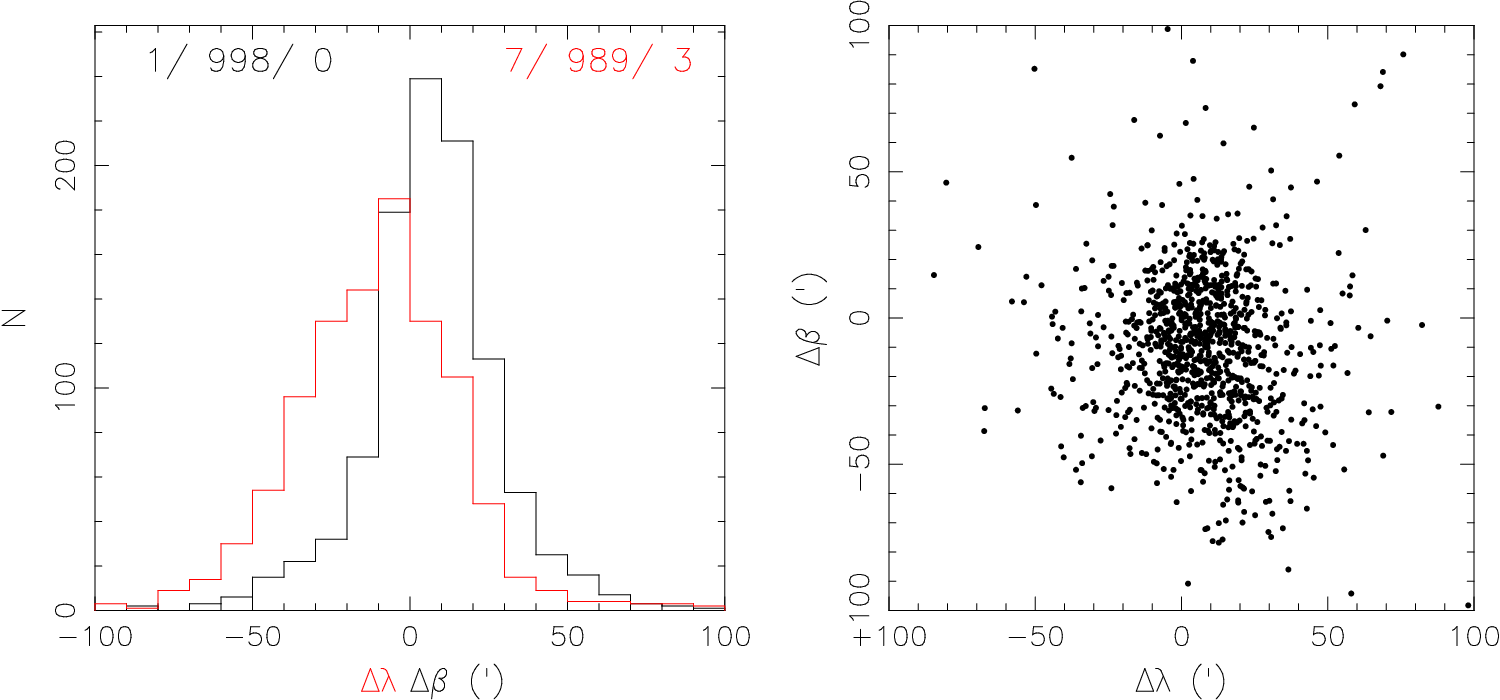} 

\includegraphics[angle=270,width=\columnwidth]{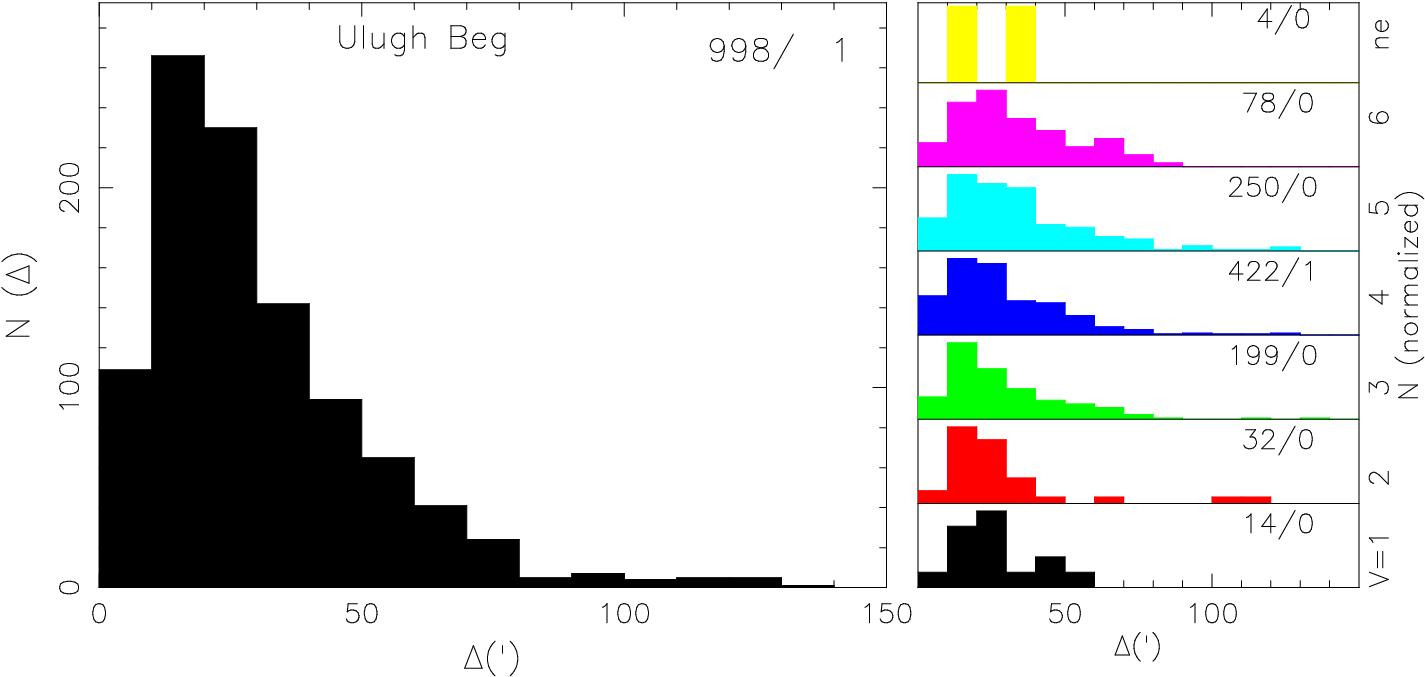} 

 \caption{Magnitude and position errors and their correlations of
 \ulughbeg. As Figure\,\ref{f:vdptol}, now for \ulughbeg.  \label{f:vdulug}}
 \end{figure}

\begin{table}
\caption{Frequency of flags K of identifications by Knobel 
(1917) and $I_P$ of identification in \ptolemaios\ as a function of our flags I. \label{t:knobel}}
\begin{tabular}{l||rrrrr||rrrrr}
 & \multicolumn{5}{c}{flag Knobel}  & \multicolumn{5}{c}{flag Ptolemaios} \\
I & 0 & 1 & 2 & 3 & all & 0 & 1 & 2 & 3 & all \\
\hline 
1 &   0 & 858 &   3 &   9 & 870 &  1 & 826 &   6 &  37 & 870 \\
2 &   0 & 121 &   0 &   8 & 129 &  0 & 121 &   0 &   8 & 129 \\
3 &   0 &  10 &   0 &   1 &  11 &   0 &   9 &   0 &   2 &  11 \\
4 &   0 &   0 &   3 &   1 &   4 &   0 &   0 &   4 &   0 &   4 \\
5 &   0 &   0 &   0 &   3 &   3 &   0 &   0 &   0 &   3 &   3 \\
6 &   0 &   0 &   0 &   0 &   0 &   0 &   0 &   0 &   0 &   0 \\
all & 0 & 989 & 6 & 22 & 1017 & 1 & 956 & 10 & 50 & 1017
\end{tabular}
\tablefoot{The meanings of flags $I$ are explained in
Sect.\,\ref{s:identification}. Those of flags $K$/$I_P$ are as follows:
0 unidentified in Knobel/\ptolemaios, 1 Toomer/\ptolemaios\ gives the same identification 
as we do, 2 Knobel/\ptolemaios\ choses one of two plausible identifications
and we the other, 3  the identification is different from that given by
Knobel/\ptolemaios.  U\,961, which has no coordinates, is excluded.}
\end{table}

\subsection{\ulughbeg \label{s:resultsub}}

We have securely identified 999 entries in \ulughbeg\ and tentatively
another 15 entries; 3 entries we leave unidentified
(Table\,\ref{t:knobel}).  In 989 cases our identification agrees with
that by Knobel (1917),\nocite{knobel} in 6 cases we consider the identification by
Knobel plausible even if different from ours, and in 22 cases we think
our identification is better. The constellation Bootes
provides an example where the identifications by Knobel (1917) are a
permutation of our preferred identifications
(Fig.\,\ref{f:bootesdet}). The Pleiades illustrate some other
differences (Fig.\,\ref{f:pleiades}): Knobel identifies U\,409 with the
stars that we choose for the corresponding entry P\,412 in
\ptolemaios; because Ulugh Beg places U\,409 almost a degree further
south than Ptolemaios does P\,412, we prefer a different counterpart.
These examples imply that the numbers given in Table\,\ref{t:knobel}
should be considered  as approximate rather than exact.

Figure\,\ref{f:vdulug} shows that the magnitudes assigned by Ulugh Beg
correlate well with the modern Hipparcos measurements; his faintest
magnitude 6 slightly underestimates the actual brightness.  Note that
in the figure we ignore the qualifiers fainter and brighter for the
magnitudes. The errors $\Delta\lambda$ in longitude show a trend with
longitude similar to that in Ptolemaios, with a maximum at the largest
distance from the vernal equinox, i.e.\ near the autumnal equinox.
Due to an overall offset, however, the absolute value of the errors are
actually smallest near the autumnal equinox and largest near the vernal
equinox; this is in contrast to the situation in Ptolemaios.
The errors $\Delta\beta$ in
latitude show a smaller correlation with longitude. The errors in
longitude and latitude are not correlated. Using maximum likelihood,
we fit gaussians to the
histograms of the errors for all errors in the range $-$100\arcmin\ to
+100\arcmin, and find an offset of about $-10$\arcmin\ for
$\Delta\lambda$ and $\sigma\simeq$26\arcmin; and for $\Delta\beta$ an
offset of about 7\arcmin\ and $\sigma\simeq$21\arcmin.  Limiting the
fits to errors with absolute values less than about 50\arcmin\ we find
the same offsets, but $\sigma$'s reduced to 22\arcmin\ for
$\Delta\lambda$ and 18\arcmin\ for $\Delta\beta$.  These latter widths
are comparable to those given by Shevchenko (1990, for zodiacal stars
only), by Krisciunas (1993) and by Schwan (2002); \nocite{schwan} their
subtraction of the systematic trends with longitude in the errors has
only a small effect. The total error peaks between 10\arcmin\ and
20\arcmin. 

 Krisciunas (1993) assumes that Ulugh Beg used the same principal
  reference stars, Spica and Regulus, as Ptolemaios, to explain that
  the errors in longitude as smallest near the autumnal equinox.
  However, our analysis shows that the longitude errors of Ptolemaios
  are smallest near the vernal equinox (Fig.\,\ref{f:vdptol}). The
  difference is affected by the different overall offsets of the
  longitudes (+7\arcmin\ in Ptolemaios and $-$11\arcmin\ in Ulugh
  Beg). Clearly, the correct interpretation of the trend in the
  longitude errors depends on a correct understanding of the overall
  offset.

\subsection{Comparison of  Ptolemaios and  Ulugh Beg}

Since Ulugh Beg chose to observe the same stars that Ptolemaios 
lists in his catalogue, one expects a large agreement between 
the identifications we have produced for corresponding pairs.
Table\,\ref{t:knobel} shows that  in 956 cases the identifications
for the corresponding pairs are identical; and in 10 cases we 
consider identical identifications possible but prefer different ones.
In 50 cases we think the star observed by Ulugh Beg is different from
the corresponding entry in \ptolemaios. This may be based
on a different position, as is illustrated by the pair P\,98/U\,98
(Fig.\,\ref{f:bootesdet}), or the pairs P\,409/U\,406 and
P\,412/U\,409 (Fig.\,\ref{f:pleiades}). 

The magnitudes assigned by al-Sufi correlate very well with those
assigned by Ptolemaios, as illustrated in Table\,\ref{t:alsufi}. Stars
called faint by Ptolemaios, which have magnitude 7 in the
machine-readable table, are distributed around magnitude 5 by Ulugh
Beg, which confirms our conclusion based on Figure\,\ref{f:vdptol}
that the term faint in Ptolemaios refers to his magnitudes 5 and 6. There
are ten cases where the magnitude as given by al-Sufi differs by two
or more from that in Ptolemaios. In two cases, P\,289/U\,287 and
P\,634/U\,631, our identifications 
indicate that al-Sufi referred to a different star than
Ptolemaios; in seven cases the fainter magnitude given by al-Sufi
corresponds better to the actual magnitude; in one case,
P\,989/U\,986, the brighter magnitude given by Ptolemaios corresponds
better to the actual magnitude.

All four nebulous stars in \ulughbeg\ correspond
to one of the five nebulous stars in \ptolemaios.
P\,567, near 12.3,$-$4.5 in Fig.\,\ref{f:scorpius},
is nebulous in \ptolemaios, but the corresponding
entry in \ulughbeg\ U\,564 has magnitude 4--5 (4f).
The nebulous entries common to \ptolemaios\ and 
\ulughbeg\ are the open clusters h\,Per (P\,191, U\,190),
and Praesepe (P\,449, U\,446), the close pair
$\nu^{1,2}$\,Sgr (P\,577, U\,574) and $\lambda$\,Ori
(P\,734, U\,731, Fig.\,\ref{f:lamori}).
The globular cluster $\omega$\,Cen (P\,955, U\, 952)
is present in both catalogues, but was not recognised as
a nebulous object.

Our analysis of the positional errors in Sections \ref{s:results} and
\ref{s:resultsub} show that the catalogue of Ulugh Beg is more
accurate than the catalogue of Ptolemaios, and thus confirm that his
measurements are largely independent.

\begin{table}
\caption{Correlation of magnitudes by Ulugh Beg / al-Sufi
with those by Ptolemaios\label{t:alsufi}}
\begin{tabular}{c|cccccc|cc}
 & 1 & 2 & 3 & 4 & 5 & 6 & 7 &  9 \\
\hline
 1& 15 &     0 &     0 &     0 &     0 &     0 &     0 &   0 \\
  2& 0 &    32 &     3 &     0 &     0 &     0 &     0 &   0 \\
 3&  0 &    10 &   178 &    13 &     0 &     0 &     0 &   0 \\
 4&  0 &     3 &    20 &   394 &     6 &     0 &     2 &   1 \\
 5&  0 &     0 &     2 &    61 &   184 &     3 &     6 &   0 \\
 6&  0 &     0 &     1 &     4 &    25 &    45 &     4 &   0 \\
\hline
 8&   0 &     0 &     0 &     0 &     1 &     0 &     0 &   0 \\
 9&  0 &     0 &     0 &     0 &     0 &     0 &     0 &   4 \\
\end{tabular}
\tablefoot{In the machine-readable tables magnitude 7
refers to stars called faint by Ptolemaios; magnitude 8
refers to the entry for which Ulugh Beg remarks `no star
is visible in that location'; magnitude 9 to entries called
nebulous.}
\end{table}

\subsection{Non-stellar sources and double stars\label{s:doubles}}

 \begin{figure}
 \centerline{\includegraphics[angle=270,width=\columnwidth]{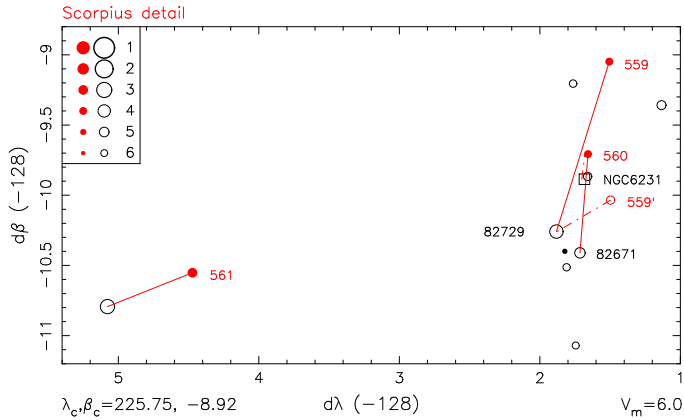}}
 \caption{Detail of Scorpius (see Fig.\,\ref{f:scorpius})
showing the relative positions of
P\,559 and P\,560 (red) and the open cluster NGC\,6231 (black $\Box$), together with the
double star HIP\,82729 / HIP\,82671 (open circles). A red circle
marked 559\arcmin\
denotes the position of P\,559 when its latitude is corrected 
from $-$18\degr\ to $-$19\degr. HIP\,82729 is a high-proper-motion
star: the black $\bullet$ shows the position one would find for $-$128
if proper motion is ignored. Solid lines indicate our identifications,
the dash-dotted lines those by Ashworth (1981).
   \label{f:scodet}}
 \end{figure}
\nocite{ashworth}

We identify two entries in \ptolemaios\ with open clusters:
P\,191 with h\,Per, or possibly with the pair h and $\chi$ Per  
(Fig.\,\ref{f:perseus}), and  P\,449 with Praesepe in Cancer 
(Fig.\,\ref{f:cancer}); and we identify one entry with a globular cluster:
P\,955 with $\omega$\,Cen  (Fig.\,\ref{f:centaurus}).
P\,955 is given magnitude 5 by Ptolemaios, P\,191 and
P\,449 are indicated as nebulous.

The third object indicated nebulous is P\,567, in Scorpius.  On the
basis of the description of P\,567 by Ptolemaios {\em the nebulous
star to the rear of the sting} (translation Toomer),
Kunitzsch (1974b) argued that this entry corresponds to the open
cluster M\,7. Ashworth (1981) repeats this suggestion.
\nocite{kunitzsch74}\nocite{ashworth}
The position of P\,567 is about 3\degr\ from M\,7, rather further
than the offsets of other identifications in Scorpius
(Fig.\,\ref{f:scorpius}), which leads us to agree with earlier
authors and prefer HIP\,87261 as the counterpart.

Ashworth (1981) also makes the interesting suggestion that Ptolemaios
may have catalogued NGC\,6231. Of the two stars in the third joint of
Scorpius the one with the more northern coordinates is called 
south of the other, and vice versa.  Toomer (1998) keeps the
descriptions for P\,559 and P\,560, and switches the coordinates so
that P\,559 is the northern star both according to position and to
description.  We follow him in identifying P\,559 and
P\,560 with HIP\,82729 and HIP\,82671, respectively. However, the
positions shown in Fig.\,\ref{f:scodet} suggest that an identification
of P\,559 with NGC\,6231 and P\,560 with HIP\,82729 is also possible.
Ashworth does not switch the coordinates, but emends the latitude of
P\,559 from $-18$\degr\ to $-$19\degr. We denote the star with the
emended position P\,559', and show its position in
Fig.\,\ref{f:scodet}.  Ashworth (1981) identifies P\,560 with
NGC\,6231 and P\,559' with the pair HIP\,82729 + HIP\,82671
($\zeta^{2,1}$\,Sco). We note that the angular separation between
these two stars was 13\farcm6 in $-128$ compared to only 6\farcm5
today, as shown in Fig.\,\ref{f:scodet}, and consider the pair
HIP\,82729 + HIP\,82671 an excellent match for P\,559' and P\,560.

 \begin{figure}
 \centerline{\includegraphics[angle=270,width=\columnwidth]{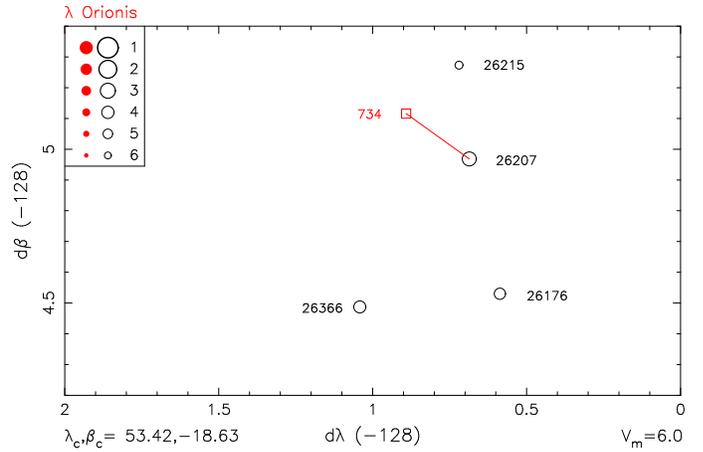}}
 \caption{Detail of Orion (see Fig.\,\ref{f:orion}) showing the positions of
P\,734 (red) and various stars from the modern Hipparcos catalogue
(open circles). 
   \label{f:lamori}}
 \end{figure}

 P\,577 is described by Ptolemaios as `the nebulous and double
 (\textgreek{diplo\~us}) star at the eye [of Sagittarius]'; it corresponds
to HIP\,92761 ($V$=4.86) and HIP\,92845 ($V$=5.00) , separated
by 12\farcm2 in $-$128. 
P\,734 is also described as nebulous by Ptolemaios; it corresponds
to HIP\,26207, and it is not clear why Ptolemaios would call it
nebulous, as the nearest star, HIP\,26215, is 18\farcm5 from
HIP\,26207 and is rather faint ($V$=5.6, see Fig.\,\ref{f:lamori}).

It is interesting to see which close pairs of stars were noted by
Ptolemaios as double, and which ones were not. Another
entry explicitly denoted as corresponding to two stars is
P\,150, truely remarkable as the corresponding pair
HIP\,91919  / HIP\,91926 ($\epsilon^{1,2}$\,Lyr) was separated
by only 3\farcm2 in $-$128, and the stars are not very bright
at $V$=4.46 and 4.59, respectively.
In contrast, the pair HIP\,72603 / HIP\,72622 ($\alpha^{1,2}$\,Lib)
is not denoted as double by Ptolemaios, perhaps because of its
even smaller separation (2\farcm9) or because of the larger
brightness contrast ($V$=5.15 and 2.75). P\,601 corresponds to
the pair HIP\,100027 / HIP\,100064 ($\alpha^{1,2}$\,Cap) is also
not denoted as double by Ptolemaios, even though the pair is
bright ($V$=4.30 and 3.58) and well separated (5\farcm1).

\begin{acknowledgements}
  Our research was much stimulated by the article of Heiner Schwan (2002) on the
  star catalogue of Ulugh Beg and its accuracy.  It has made
  use of the SIMBAD database, operated at CDS, Strasbourg, France, and
  was supported by the Netherlands Organisation for Scientific
  Research under grant 614.000.425.
\end{acknowledgements}


\begin{appendix}
\section{Emendations and annotations}

\subsection{Emendations to Toomer}


\noindent P\,138 -- P\,140. Toomer (1998) in his corrigenda on p.\,xii cites
Kunitzsch (1986) to identify P\,138 -- P\,140 with 77(x)\,Her,
82(y)\,Her and 88(z)\,Her. 88(z)\,Her = HIP\,87280 has $V$=6.8, too
faint for a plausible counterpart.  We emend the
identifications to 74\,Her for P\,138, 77(x)\,Her for
P\,139, and 82(y)\,Her for P\,140, in agreement with Kunitzsch (1986).

\noindent P\,166: identification $\iota$\,Cyg emended to
$\iota^2$\,Cyg.

\noindent P\,332: identification $\pi$\,Peg emended tp
$\pi^2$\,Peg.

\noindent P\,371: identification $\tau$\,Ari emended to $\tau^2$\,Ari.

\noindent P\,404: identification $\omega$\,Tau emended to $\omega^2$\,Tau.

\noindent P\,456: identification $\mu$\,Cnc emended to $\mu^2$\,Cnc.

\noindent P\,458: identification $\pi$\,Cnc emended to $\pi^2$\,Cnc.

\noindent P\,570: identification $\gamma$\,Sgr emended to $\gamma^2$\,Sgr.

\noindent P\,647: identification $\tau$\,Aqr emended to $\tau^2$\,Aqr.

\noindent P\,749: identification $\psi$\,Ori emended to $\psi^2$\,Ori.

\noindent P\,910: identification $\phi$\,Hya emended to $\phi^3$\,Hya.

\subsection{Emendations to Manitius; comparison of editions by
  Manitius and Toomer\label{a:knobtoom}}

Having completed our machine-readable catalogue of Ptolemaios
according to the edition by Toomer (1998), we may compare it with
the (machine-readable version of the) edition by Manitius (1913), as made
available by Jaschek (1987). We compare the longitude, latitude and
magnitude of each entry. In a number of cases, the difference
we obtain is due to a difference between Jaschek (1987) and the
Manitius edition by Neugebauer (1963). Some are due to differences
between Neugebauer and Manitius, some due to differences between
Jaschek and Manitius, as follows:

{\small
\begin{tabular}{cccrrrr}
P & star &  & Man.  & Jas. & Neu. & Toom.\\
\hline
305 & Del\,5  &   $\beta$ &   33~50  &   33~00  &    33~50 & 33~50\\
501 & Vir\,5  &   $\beta$ &   00~20  &   00~20  &    00~10 & 00~10\\
505 & Vir\,9  & $\lambda$ &  171~00  &  161~00  &   171~00 & 171~00 \\
703 & Psc\,30 & $\lambda$ &  356~20  &  256~20  &   356~20 & 356~20 \\
918 & Hya\,25 &   $\beta$ & $-$13~40 & $-$13~40 & $-$17~40 & $-$17~40
\\
\end{tabular}
}


We emend Jaschek (1987) to agree with Manitius (1913), and emend
in both the latitude of P\,918. Table\,\ref{t:mantoo} gives the differences
that remain, between the Manitius (1913)  and Toomer (1998) editions
of the star catalogue of Ptolemaios. 
Positions in Manitius that are very different from those in Toomer
are indicated in the Figures in Sect.\,\ref{s:figures} in brown.

The magnitudes are identical in both catalogues, with the exception
of the magnitude of P\,568: 5b in Toomer, 5 in Manitius.

\begin{table}
\caption{Differences between the Manitius (1913) edition and 
the machine-readable table \ptolemaios.\label{t:mantoo}}
\begin{tabular}{cccc|cccc}
P & $\Delta M_\lambda$ & $\Delta M_\beta$ & 
HIP$_\mathrm{Man}$ & P & $\Delta M_\lambda$ & $\Delta M_\beta$ & 
HIP$_\mathrm{Man}$ \\
\hline
   3 &   350 &     0 & &       518 &     0 &   -20 & \\ 
  42 &     0 &   160 &  46952 & 569 &  -240 &     0 &  80473 \\
  61 &     0 &   240 & &        596 &  -180 &     0 & \\       
  69 &  -160 &     0 & &        641 &     0 &    10 & 109857 \\
 129 &     0 &  -150 & &        642 &     0 &   -10 & \\       
 233 &     0 &   340 &  27949 & 645 &     0 &   -30 & \\       
 268 &   180 &     0 & &        667 &     0 &   -35 & \\       
 347 &    10 &     0 & &        668 &     0 &    35 & \\       
 348 &   -10 &     0 & &        822 &  -240 &     0 & \\       
 366 &    20 &    20 & &        852 &     0 &   -10 & \\       
 375 &     0 &    30 & &        855 &     0 &   -15 & \\       
 389 &  -160 &     0 & &        922 &     0 &    10 & \\       
 395 &    20 &     0 & &        931 &   -10 &     0 & \\       
 448 &   140 &     0 &  40167 & 933 &     0 &     5 & \\       
 501 &     0 &    10 && 954 &     0 &  -160 &  68523 \\
 502 &     0 &    20 & &        958 &     0 &  -160 &  59449 \\
 504 &     0 &   -20 &  63414 & 961 &   -10 &     0 &     25 \\
\end{tabular}
\tablefoot{$\Delta M_\lambda$ and $\Delta M_\beta$ give
$60(\lambda_\mathrm{Manitius}-\lambda_\mathrm{Toomer})$ and
$60(\beta_\mathrm{Manitius}-\beta_\mathrm{Toomer})$, respectively;
the identification by Manitius is indicated only when different
from our identification.}
\end{table}

\subsection{Annotations to Knobel}

In many cases Knobel identifies a catalogue entry with a close pair of
stars. If the pair can be separated with the naked eye, we choose
the brightest of the pair as identification. 
 
\noindent U\,879 is identified by Knobel as [Piazzi] VII 235, P\,Pup. These are
two different stars: HIP\,38020 and HIP\,38164, respectively. The
latter is the correct identification.

\noindent U\,882 is identified by Knobel as o\,Pup; we amend to o\,Vel.

\subsection{Annotations to individual identifications in \ptolemaios
\label{s:idptol}}

\begin{figure*}
 \centerline{\includegraphics[angle=270,width=\columnwidth]{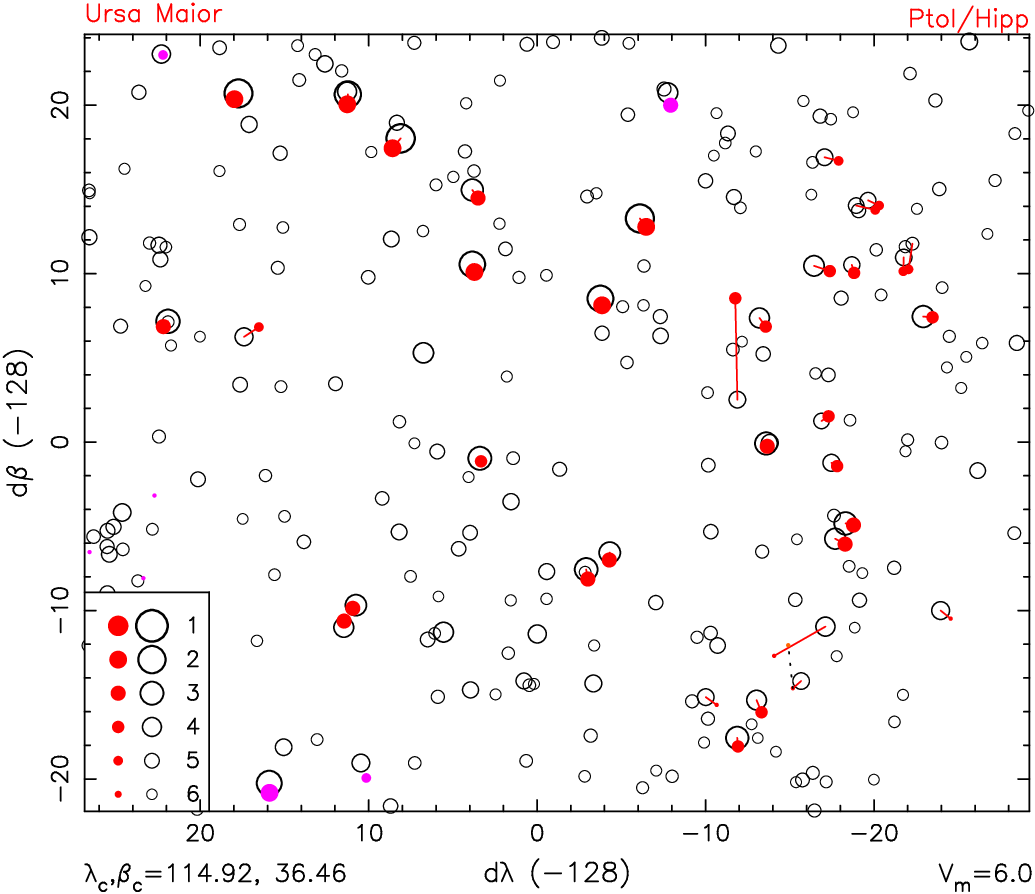}
 \hfill           \includegraphics[angle=270,width=\columnwidth]{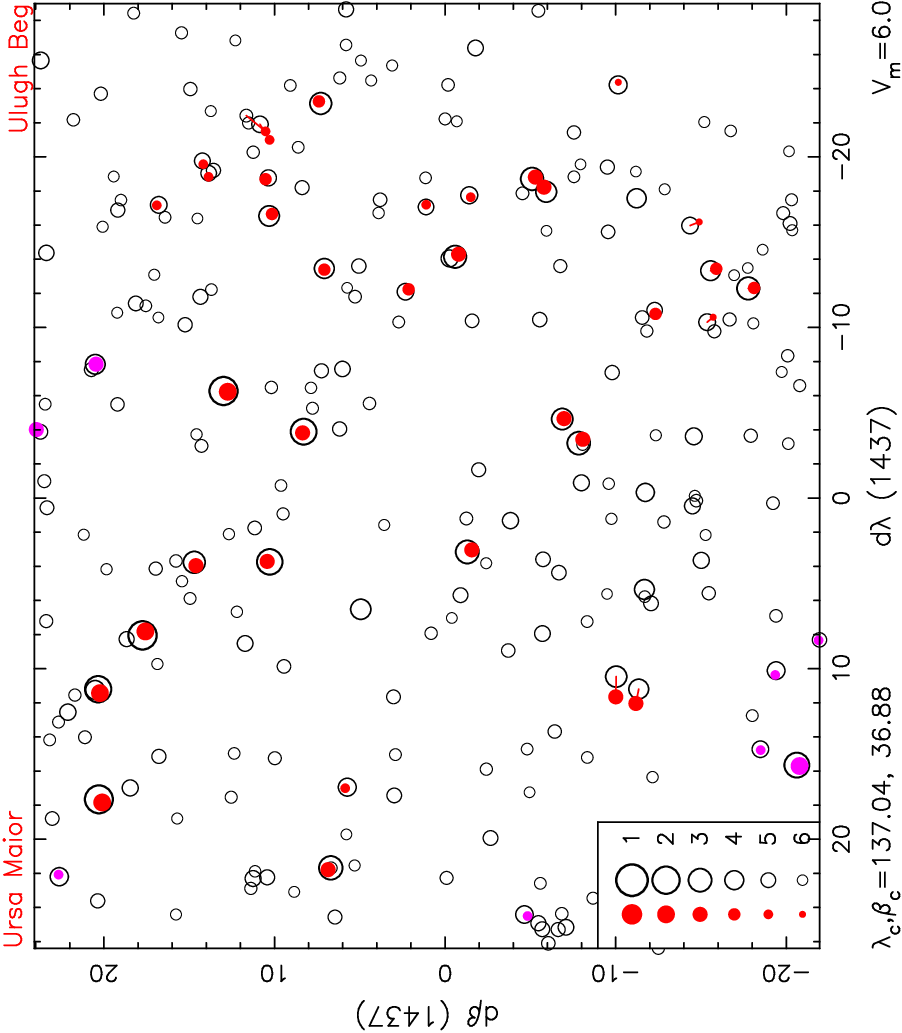}}
 \caption{Ursa Maior   \label{f:ursamaior}}
 \end{figure*}

\noindent {\bf P\,17-18}, near $-$13.6,6.9 and $-$11.8,8.5 in Fig.\,\ref{f:ursamaior}
respectively, are described by Ptolemaios as the northern and southern
one of the pair in the breast of the Bear. This implies a position for
P\,18 further south than the catalogue value, and Toomer identifies it
with HIP\,48402, near $-$11.9,2.5, which implies an error
of 6\degr. We follow him (and earlier authors) in doing so.

\noindent {\bf P\,36}, near 22.2,6.9 in Fig.\,\ref{f:ursamaior}, corresponds to the
combined light of the close (1\farcm9)  pair HIP\,63121, HIP\,63125.

\noindent {\bf P\,41}, near $-$14.1,$-$12.7 in Fig.\,\ref{f:ursamaior}, is equally far
removed from HIP\,47029, near $-$10.7,$-$12.1, to the east and
HIP\,44248 to the west; we choose the latter, slightly brighter star;
Toomer mentions the former as an uncertain possibility.

\noindent {\bf P\,60}, near 5.1,15.8 in Fig.\,\ref{f:draco} is the combined light of
the close (38\farcs5) pair HIP\,86614/HIP\.86620.

\begin{figure*}
 \centerline{\includegraphics[angle=270,width=\columnwidth]{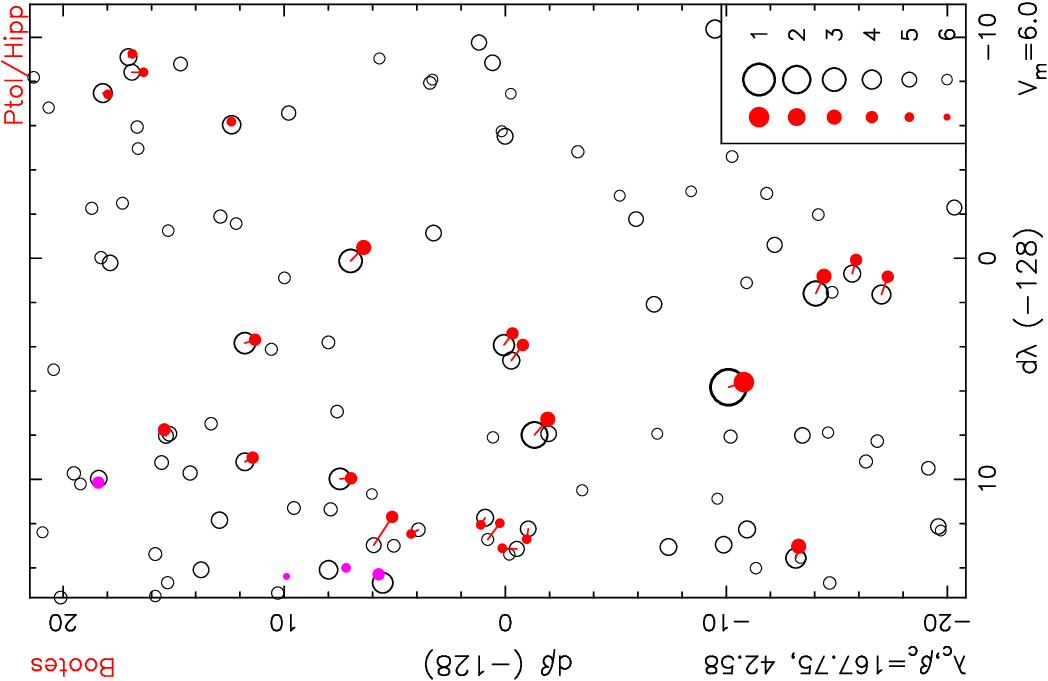}
\hfill\includegraphics[angle=270,width=\columnwidth]{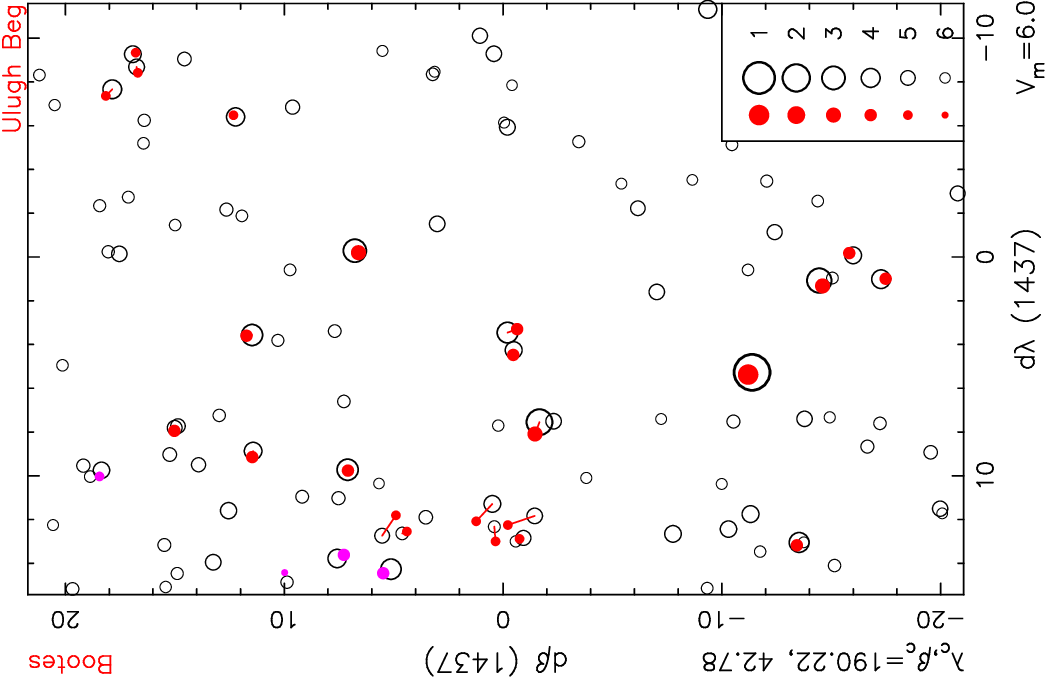}}
 \caption{Bootes       \label{f:bootes}}
 \end{figure*}

\noindent {\bf P\,96}, near 7.8,15.4 in Fig.\,\ref{f:bootes}, is almost at the same
distance to HIP\,75973 and HIP\,76041 ($V$=5.04 and 4.98, 
$d$=18\farcm0 and 17\farcm3, respectively). We  choose the marginally
brighter and closer star, Toomer prefers the other.

\noindent {\bf P\,98}, near 12.5,4.3 in Fig.\,\ref{f:bootes} is identified by Toomer
with HIP\,75049, near 13.0,5.0; we prefer the nearer, brighter star.

\noindent {\bf P\,100-101}, near 12.0,0.2 and 12.1,1.1, respectively in
Fig.\,\ref{f:bootes}: the identification of P\,100 by Toomer
is our identification for P\,101 and vice versa.

\begin{figure*}
 \centerline{\includegraphics[angle=270,width=\columnwidth]{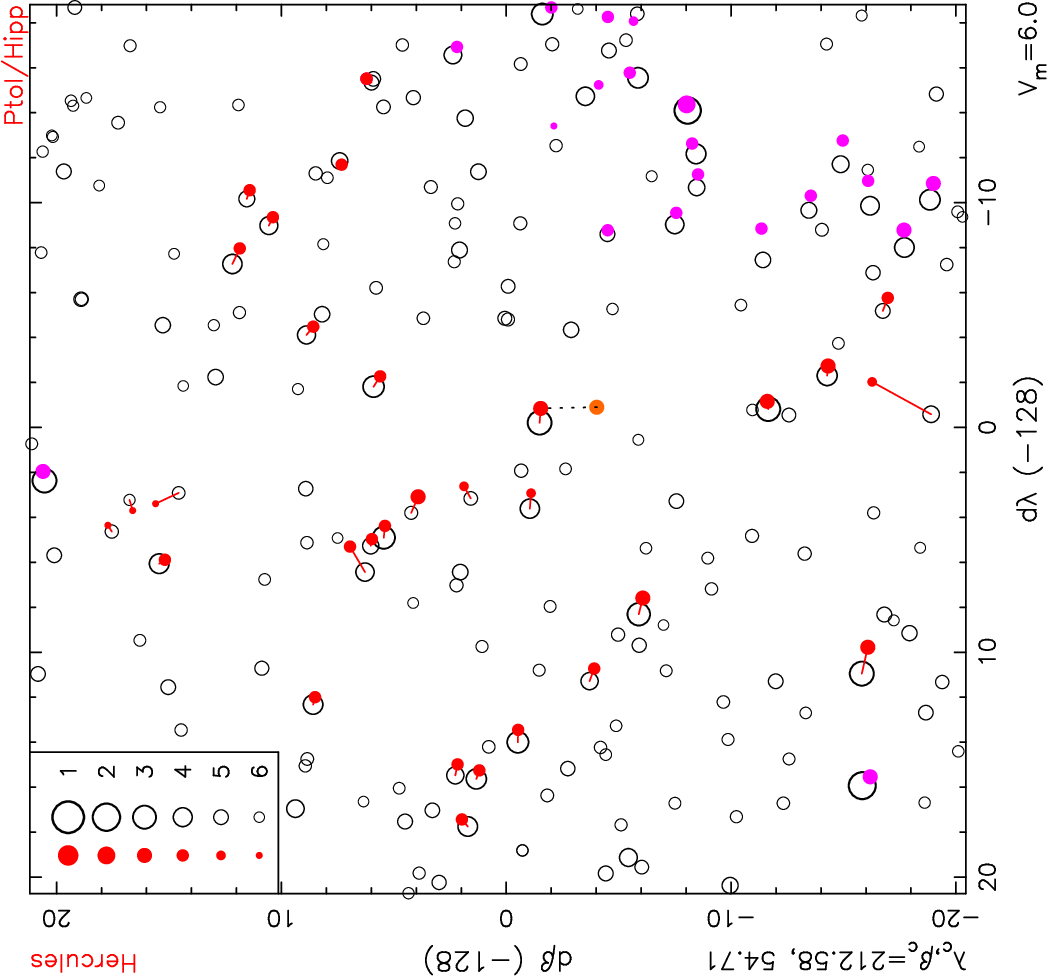}
\hfill\includegraphics[angle=270,width=\columnwidth]{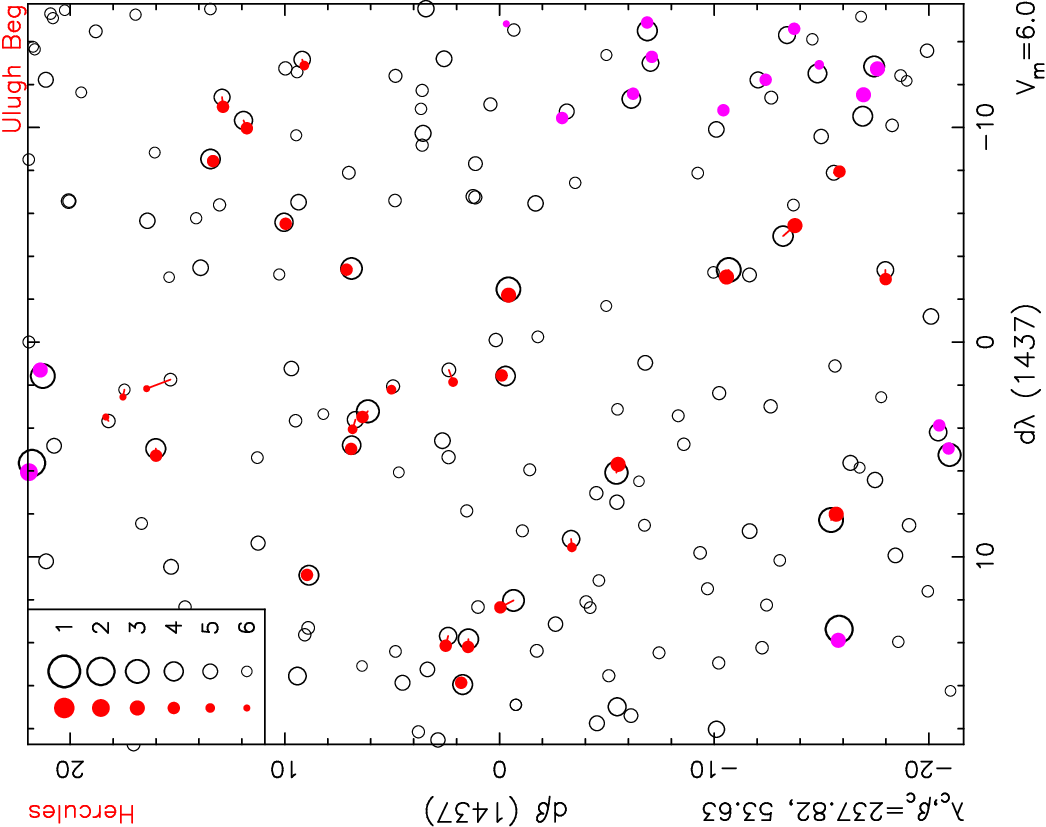}}
 \caption{Hercules     \label{f:hercules}}
 \end{figure*}

\noindent {\bf P\,132}, near 3.1,3.9 in Fig.\,\ref{f:hercules}, is identified by
Toomer with a closer ($d$=26\farcm6) but rather fainter ($V$=6.2) star.

\begin{figure*}
 \centerline{\includegraphics[angle=270,width=\columnwidth]{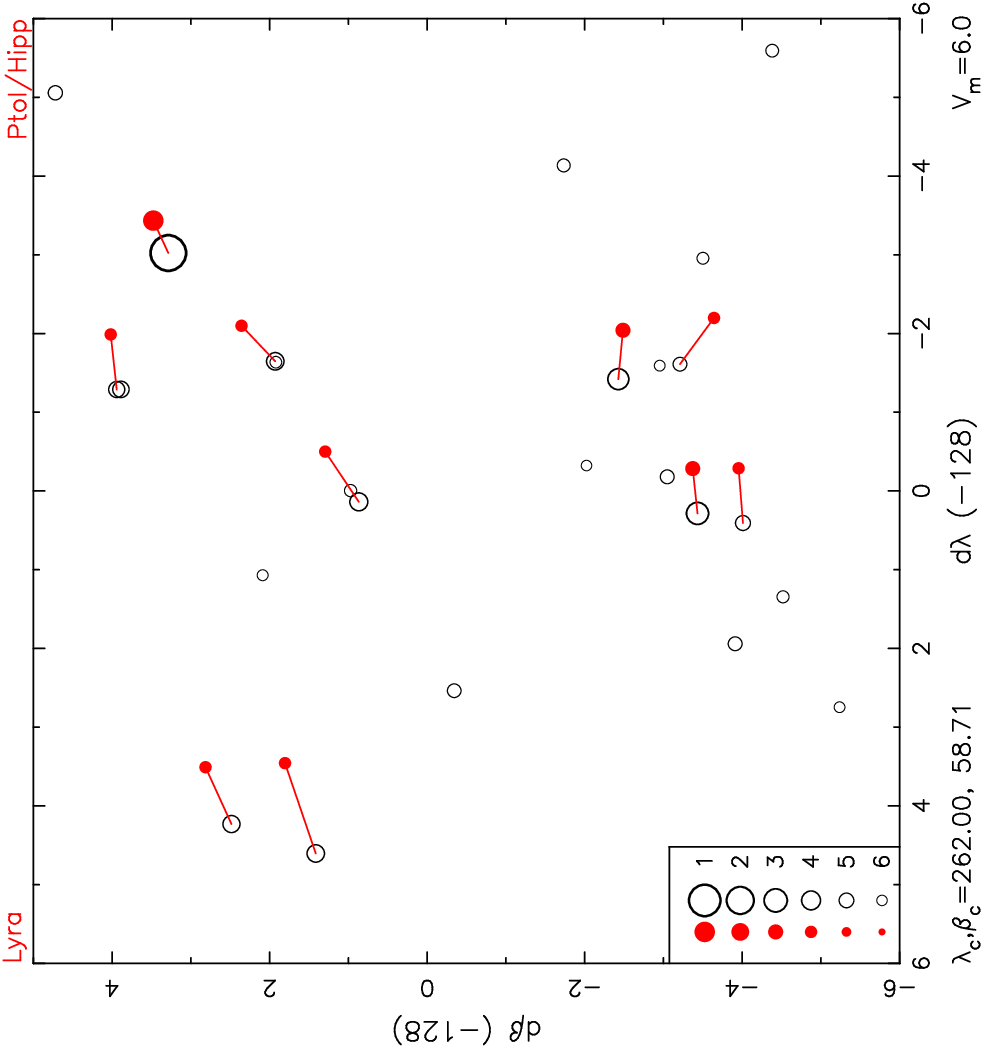}
\hfill\includegraphics[angle=270,width=\columnwidth]{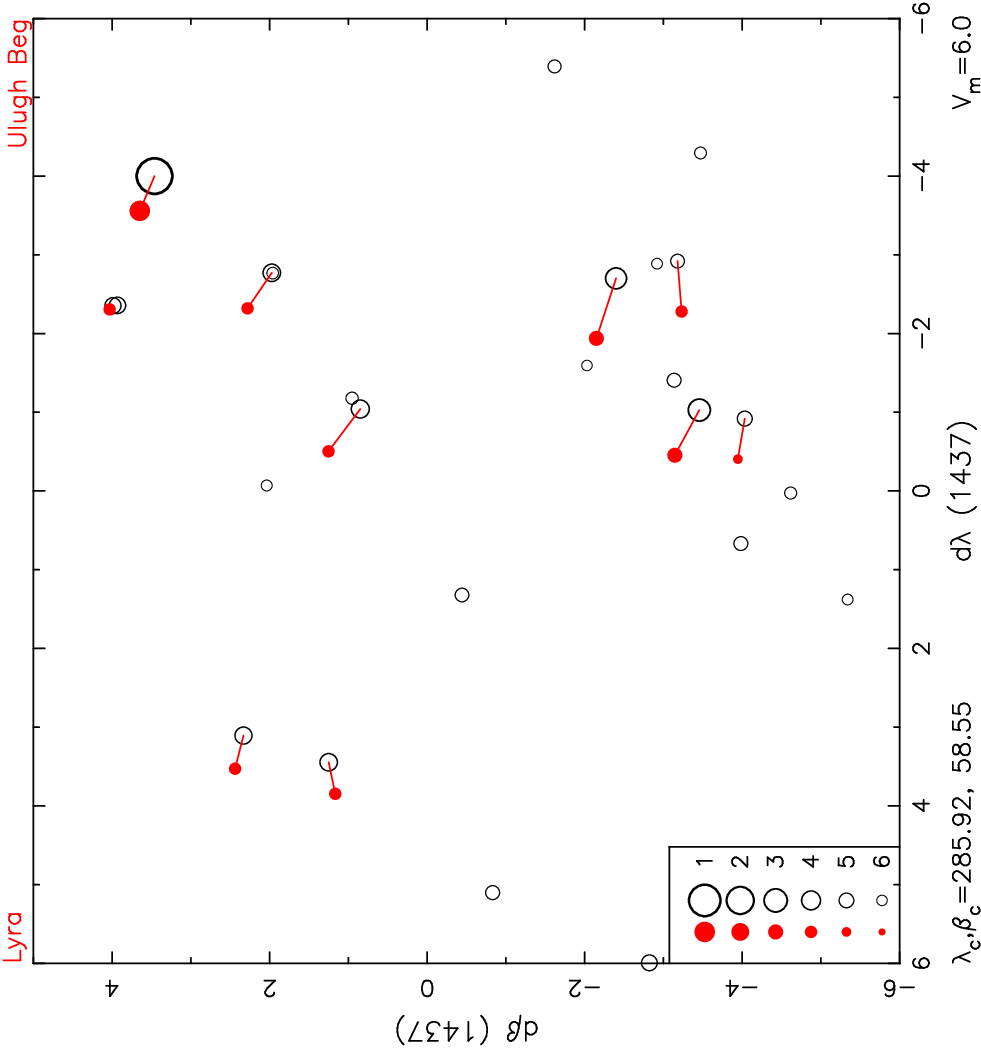}}
 \caption{Lyra         \label{f:lyra}}
 \end{figure*}

\noindent {\bf P\,150}, near $-$2.0,4.0 in Fig.\,\ref{f:lyra}, the combined light
of the close (3\farcm2) pair HIP\,91926 and HIP\,191919. Remarkably,
Ptolemaios notes that this entry corresponds to two stars, i.e.\ he was
able to see them separately. 

\noindent {\bf P\,151}, near $-$2.1,2.4 in Fig.\,\ref{f:lyra}, combined light of
the close (50\farcs1) pair HIP\,91971 and HIP\,91973

\noindent {\bf P\,152}, near $-$0.5,1.3 in Fig.\,\ref{f:lyra}, is identified by Toomer
with the nearest star HIP\,92728, but we prefer the brighter star next to
it.

\noindent {\bf P\,159}, near $-$12.9,$-$8.0 in Fig.\,\ref{f:cygnus}, is the combined
light of the close (19\farcs9) pair HIP\,95947/HIP\,95951.

\begin{figure*}
 \centerline{\includegraphics[angle=270,width=\columnwidth]{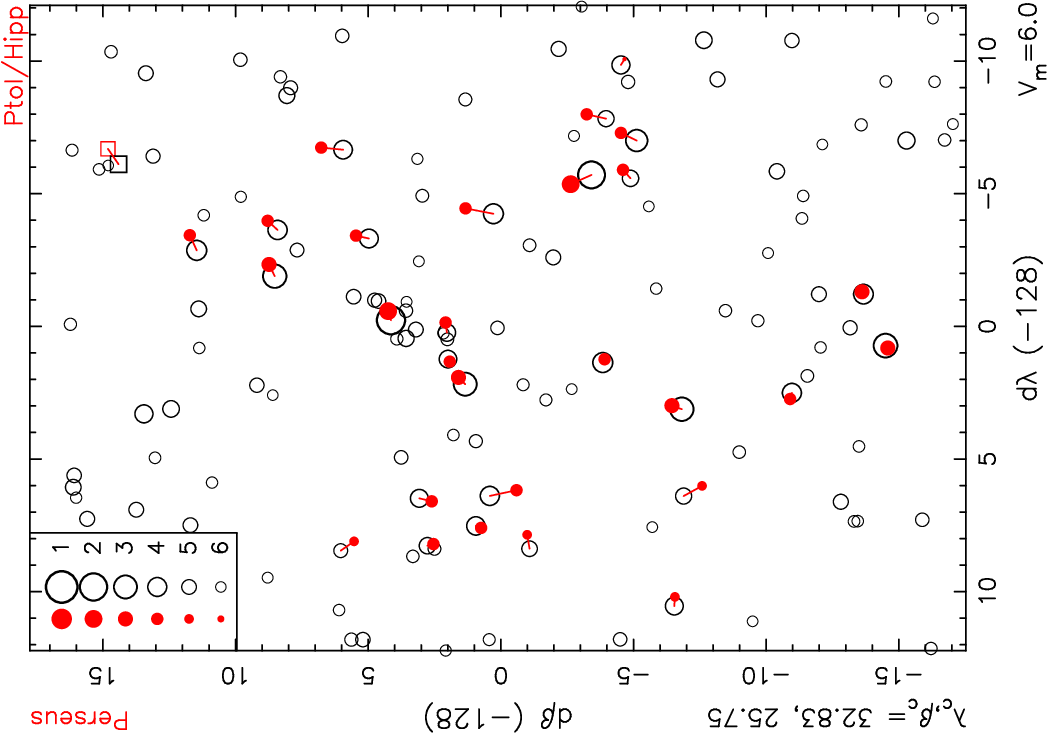}
\hfill\includegraphics[angle=270,width=\columnwidth]{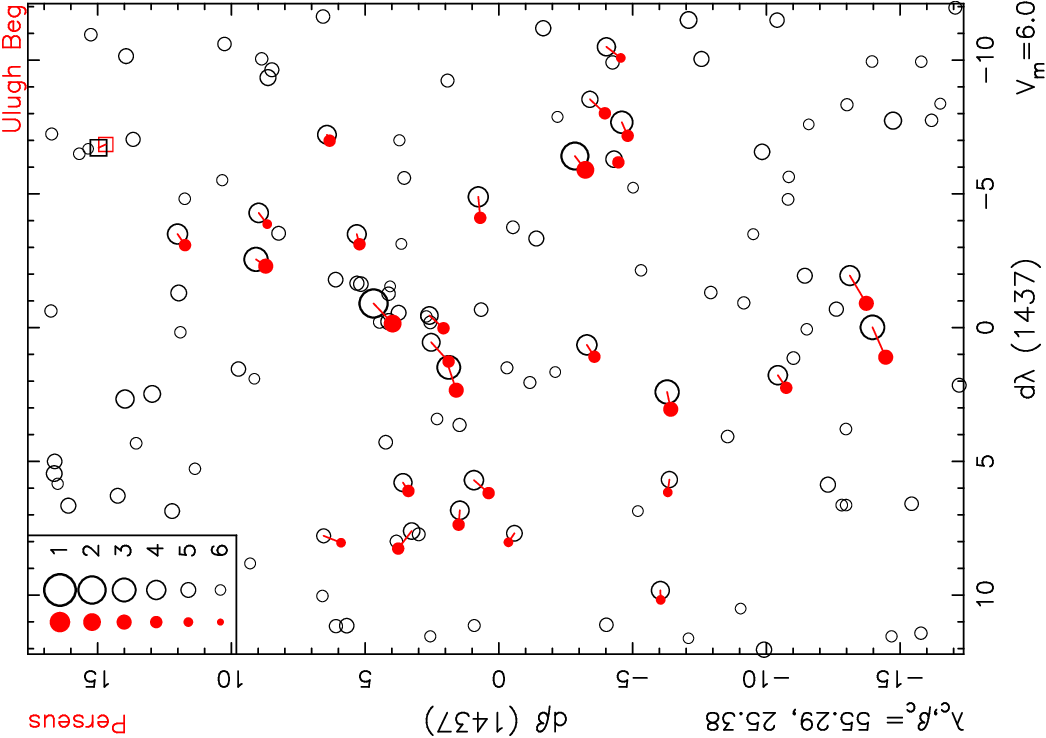}}
 \caption{Perseus. The open cluster h\,Per is indicated with $\Box$
   near $-$6,14      \label{f:perseus}}
 \end{figure*}

\noindent {\bf P\,191}, near $-$6.7,14.8 in Fig.\,\ref{f:perseus}, is the
open cluster h Per. The center of this cluster is $\alpha=34\fdg8425,
\delta=57\fdg15$ (J2000.0, Slesnick et al.\ 2002).

\noindent {\bf P\,409-411}, the Pleiades. Our identifications are based on
the description by Ptolemaios as {\em the northern end of the advance
  side} (P\,409), {\em the southern end of the advance side} (P\,410),
and {\em the rearmost and narrowest end of the Pleidades} (P\,411).
(Translation by Toomer, who has the same identification for P\,409 and
P\,411; but prefers HIP\,17608 for P\,410; see Fig.\,\ref{f:pleiades}).

\begin{figure*}
 \centerline{\includegraphics[angle=270,width=\columnwidth]{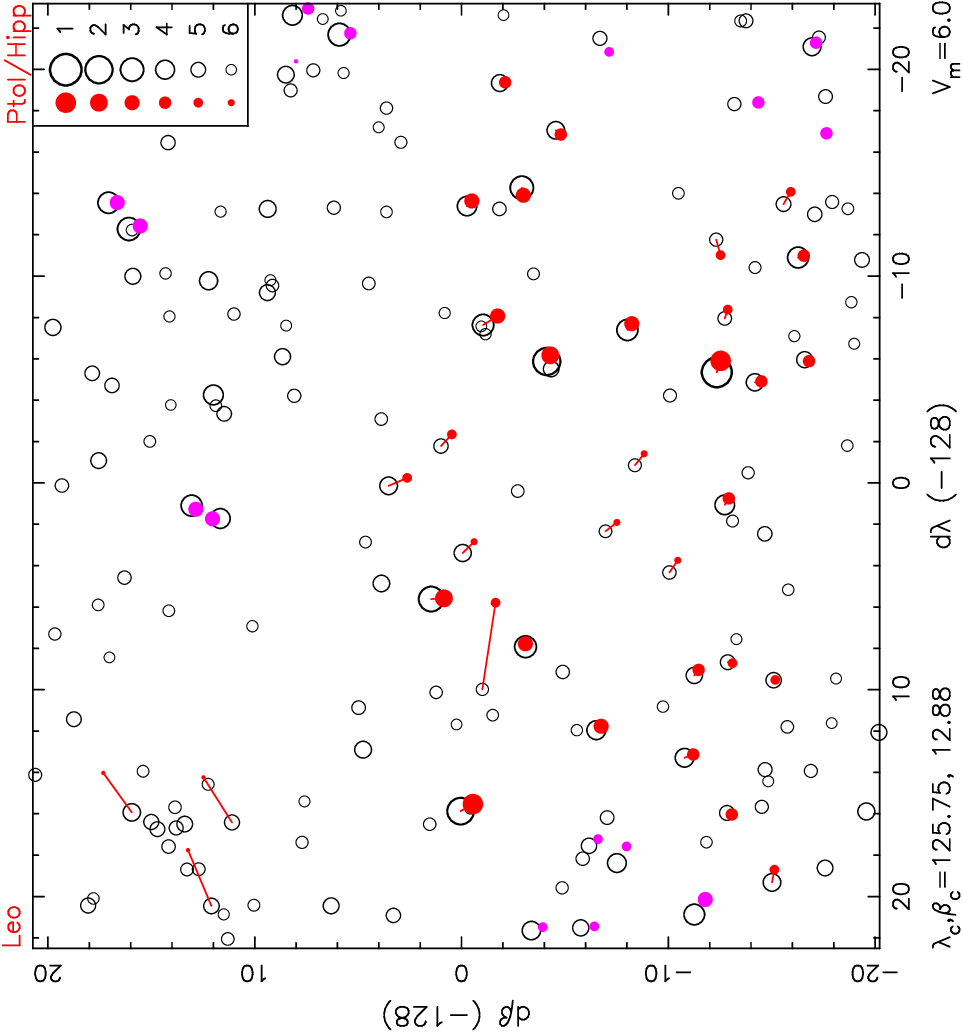}
\hfill\includegraphics[angle=270,width=\columnwidth]{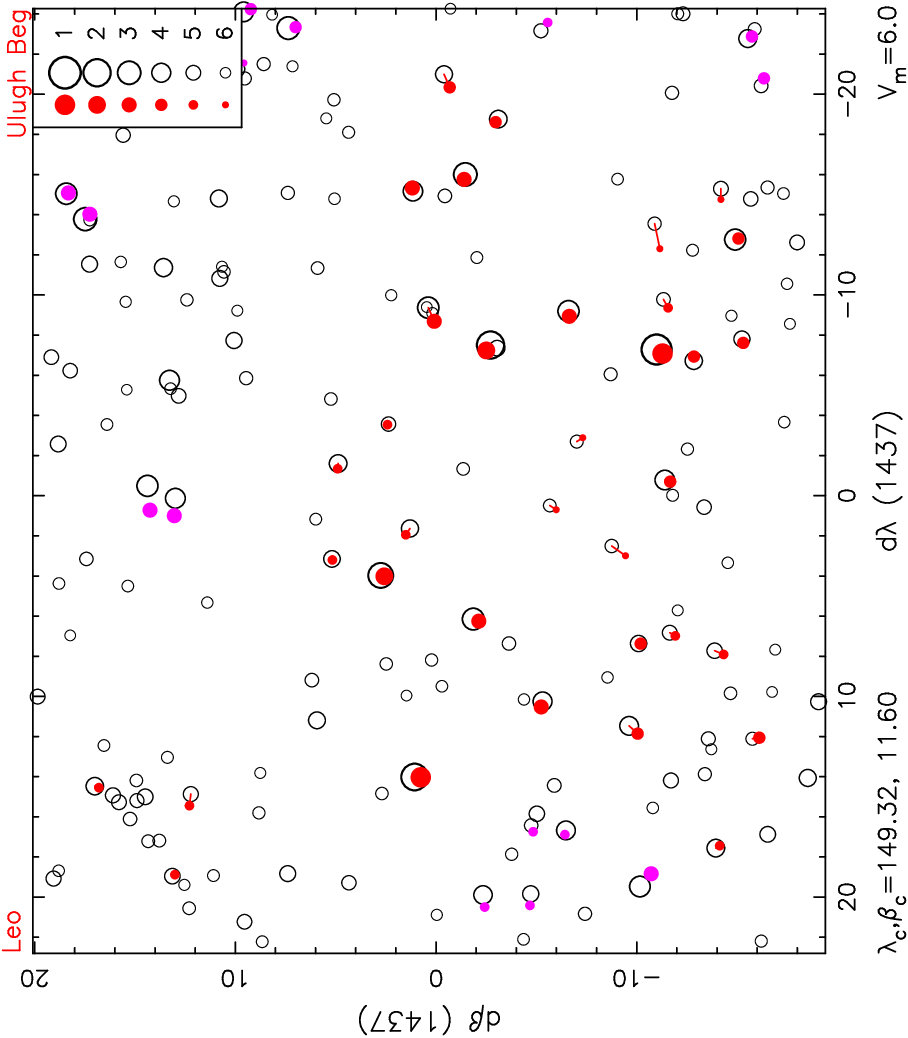}}
 \caption{Leo          \label{f:leo}}
 \end{figure*}

\noindent {\bf P\,494-496}, in the north-east part of Fig.\,\ref{f:leo} are
identified with help of their descriptions: {\em the northernmost of 
the nebulous mass between the edges of Leo and Ursa, called Coma},
{\em the most advanced of the southern outrunners of Coma}, and
{\em the rearmost of them, shaped like an ivy leaf} (translation by
Toomer).

\begin{figure*}
 \centerline{\includegraphics[angle=270,width=0.8\columnwidth]{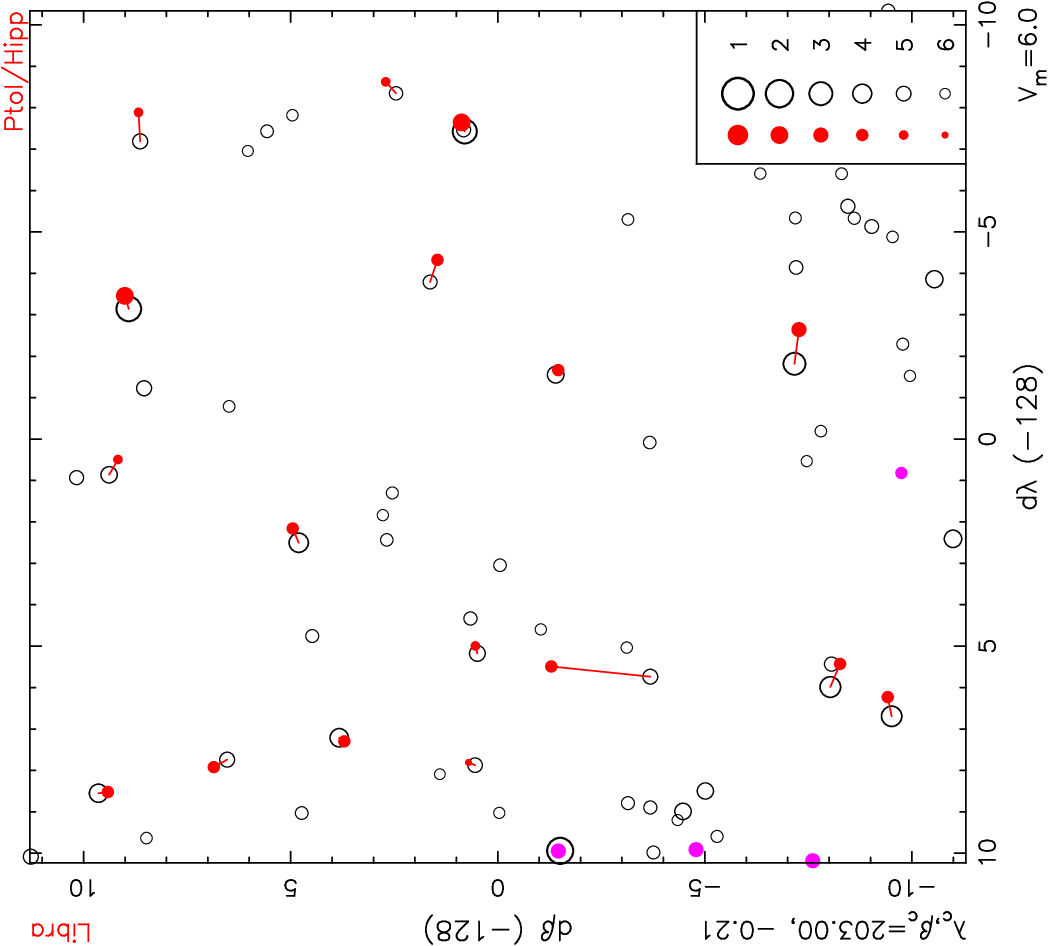}
\hfill\includegraphics[angle=270,width=0.8\columnwidth]{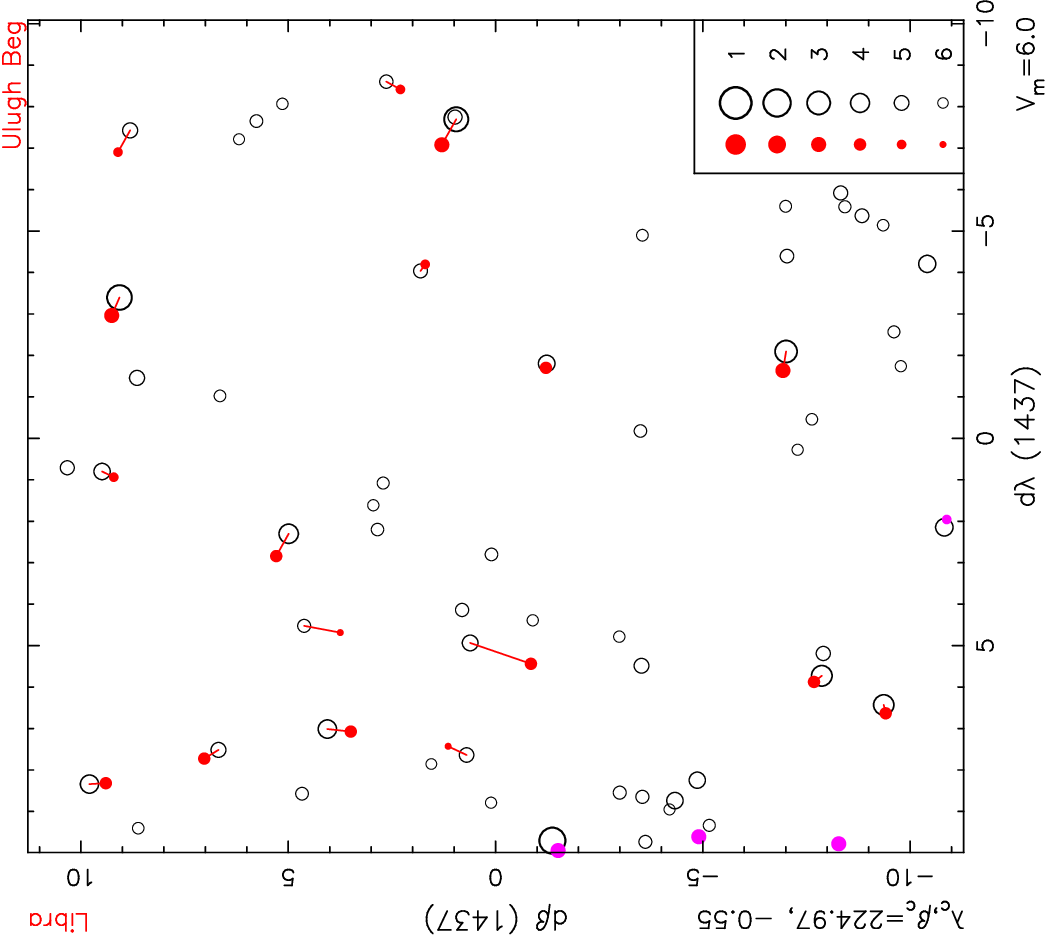}}
 \caption{Libra        \label{f:libra}}
 \end{figure*}

\noindent {\bf P\,529}, near $-$7.6,0.9 in Fig.\,\ref{f:libra}, is
identified with the close (2\farcm9) pair HIP\,72622/HIP\,72603
($V$=2.75/5.15).

\noindent {\bf P\,541}, near 5.0,0.5 in Fig.\,\ref{f:libra}, is identified
by Toomer with HIP\,76628; our counterpart is closer and brighter.

\noindent {\bf P\,542}, near 5.5,$-$1.3 in Fig.\,\ref{f:libra}, is identified
by Toomer with the nearest star HIP\,76569; we prefer the brighter
but rather further HIP\,76742.

\begin{figure*}
 \centerline{\includegraphics[angle=270,width=\columnwidth]{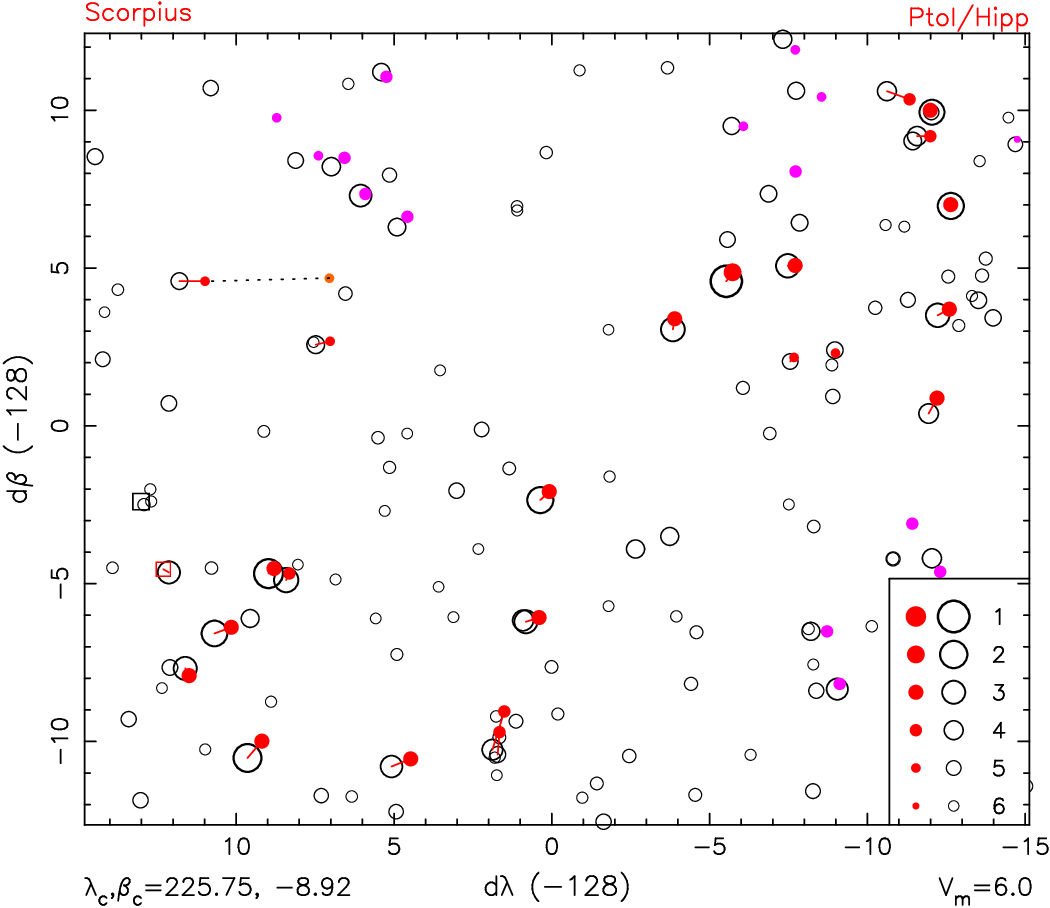}
\hfill\includegraphics[angle=270,width=\columnwidth]{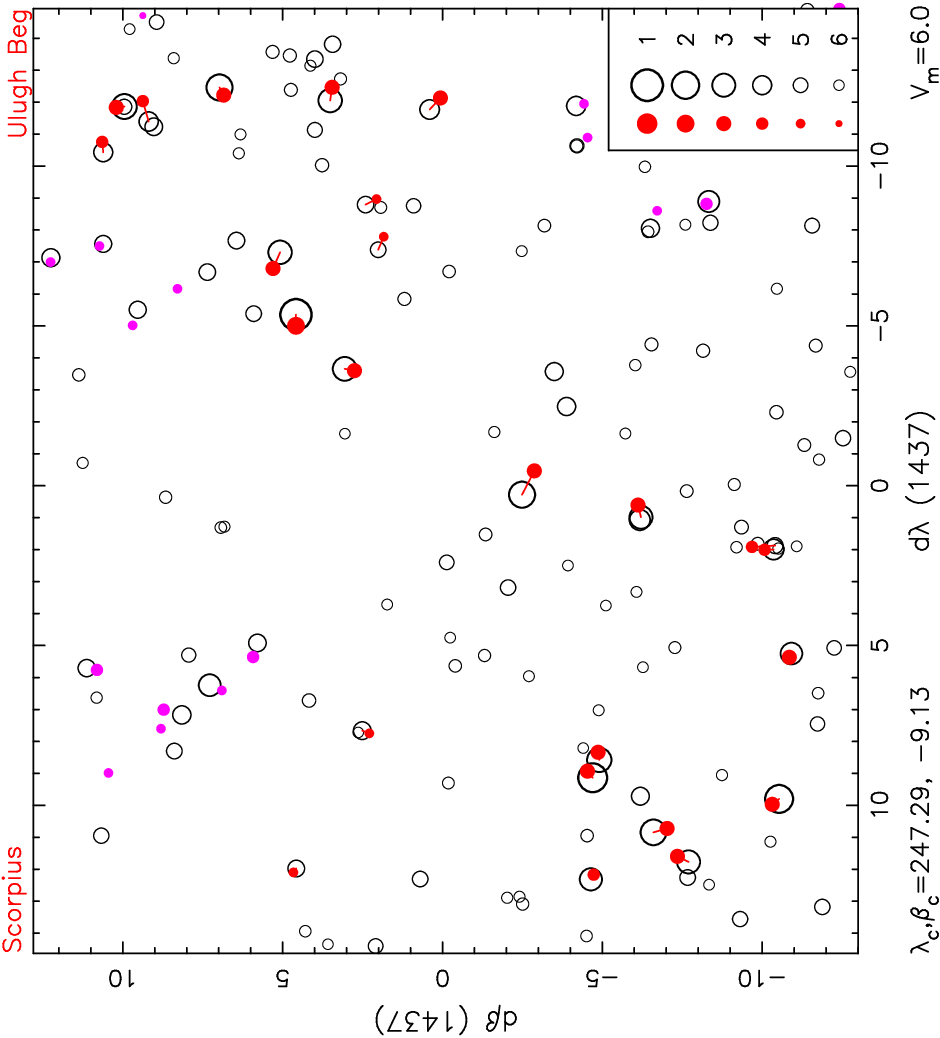}}
 \caption{Scorpius. In the left image, the $\Box$  near 13.0,$-2.4$
   indicates M\,7   \label{f:scorpius}}
 \end{figure*}

\noindent {\bf P\,567}, near 12.3,$-$4.5 in Fig.\,\ref{f:scorpius}, is
called nebulous by Ptolemaios. Toomer suggests that this is due
to the proximity of NGC\,6441. We consider this unlikely, as this
globular cluster has integrated magnitude $V$=7.15. See also the
discussion in Sect.\,\ref{s:doubles}.

\begin{figure*}
 \centerline{\includegraphics[angle=270,width=\columnwidth]{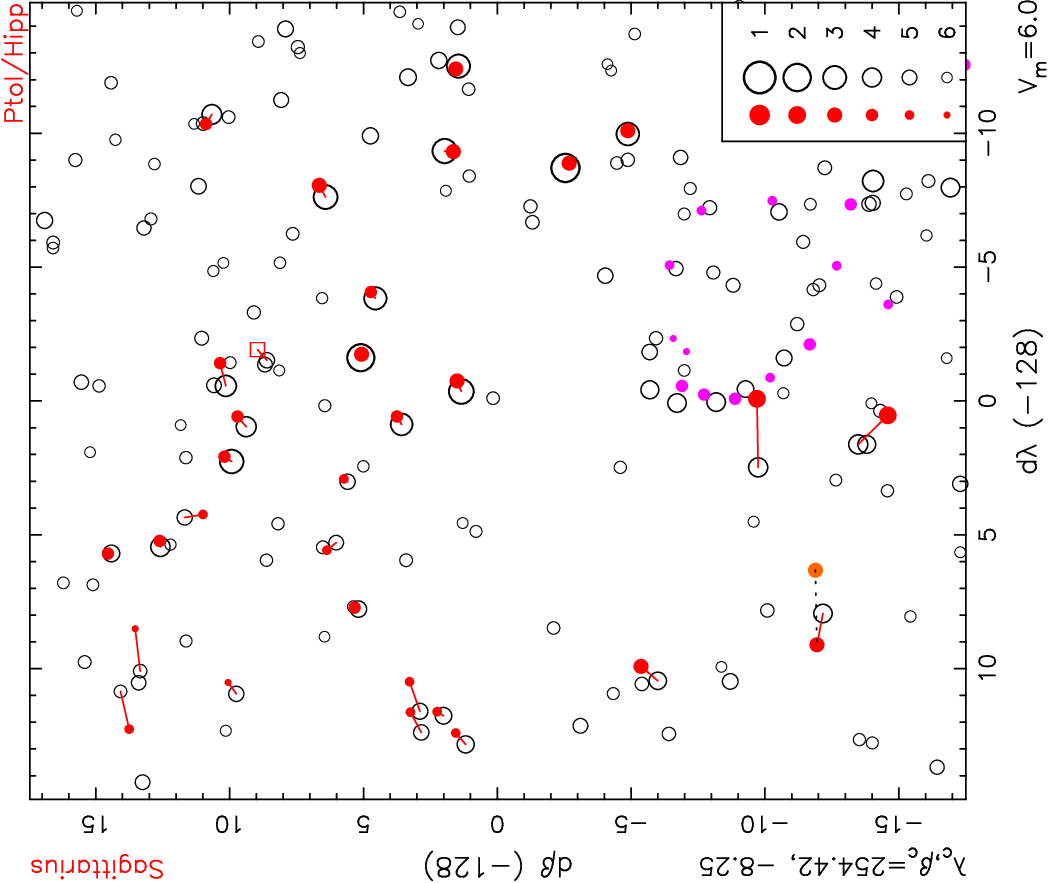}
\hfill  \includegraphics[angle=270,width=\columnwidth]{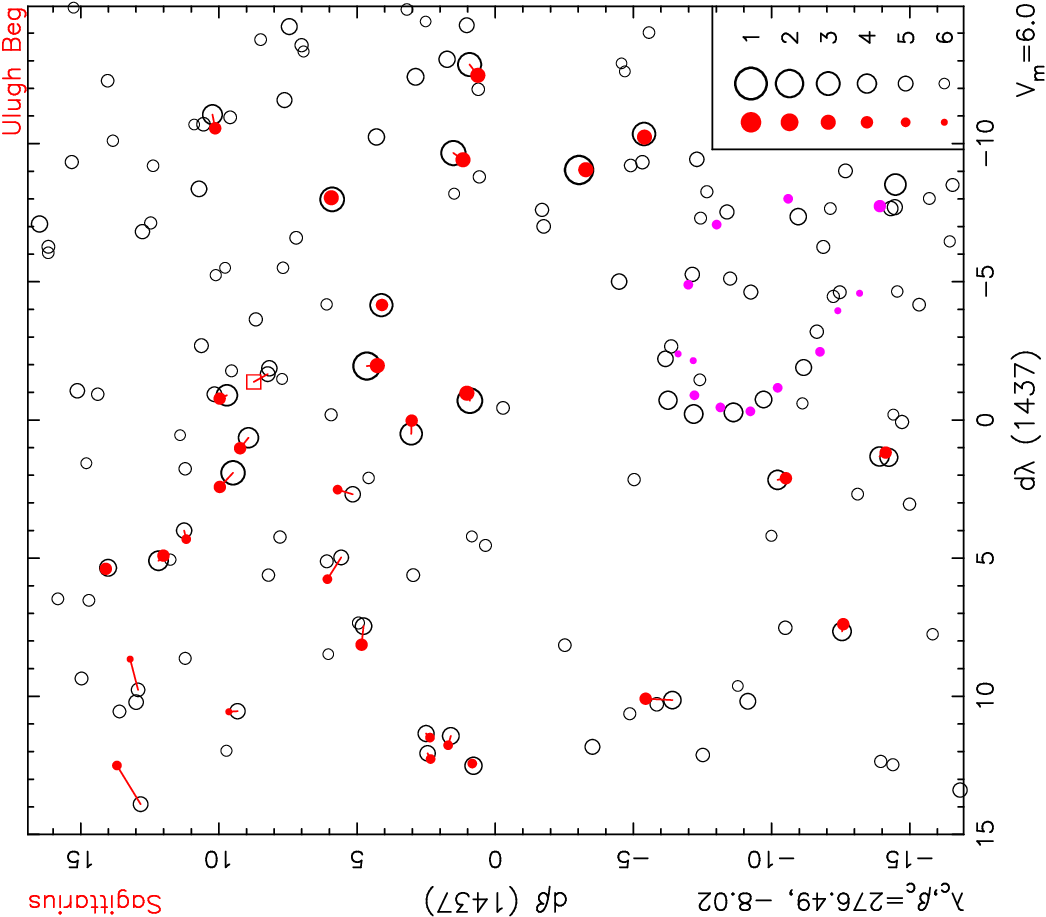}}
 \caption{Sagittarius  \label{f:sagittarius}}
 \end{figure*}

\noindent {\bf P\,584}, near 8.5,13.5 in Fig.\,\ref{f:sagittarius},
is identified by us with the nearest star, by Toomer with the
marginally brighter star just beyond it, HIP\,96950.

\noindent {\bf P\,585}, near 12.3,13.8 in Fig.\,\ref{f:sagittarius}, is
identified with Toomer with HIP\,98258 ($V$=5.0, $d$=124\farcm8),
a plausible alternative to our closer but fainter counterpart.

\noindent {\bf P\,588}, near 7.7,5.3 in Fig.\,\ref{f:sagittarius}, is
identified by Toomer with the combined light of HIP\,96406/HIP\,96465.
As the distance between these two stars was 12\farcm2 in 139 BC
(it is now 13\farcm3) we prefer to identify with the brighter star
only.

\begin{figure*}
 \centerline{\includegraphics[angle=270,width=\columnwidth]{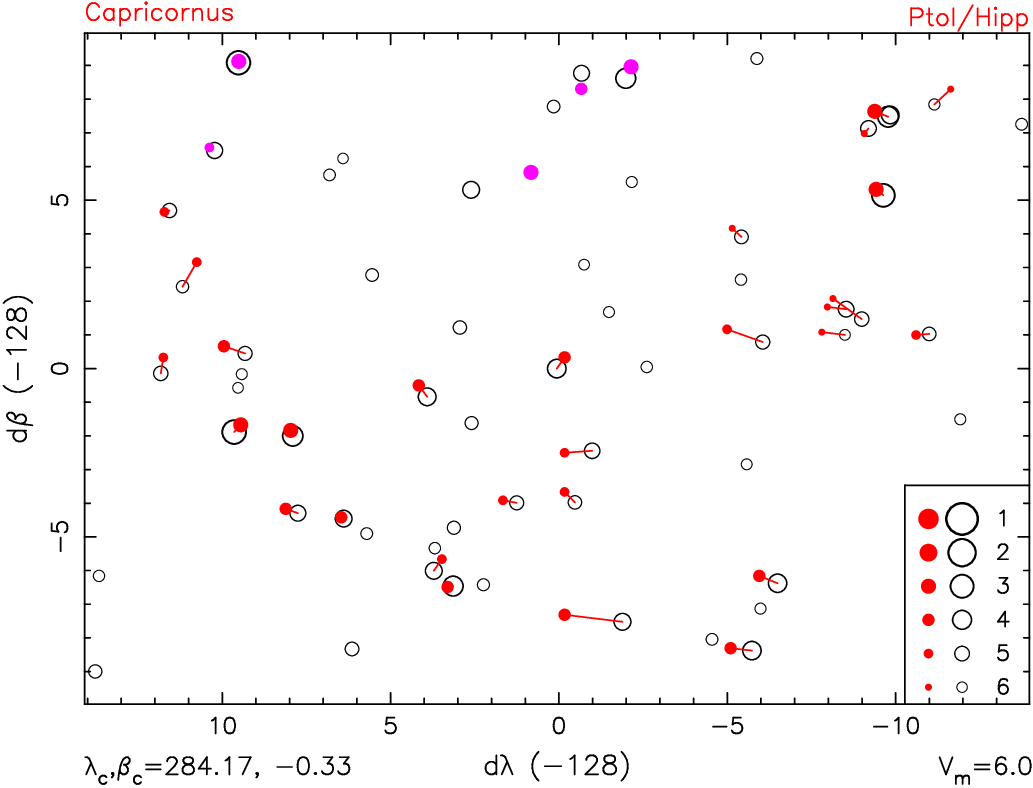}
\hfill\includegraphics[angle=270,width=\columnwidth]{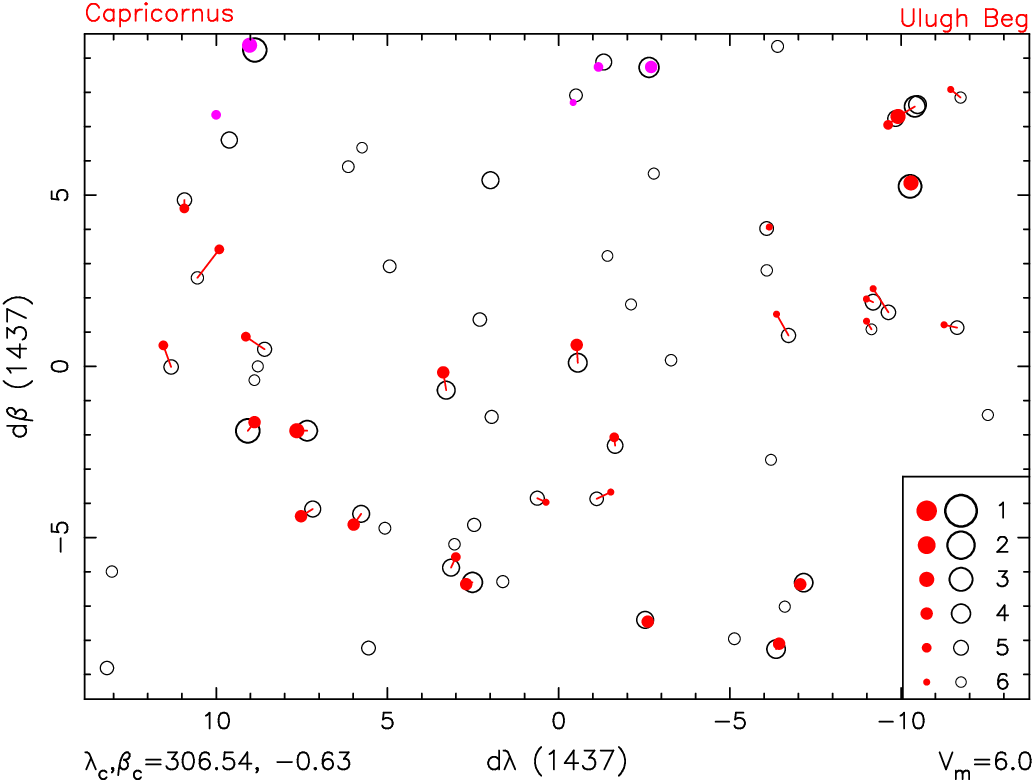}}
 \caption{Capricornus  \label{f:capricornus}}
 \end{figure*}

\noindent {\bf P\,601}, near $-$9.4,7.6 in Fig.\,\ref{f:capricornus}, is
identified by Toomer with the close (5\farcm1) pair 
HIP\,100027/HIP\,100064 ($\alpha^{1,2}$\,Cap). There is no hint in Ptolemaios
that he noticed a pair, rather than a single star.

\noindent {\bf P\,604}, near $-$11.6,8.3 in Fig.\,\ref{f:capricornus}, is
identified by Toomer with the close (7\farcm2) pair 
HIP\,99529/HIP\,99572 ($\xi^{1,2}$\,Cap). There is no hint in Ptolemaios
that he noticed a pair, rather than a single star; $\xi^1$\,Cap at
$V$=6.3 is barely visible to the naked eye.

\noindent {\bf P\,661}, near 13.9,$-$11.7 in Fig.\,\ref{f:aquarius}, is
identified by Toomer with the (10\farcm1) pair HIP\,116901/HIP\,116889
(103/104\,Aqr; $V$=4.8/5.4).  There is no hint in Ptolemaios that he
noticed a pair, rather than a single star. We identify with the
brightest of the two.

\noindent {\bf P\,786, P\,787, P\,788}, near $-$9.0,14.1, $-$11.1,14.3 and
$-$12.6,14.2 in Fig.\,\ref{f:eridanus} respectively. Toomer identifies
P\,786 with the pair HIP\,14923/HIP\,14168 ($\rho^{3,2}$\,Eri), and
P\,787 with HIP\,13701 ($\eta$\,Eri),  and thereby is forced
 to identify P\,788 with the very faint ($V$=6.3) star HIP\,13421. 
We prefer to identify P\,786 with $\rho^3$\,Eri, P\,787 with 
$\rho^2$\,Eri, and P\,788 with $\eta$\,Eri. The separation between
$\rho^3$\,Eri and $\rho^2$\,Eri is 22\farcm4.

\begin{figure*}
 \centerline{\includegraphics[angle=270,width=\columnwidth]{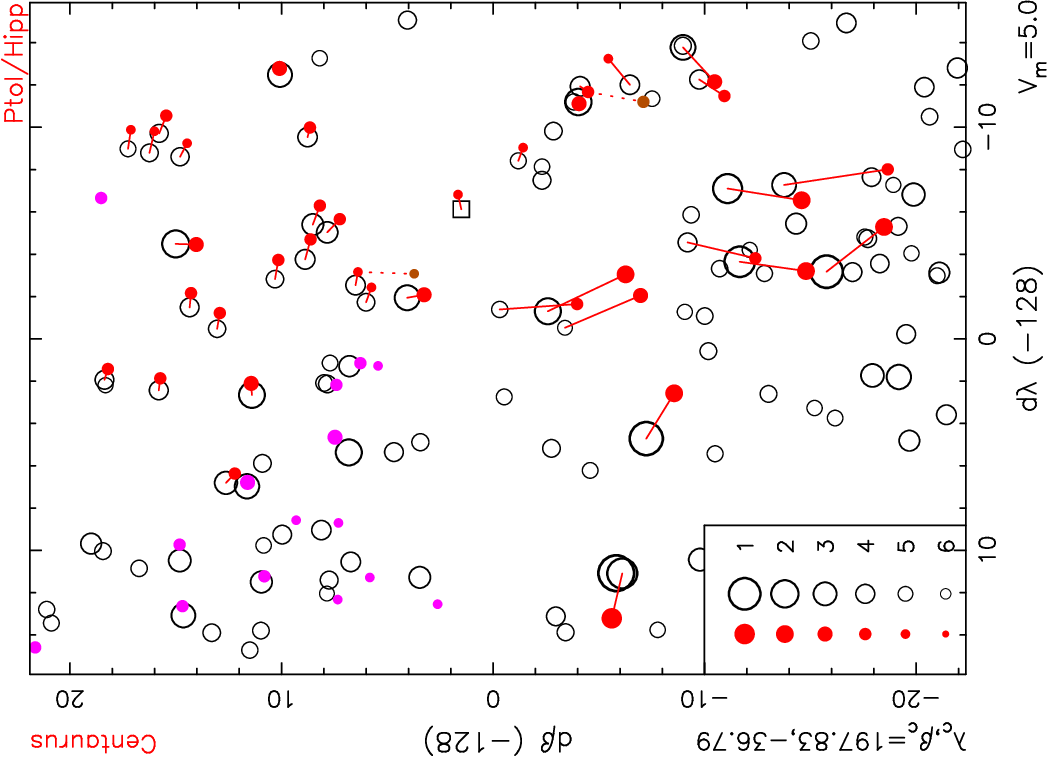}
\hfill\includegraphics[angle=270,width=\columnwidth]{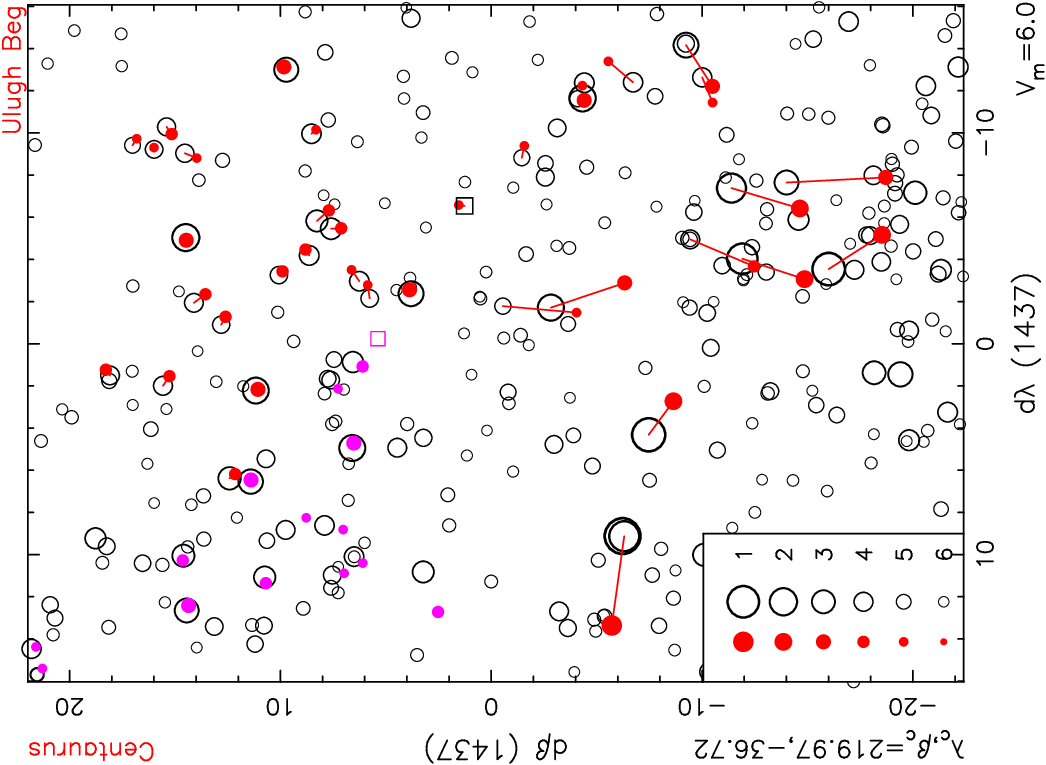}}
 \caption{Centaurus. In the plot for \ptolemaios\ only bright
   Hipparcos stars are shown. $\omega$\,Cen is indicated near $-$8,2 with $\Box$    \label{f:centaurus}}
 \end{figure*}

\noindent {\bf P\,955}, near $-$6.8,1.7 in Fig.\,\ref{f:centaurus}, is the
globular cluster $\omega$\,Cen. We take the modern position of this
cluster from Harris (1996, version of February 2003).

\begin{figure*}
 \centerline{\includegraphics[angle=270,width=\columnwidth]{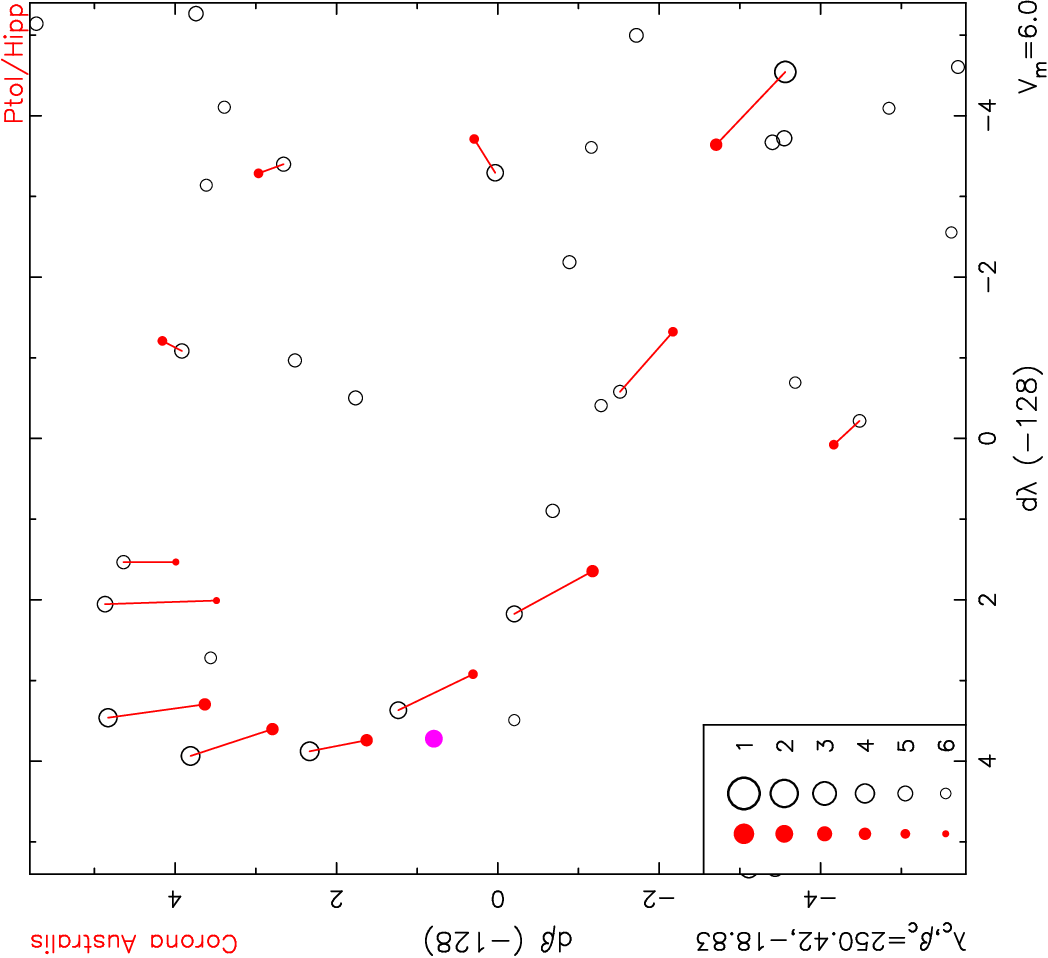}
\hfill\includegraphics[angle=270,width=\columnwidth]{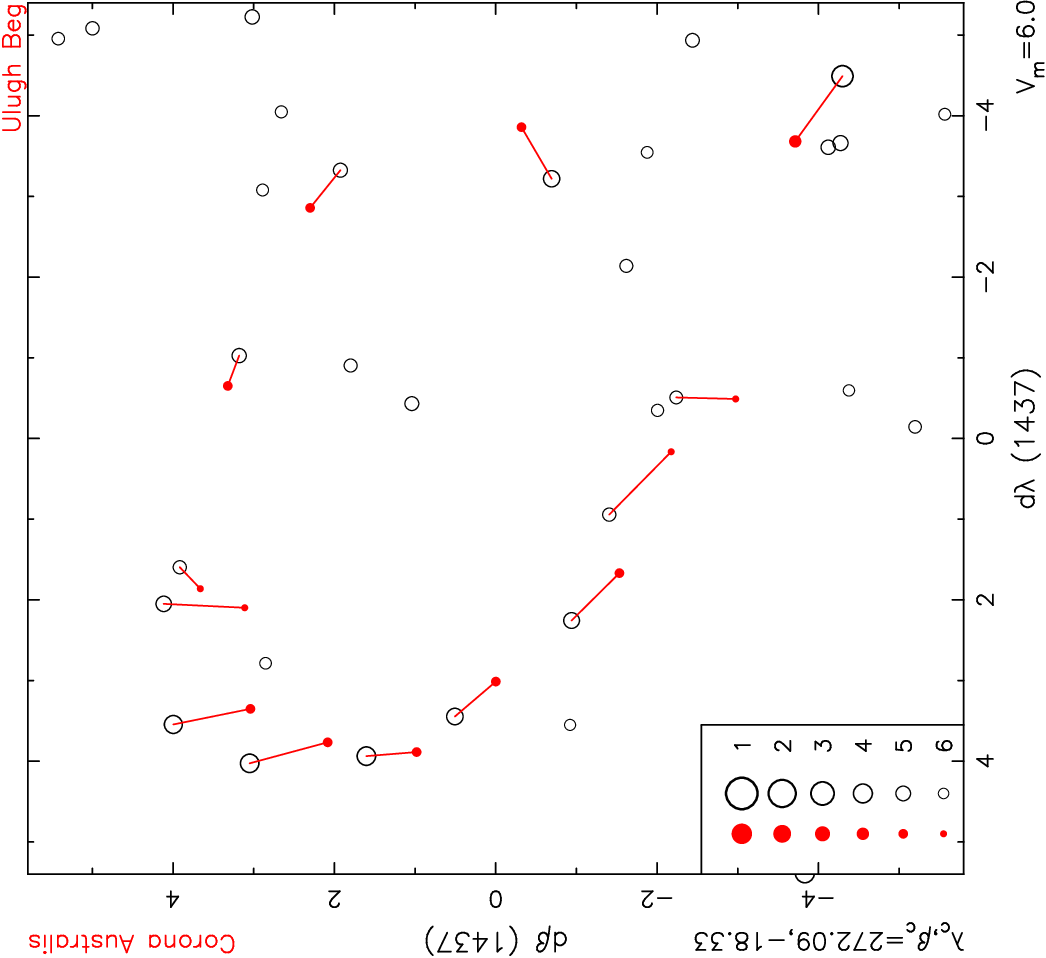}}
 \caption{Corona Australis \label{f:coraust}}
 \end{figure*}

\noindent {\bf P\,1000}, near 0.1,$-$4.2 in Fig.\,\ref{f:coraust}, is
identified by Toomer with HIP\,92953, near 0.9,$-$0.7. This implies
that the latitude of this source, $\beta=-23\degr$ ($\kappa\gamma$ in
Greek) is erroneous for $\beta=-20\degr20\arcmin$ , ($\kappa\gamma'$ in
Greek) and better fits the description {\em to the rear of this} (sc.\
the previous star).

\begin{figure*}
 \centerline{\includegraphics[angle=270,width=\columnwidth]{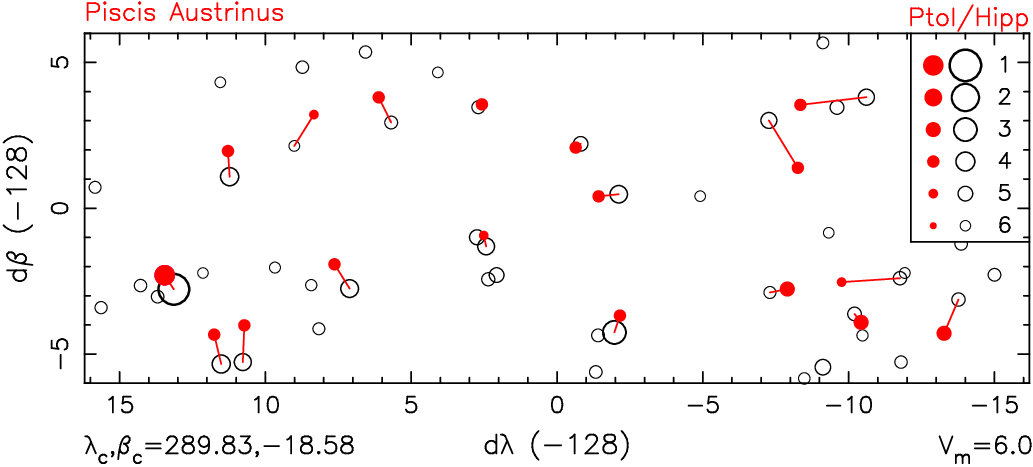}
\hfill\includegraphics[angle=270,width=\columnwidth]{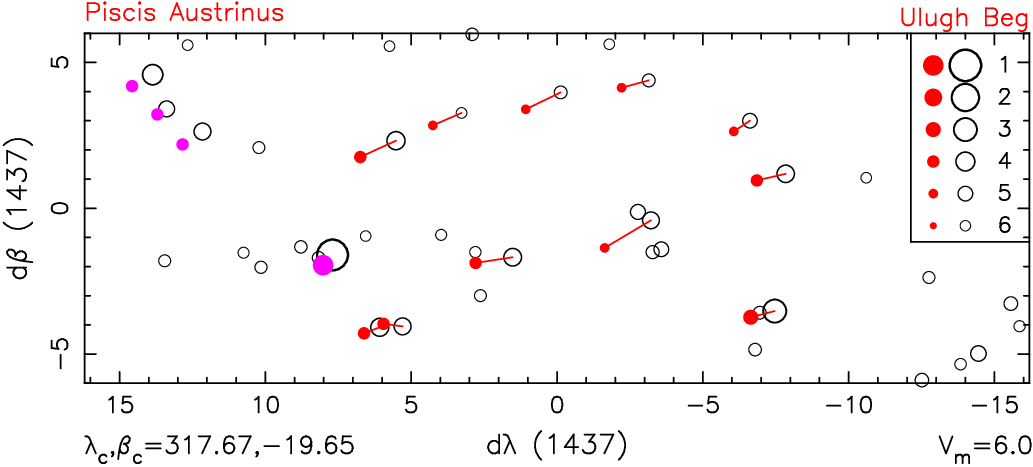}}
 \caption{Piscis Austrinus \label{f:piscaust}}
 \end{figure*}

\noindent {\bf P\,1017}, near 8.3,3.2 in Fig.\,\ref{f:piscaust}, is
identified by Toomer with HIP\,111138, which we consider too faint, at
$V$=6.4.

\subsection{Annotations to individual identifications in \ulughbeg
\label{s:idulug}}

\noindent {\bf U\,10, U\,11}, near $-$21.2,10.4 in Fig.\,\ref{f:ursamaior}.
We identify this pair with the two brightest nearby Hipparcos stars.

\noindent {\bf U\,36}, near 21.8,6.9 in Fig.\,\ref{f:ursamaior}, combines
the light of the close (44\arcsec) pair HIP\,63125/63121.

\begin{figure*}
 \centerline{\includegraphics[angle=270,width=\columnwidth]{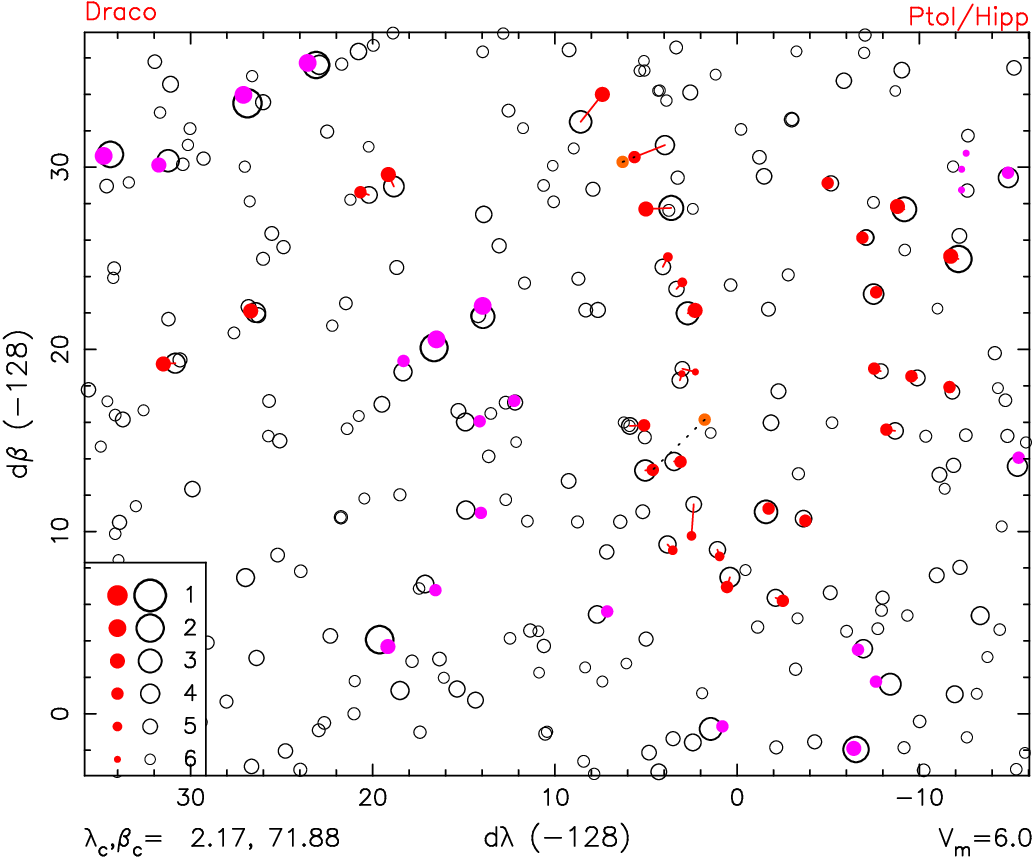}
\hfill
 \includegraphics[angle=270,width=\columnwidth]{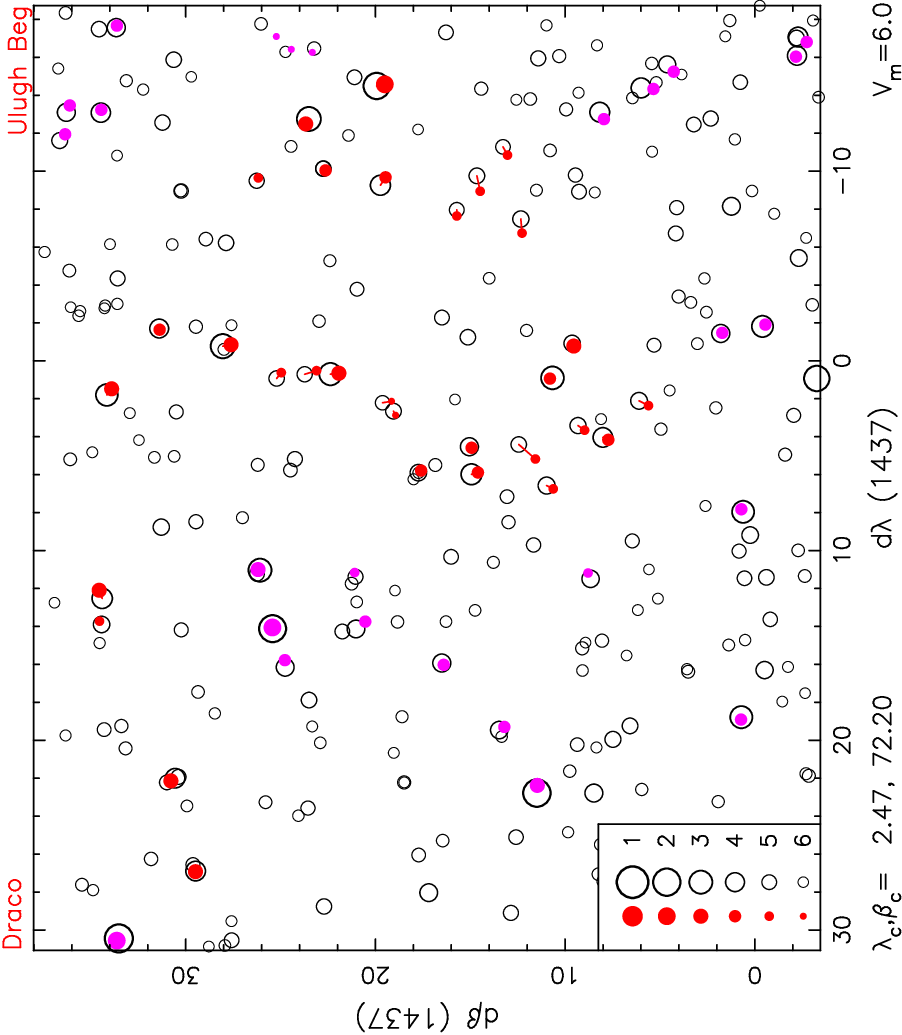}}
 \caption{Draco        \label{f:draco}}
 \end{figure*}

\noindent {\bf U\,60}, near 5.8,17.6 in Fig.\,\ref{f:draco}, combines
the light of the close (31\farcs4) pair HIP\,86614/86620.

\noindent {\bf U\,96}, near 7.8,15.0 in Fig.\,\ref{f:bootes}, is
identified by Knobel with HIP\,75973 =  $\nu^1$\,Boo, further and (marginally) fainter
(17\farcm1, $V$=5.0) than our counterpart, $\nu^2$\,Boo.

\noindent {\bf U\,98} is identified
differently in \ptolemaios, due to a slight shift in position;
see Fig.\,\ref{f:bootesdet}.

\noindent {\bf U\,99-U102}, see Fig.\,\ref{f:bootesdet}, are identified with
the same four stars in \ptolemaios\ and \ulughbeg, but in a
permutation:

\begin{tabular}{cccc}
 & \ptolemaios & \ulughbeg & Knobel \\
U\,99 &  73996 & 74087 & 73996 \\
U\,100 & 74087 & 73568 & 73745 \\
U\,101 & 73745 & 73745 & 74087 \\
U\,102& 73568 & 73996 & 73568
 \end{tabular}

\noindent where we also add the identifications of the stars in the catalogue
of Ulugh Beg according to Knobel.

\begin{figure*}
 \centerline{\includegraphics[angle=270,width=\columnwidth]{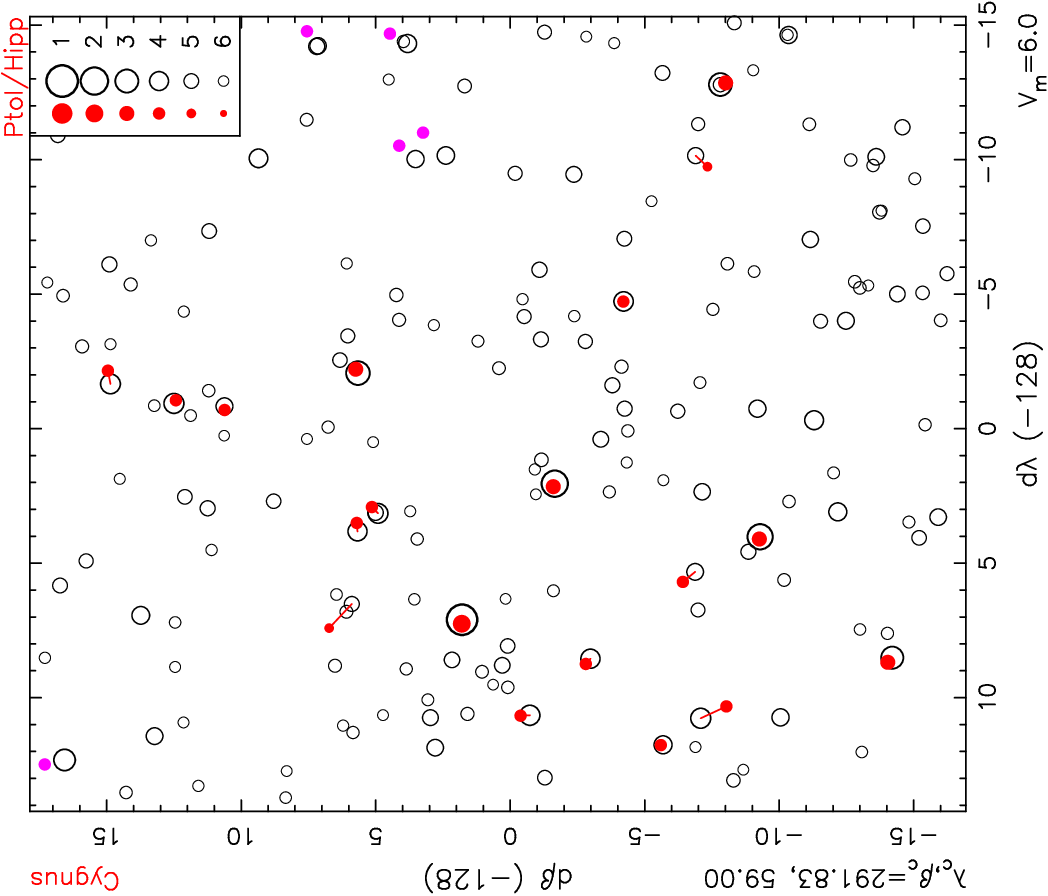}
\hfill\includegraphics[angle=270,width=\columnwidth]{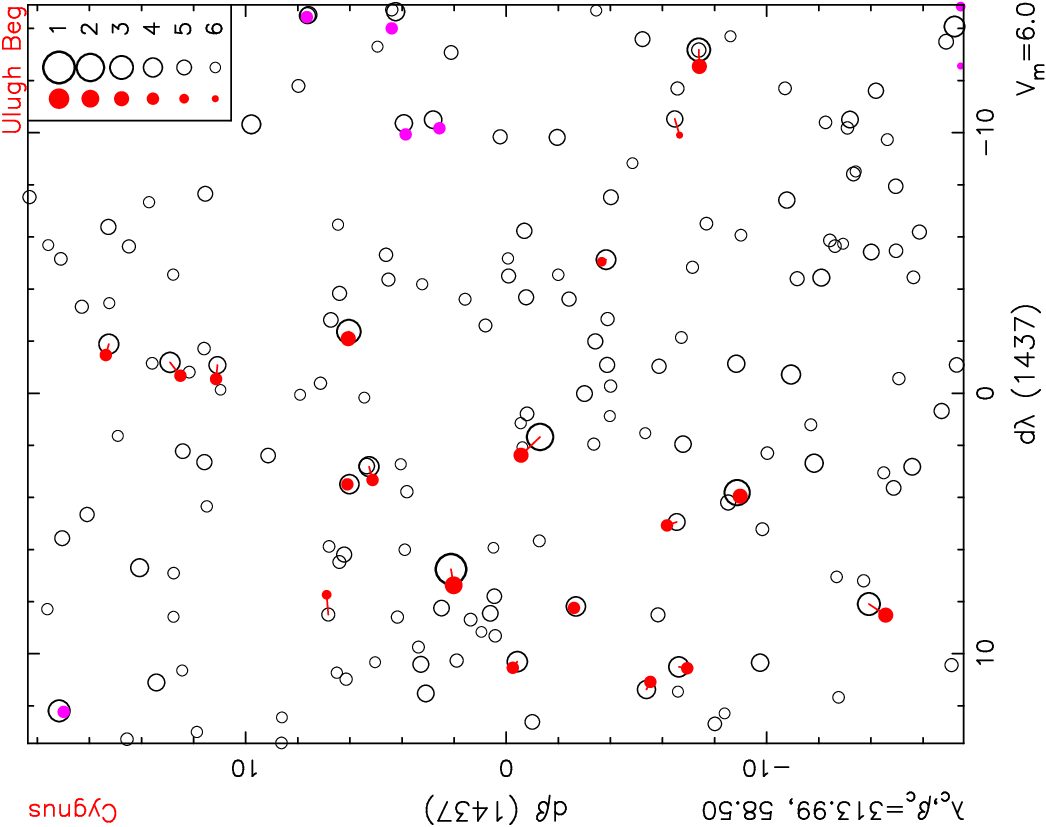}}
 \caption{Cygnus       \label{f:cygnus}}
 \end{figure*}

\noindent {\bf U\,158}, near $-$12.5,$-$7.4 in Fig.\,\ref{f:cygnus},
corresponds to the combined light of HIP\,95947 and HIP\,95951
($\beta^{1,2}$\,Cyg).

\noindent {\bf U\,172}, near 3.3,5.1 in Fig.\,\ref{f:cygnus}.
HIP\,99639 is only 5\farcm7 from our counterpart
HIP\,99675, and Knobel assigns the combined light of these two stars
as the counterpart for U\,172.

\noindent {\bf U\,174}, near 7.7,6.9  in Fig.\,\ref{f:cygnus},
is identified by us with the nearest star, but may also be identified
with HIP\,101138, brighter but further (V=4.9, $d$=102\farcm2),
the counterpart of the corresponding star in \ptolemaios, P\,175.
Knobel identifies U\,174 with HIP \,101138.

\begin{figure*}
 \centerline{\includegraphics[angle=270,width=\columnwidth]{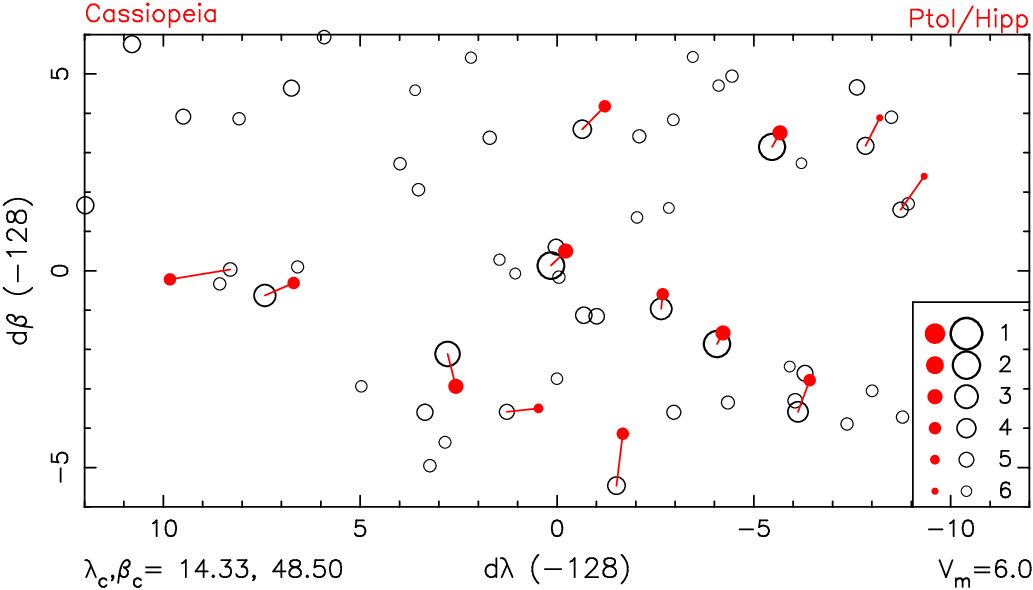}
\hfill\includegraphics[angle=270,width=\columnwidth]{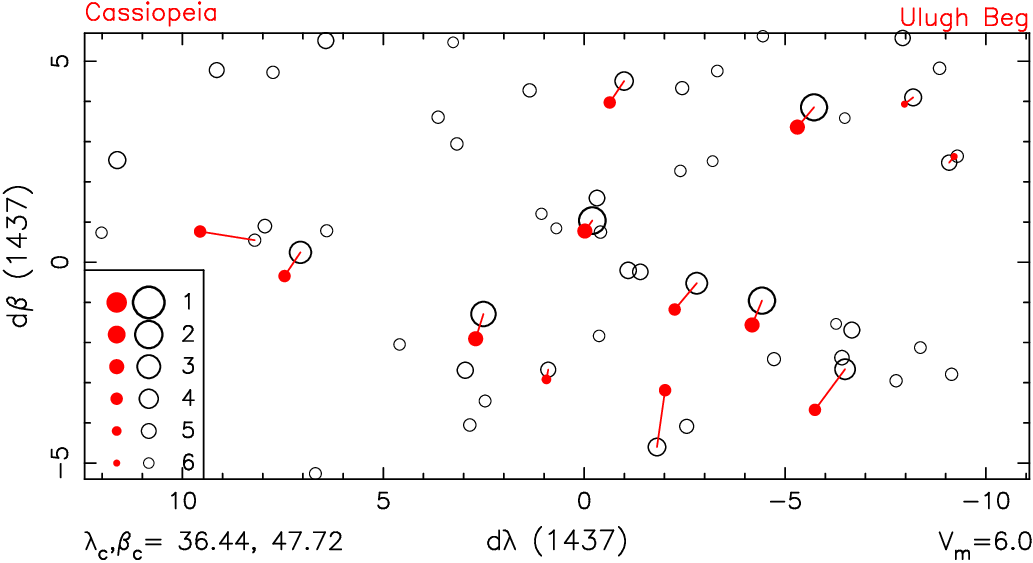}}
 \caption{Cassiopeia   \label{f:cassiopeia}}
 \end{figure*}

\noindent {\bf U\,183}, near 9.6,0.8 in Fig.\,\ref{f:cassiopeia}. An
alternative counterpart is HIP\,9312, slightly brighter, but further
($V$=5.3, $d$=98\farcm1) than our  counterpart.
Knobel identifies U\,183 with HIP\,11569 ($\iota$\,Cas), near
11.6,2.5. HIP\,9312 is our counterpart to the corresponding star
in \ptolemaios, P\,184.

\noindent {\bf U\,190}: see note with P\,191.

\begin{figure*}
 \centerline{\includegraphics[angle=270,width=0.9\columnwidth]{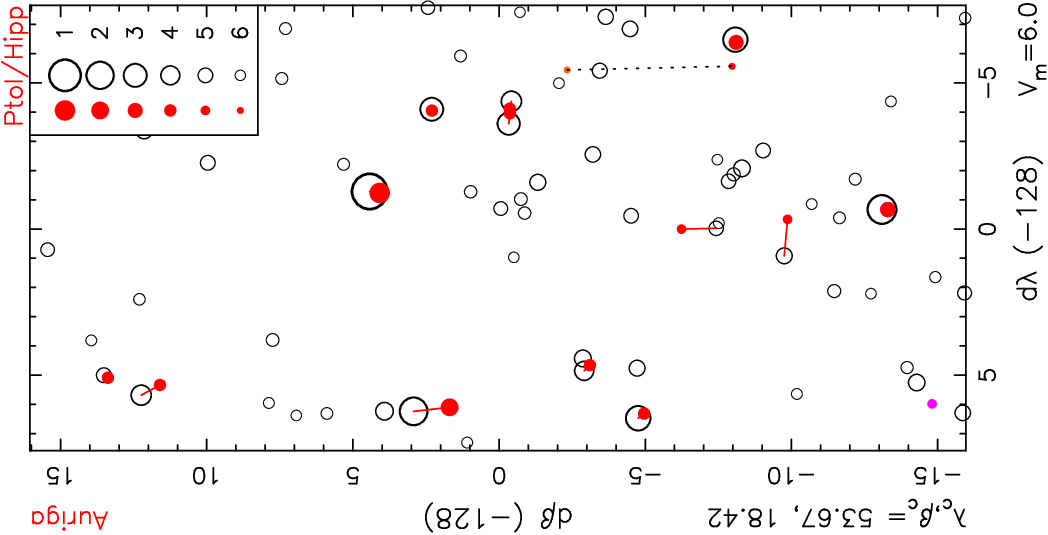}
\hfill\includegraphics[angle=270,width=\columnwidth]{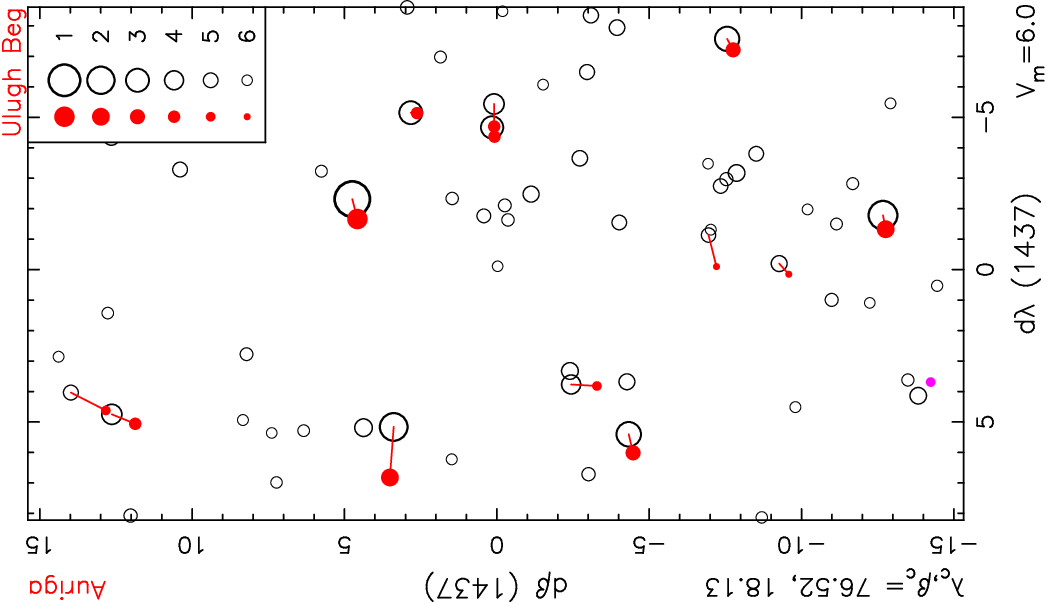}}
 \caption{Auriga       \label{f:auriga}}
 \end{figure*}

\noindent {\bf U\,226} leads {\bf U\,227}, i.e. has a smaller longitude, and we
identify these entries accordingly, with HIP\,23453 and HIP\,23767.
The pair is near $-$4.5,0.1 in Fig.\,\ref{f:auriga}.
In the corresponding pair in \ptolemaios,  P\,227 trails P\,228, and
they are identified accordingly with HIP\,23767 and HIP\,23453,
respectively. Knobel identifies U\,226 with HIP\,23767 and U\,227
with HIP\,23453.

\begin{figure*}
 \centerline{\includegraphics[angle=270,width=\columnwidth]{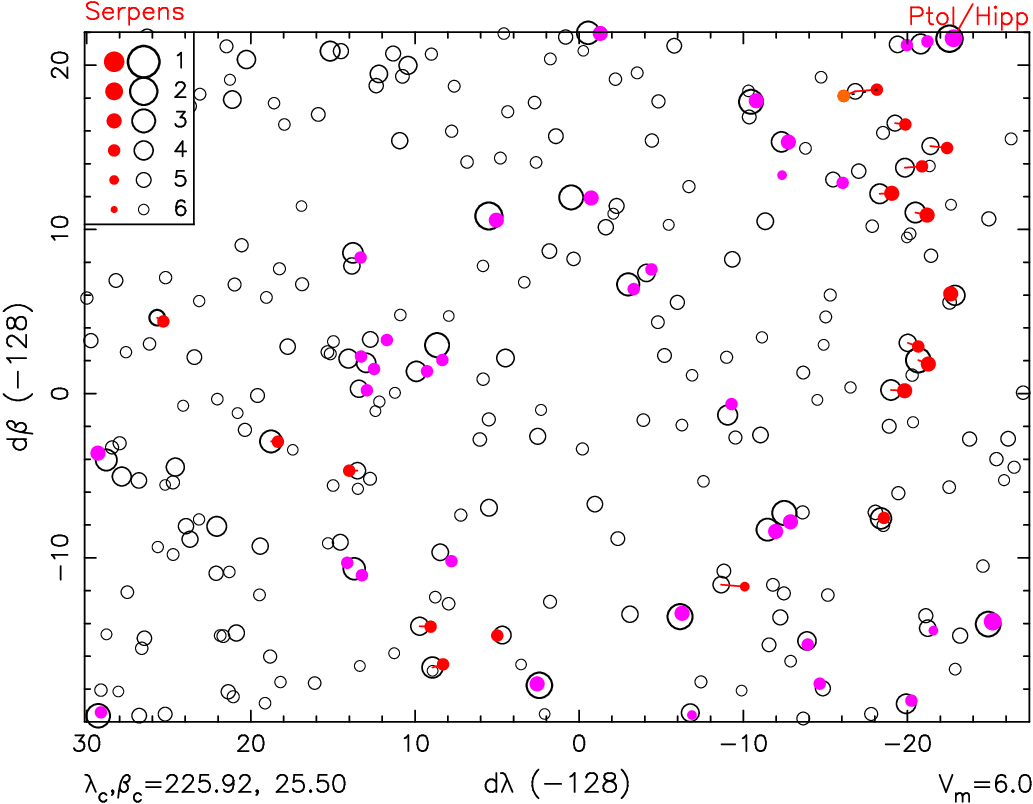}
\hfill\includegraphics[angle=270,width=\columnwidth]{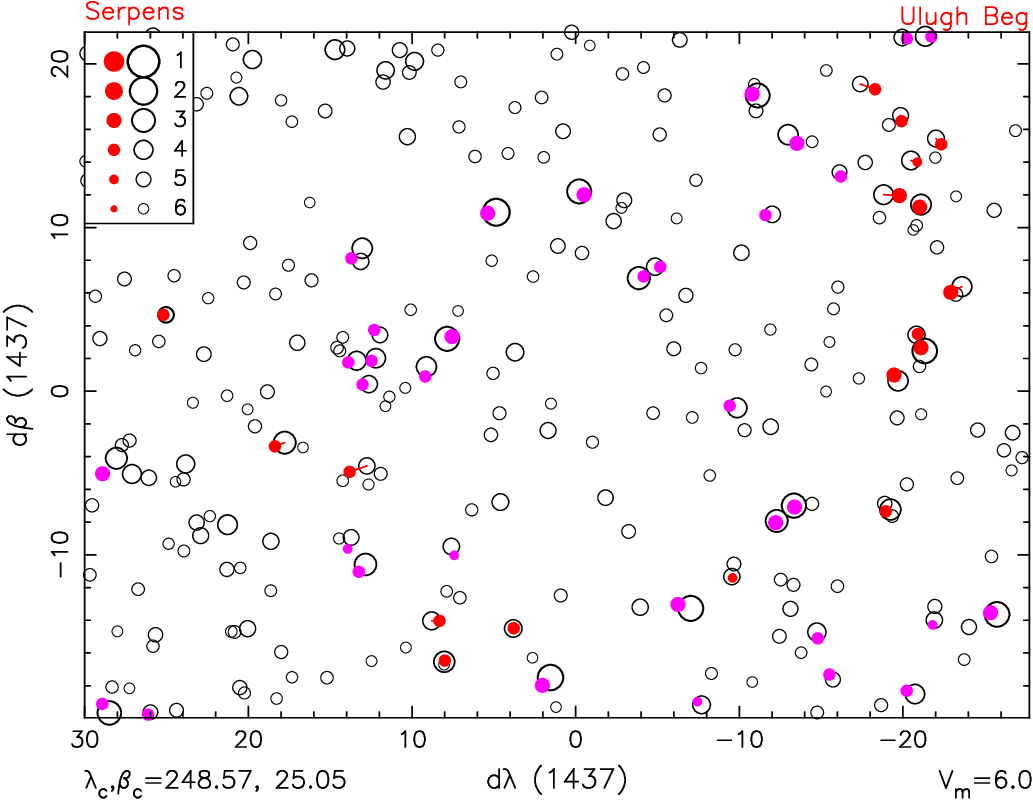}}
 \caption{Serpens      \label{f:serpens}}
 \end{figure*}

\noindent {\bf U\,278}, near 25.2,4.7 in Fig.\,\ref{f:serpens}, is the
combined light of the close (15\farcs5) pair HIP\,92946 and HIP\,92951.

\begin{figure*}
 \centerline{\includegraphics[angle=270,width=\columnwidth]{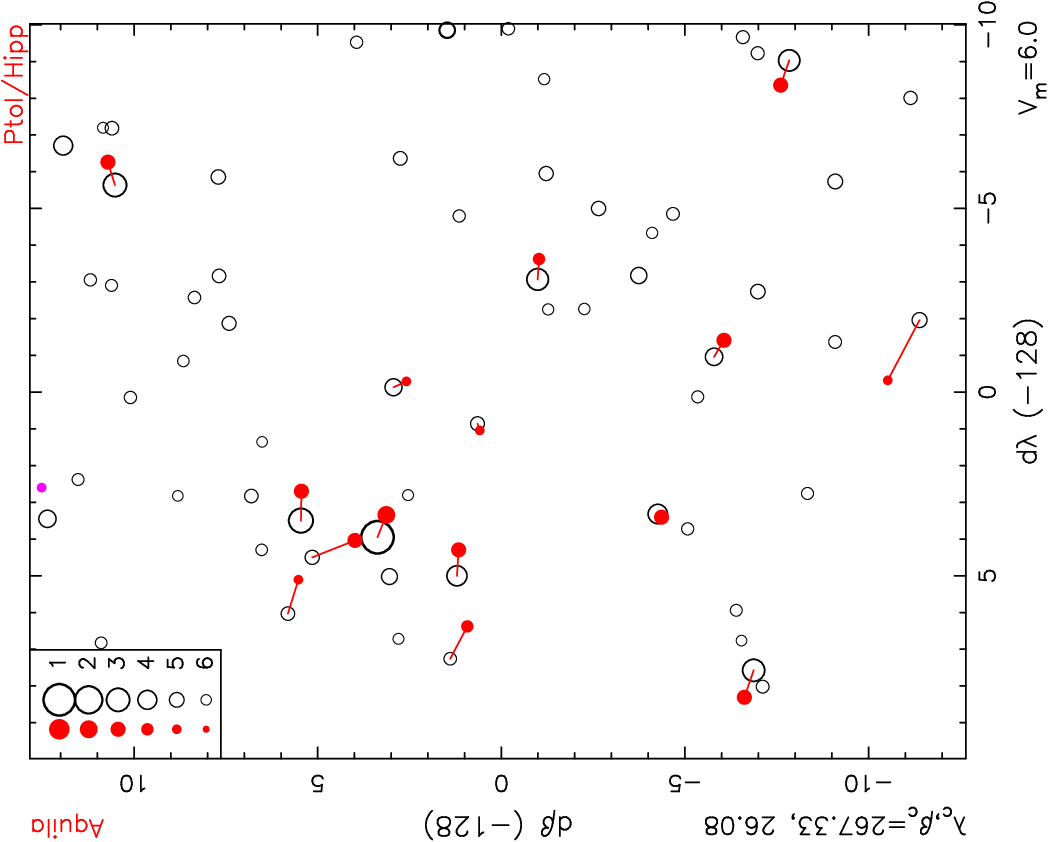}
\hfill\includegraphics[angle=270,width=\columnwidth]{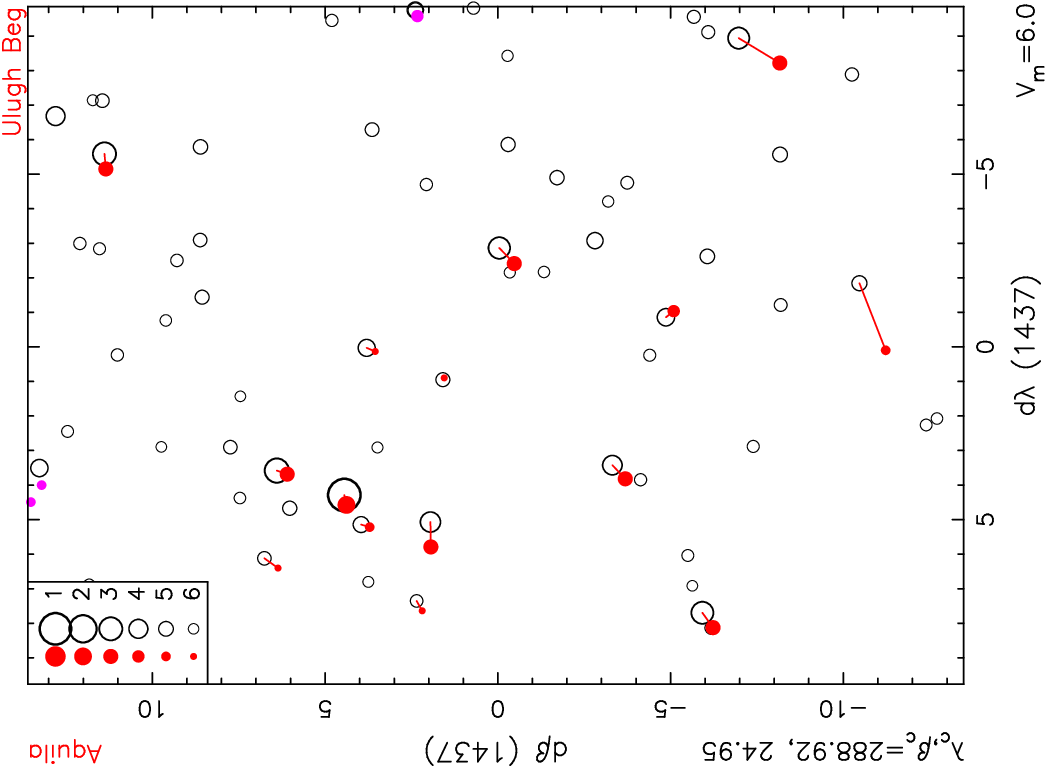}}
 \caption{Aquila       \label{f:aquila}}
 \end{figure*}

\noindent {\bf U\,287}, near 5.2,3.7 in Fig.\,\ref{f:aquila}, is south of
U\,286, just next to it. In \ptolemaios, the corresponding P\,289
is north of P\,288, and the identification differs accordingly
(Fig.\,\ref{f:aquila}).

\begin{figure*}
 \centerline{\includegraphics[angle=270,width=0.9\columnwidth]{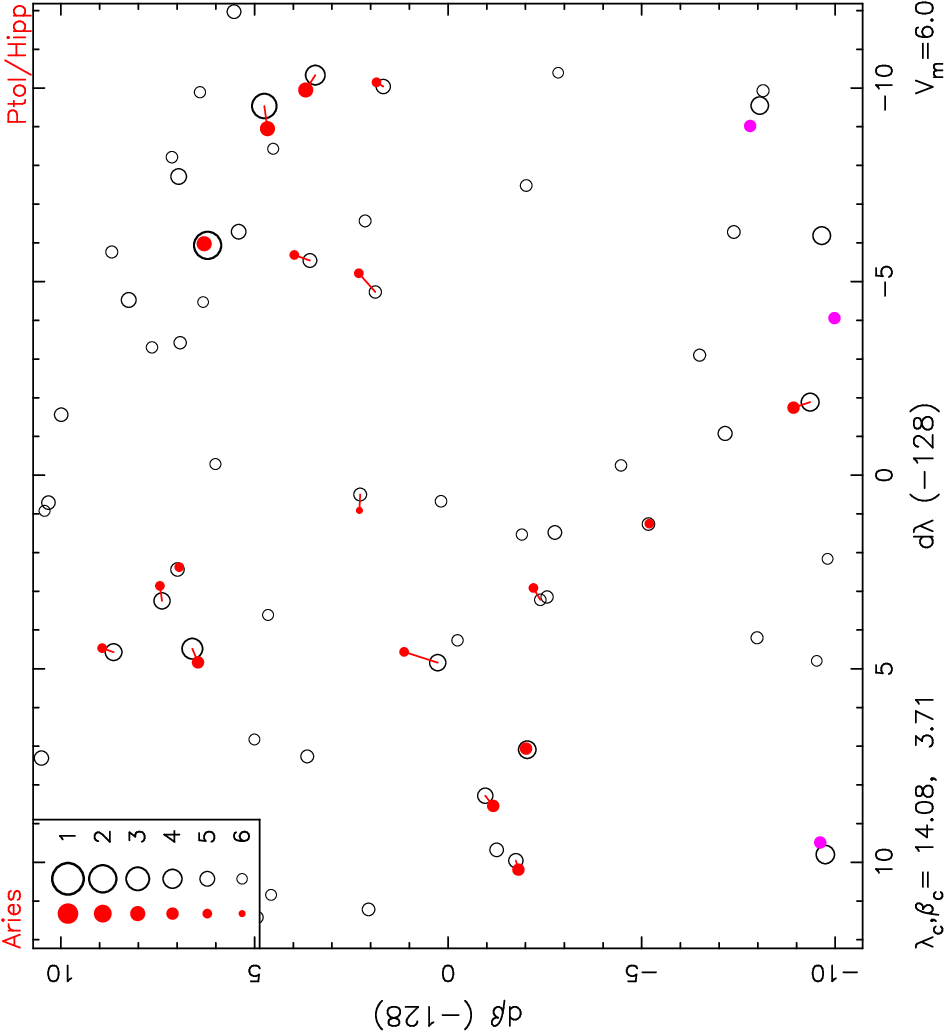}
\hfill\includegraphics[angle=270,width=0.94\columnwidth]{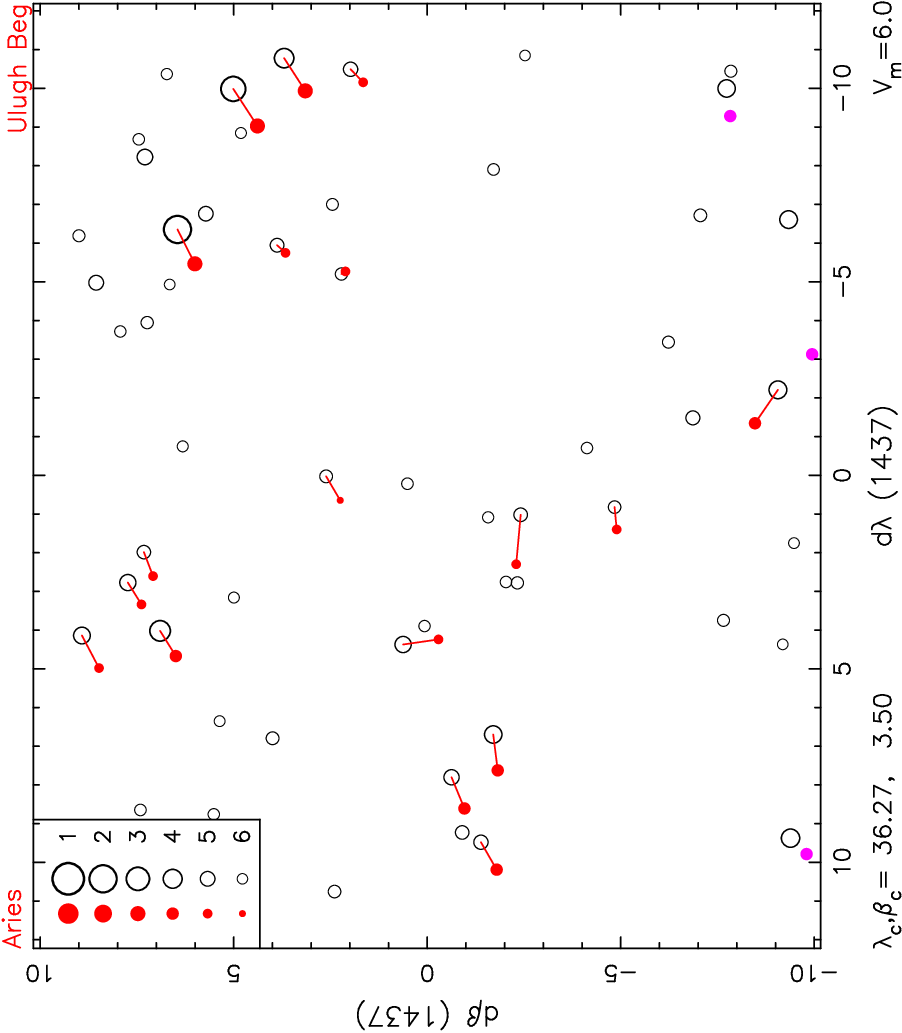}}
 \caption{Aries        \label{f:aries}}
 \end{figure*}

\noindent {\bf U\,370}, near 2.3,$-$2.3 in Fig.\,\ref{f:aries}, is
identified by Knobel with the pair HIP\,13654, HIP\,13702
($\rho^{2,3}$\,Ari), which is also the identification of the
corresponding star in \ptolemaios, P\,372. We agree with this possibility, but slightly
prefer the further but somewhat brighter HIP\,13165 ($\pi$\,Ari),
in view of the offsets of other identifications near it.

\noindent {\bf U\,406}, in Fig.\,\ref{f:pleiades}, is
identified by Knobel with HIP\,17531 (Taygeta), 
the counterpart of the corresponding star in \ptolemaios, P\,409.

\noindent {\bf U\,409},  in Fig.\,\ref{f:pleiades}, is
identified by Knobel with HIP\,17954, in accordance
with the description in the catalogue {\em an exterior and small star
  of the Pleiades}, and with the counterpart of the corresponding star
in \ptolemaios, P\,412. However, its different position in \ulughbeg,
and its magnitude of 4 in the catalogue, lead us to prefer Alcyone as
the counterpart.

\begin{figure*}
 \centerline{\includegraphics[angle=270,width=\columnwidth]{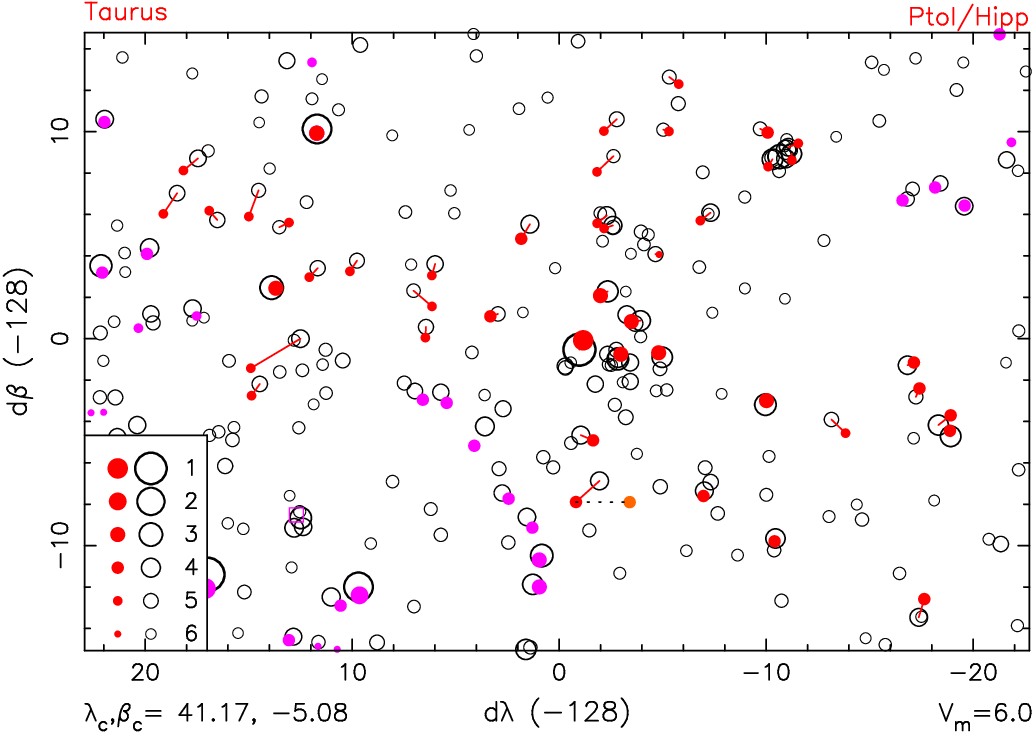}
\hfill\includegraphics[angle=270,width=0.94\columnwidth]{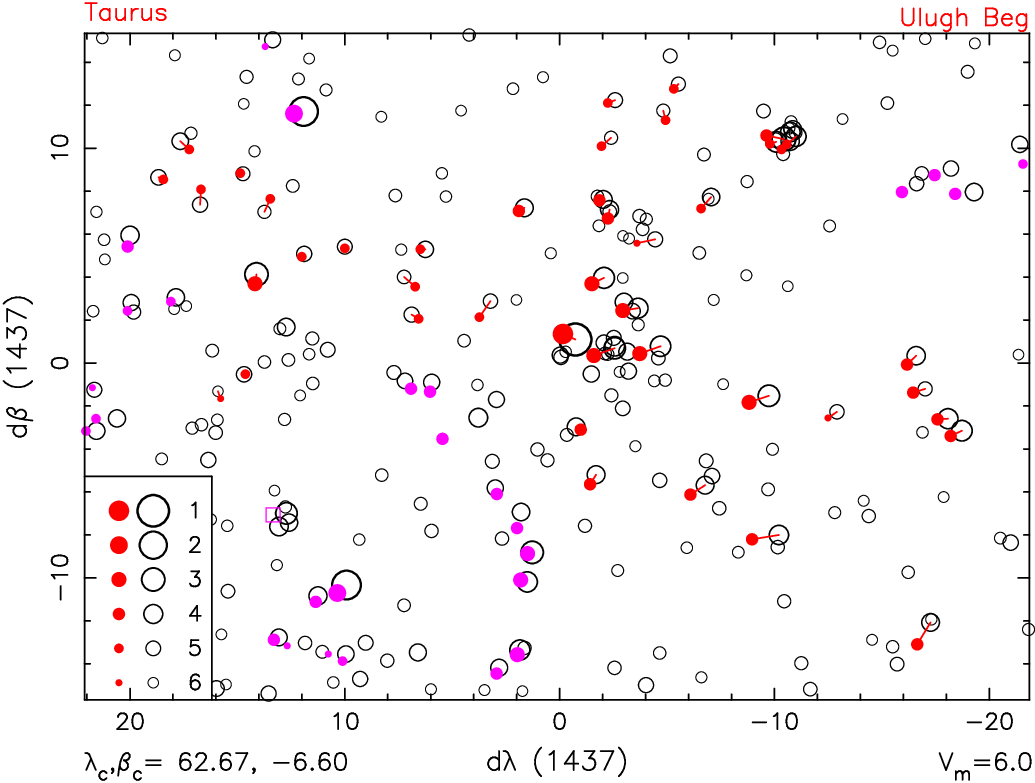}}
 \caption{Taurus       \label{f:taurus}}
 \end{figure*}

\noindent {\bf U\,414} and {\bf U\,415}, near 15,$-$1 in Fig.\,\ref{f:taurus},
have different identifications than their counterparts in \ptolemaios,
P\,417, P\,418, due to different positions. The counterpart of P\,418
is that of U\,414.

\begin{figure*}
 \centerline{\includegraphics[angle=270,width=\columnwidth]{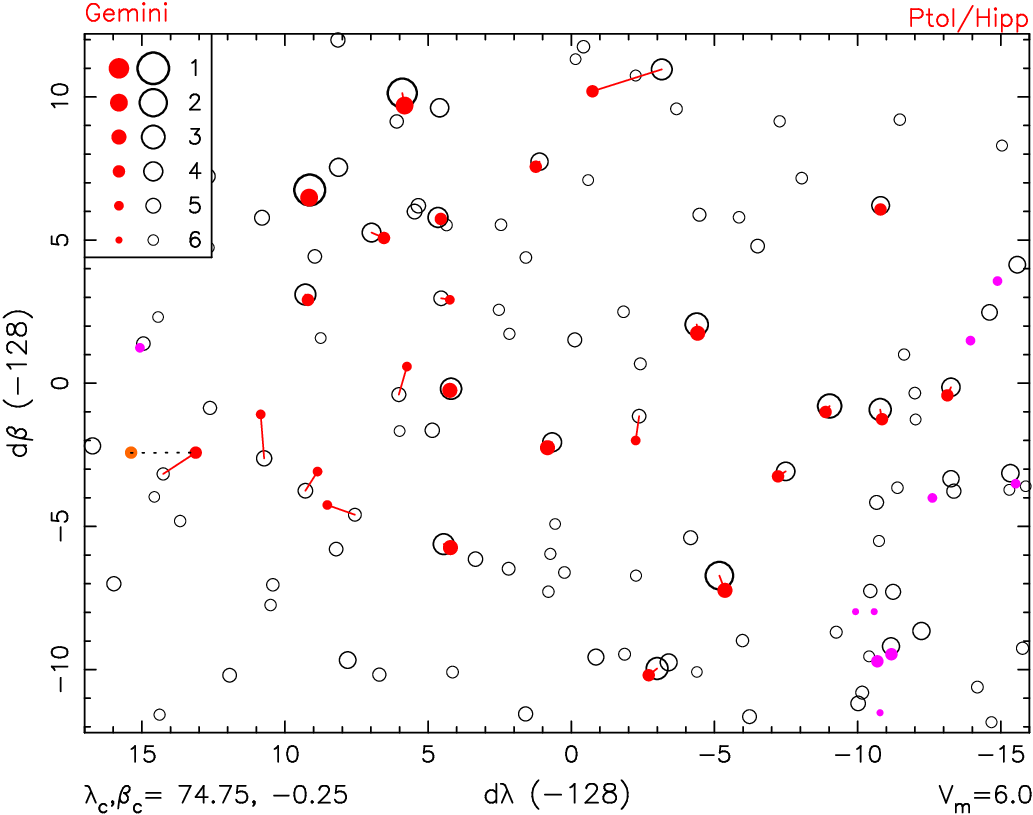}
\hfill\includegraphics[angle=270,width=\columnwidth]{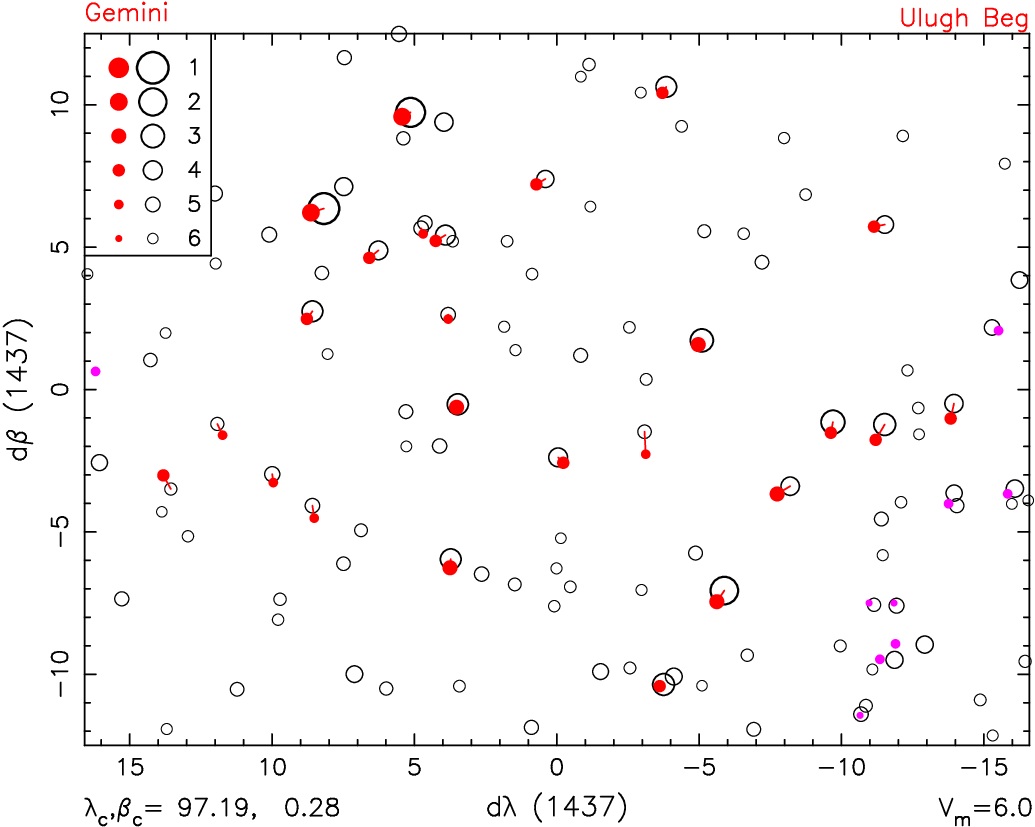}}
 \caption{Gemini       \label{f:gemini}}
 \end{figure*}

\noindent {\bf U\,429}, near 4.7,5.5 in Fig.\,\ref{f:gemini}, has a very
different latitude than its counterpart in \ptolemaios, P\,431, and
accordingly a different counterpart.

\noindent {\bf U\,442, U\,443} and {\bf U\,444}, near 11.7,$-$1.6, 10.0,$-$3.3
and 8.5,$-$4.5 in Fig.\,\ref{f:gemini} are shifted in longitude
with respect to their counterparts in \ptolemaios, P\,445-447, such
that the counterpart of P\,445 is that of U\,443 and that of P\,446
that of U\,444.

\noindent {\bf U\,446} is Praesepe; the distance given in the catalogue is
to $\epsilon$\,Cnc.

\begin{figure*}
 \centerline{\includegraphics[angle=270,width=\columnwidth]{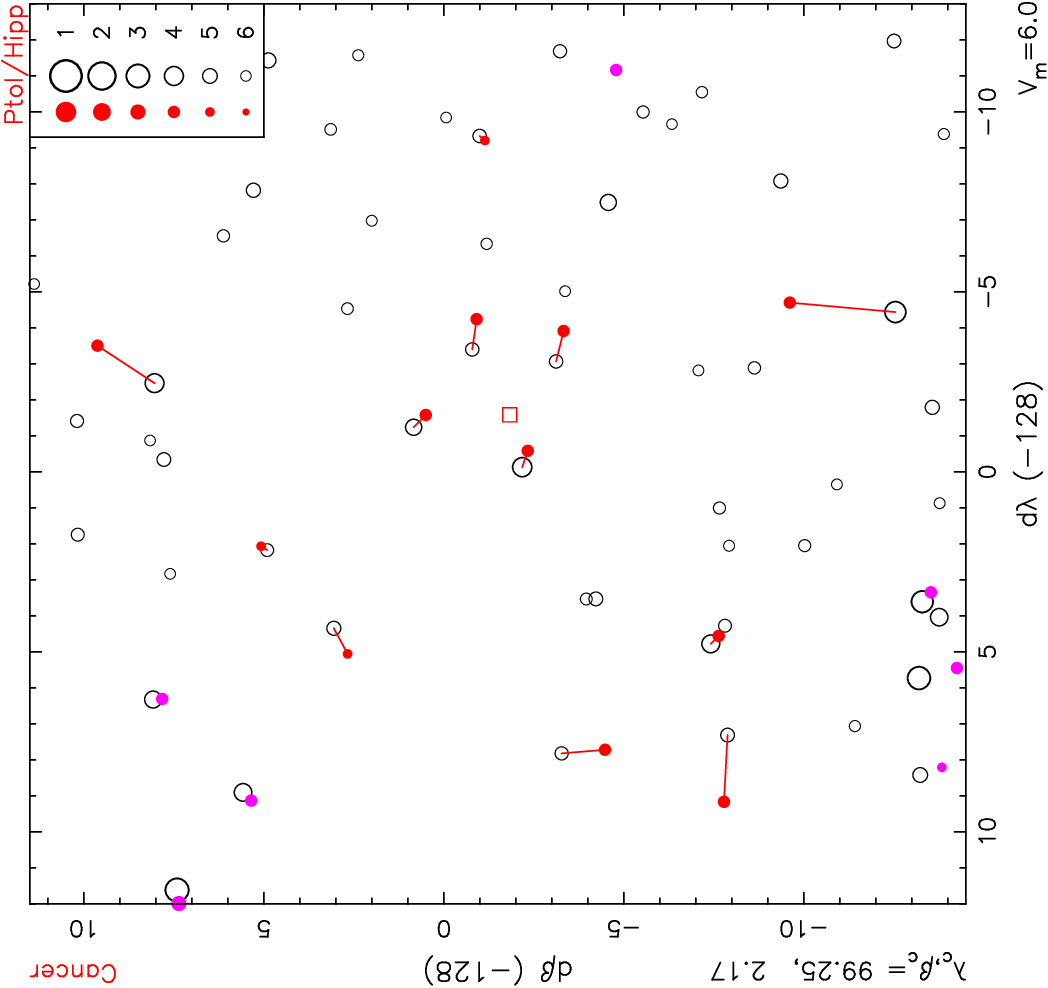}
\hfill\includegraphics[angle=270,width=\columnwidth]{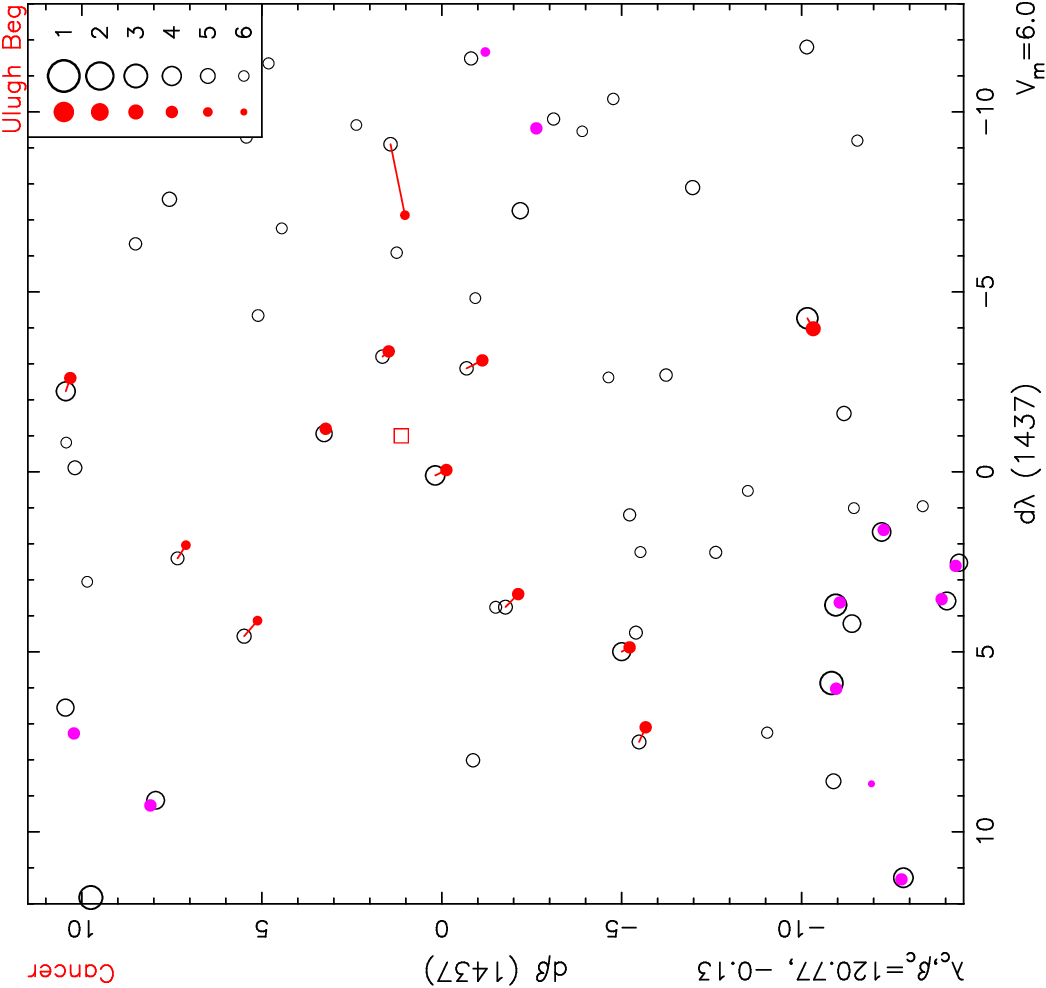}}
 \caption{Cancer       \label{f:cancer}}
 \end{figure*}

\noindent {\bf U\,455}, near 3.4,$-$2.1 in Fig.\,\ref{f:cancer}, corresponds
to P\,458 in \ptolemaios, which has a position closer to  HIP\,45410,
and is identified accordingly in \ptolemaios.

\noindent {\bf U\,492, U\,493}: Knobel's identifications are ambiguous between
7\,Leo and 23\,Leo or 7\, Com and 23\,Com; Com agrees with our identifications.

\begin{figure*}
 \centerline{\includegraphics[angle=270,width=\columnwidth]{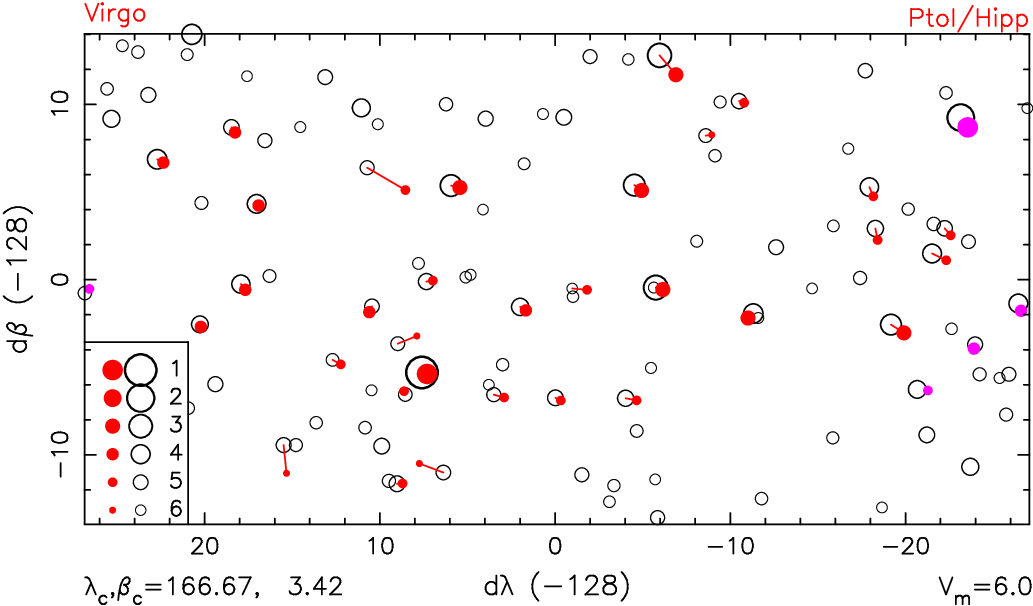}
\hfill\includegraphics[angle=270,width=\columnwidth]{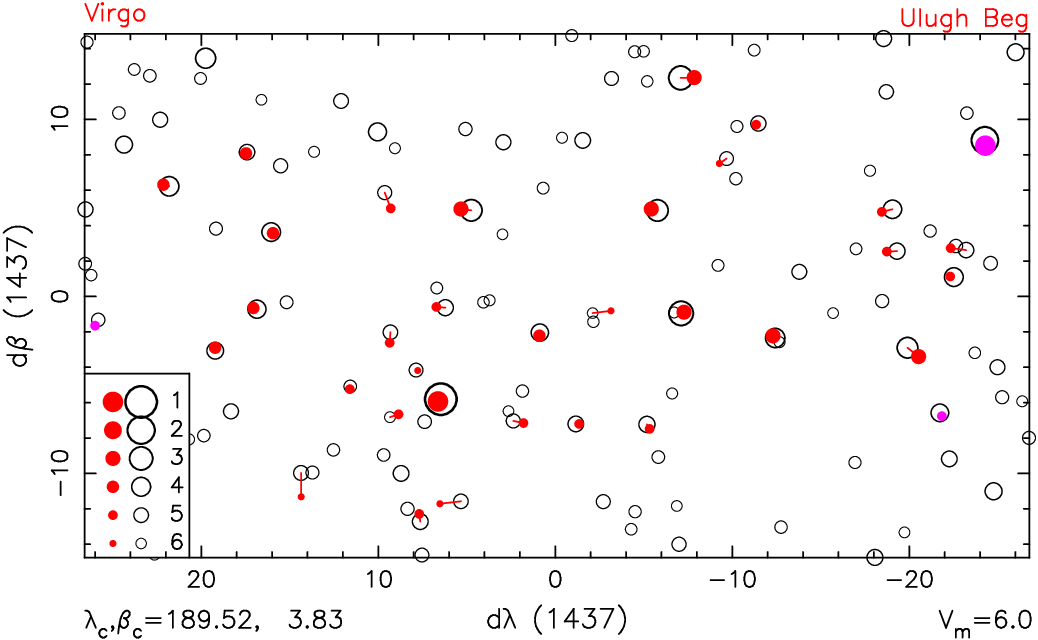}}
 \caption{Virgo        \label{f:virgo}}
 \end{figure*}

\noindent {\bf U\,512, U\,513}, near 11.6,$-$5.2 and 8.9,$-$6.7 in
Fig.\,\ref{f:virgo}. P\,516 has the same counterpart as U\,512,
but P\,515 is identified with HIP\,65581, near 7.4,$-$7.1, so that the
pair P\,515, P\,516 is shifted with respect to the corresponding pair
U\,512, U\,513.

\noindent {\bf U\,592}, near 10.1,$-$5.5 in Fig.\,\ref{f:sagittarius}, is
identified by Knobel with the pair HIP\,100469, HIP\,100591
($\kappa^{1,2}$\,Sgr) near 12.4,$-$14.2, which must be wrong.

\begin{figure*}
 \centerline{\includegraphics[angle=270,width=\columnwidth]{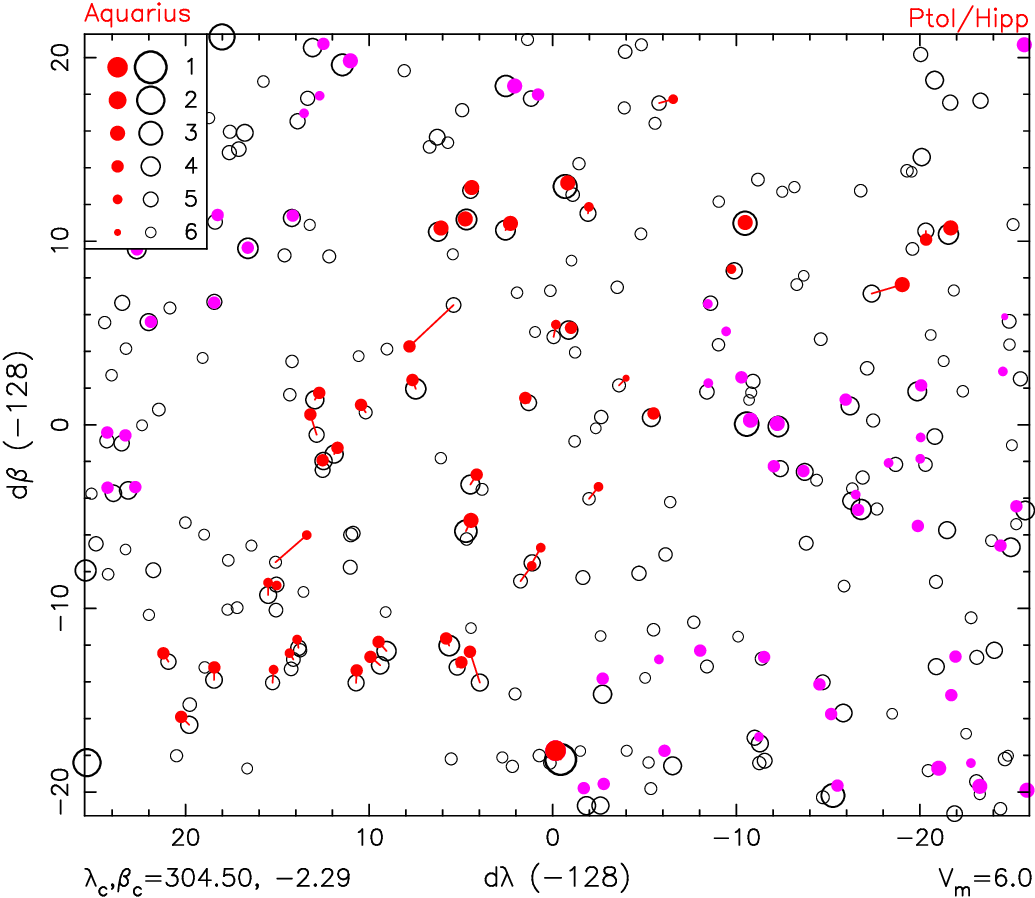}
\hfill\includegraphics[angle=270,width=\columnwidth]{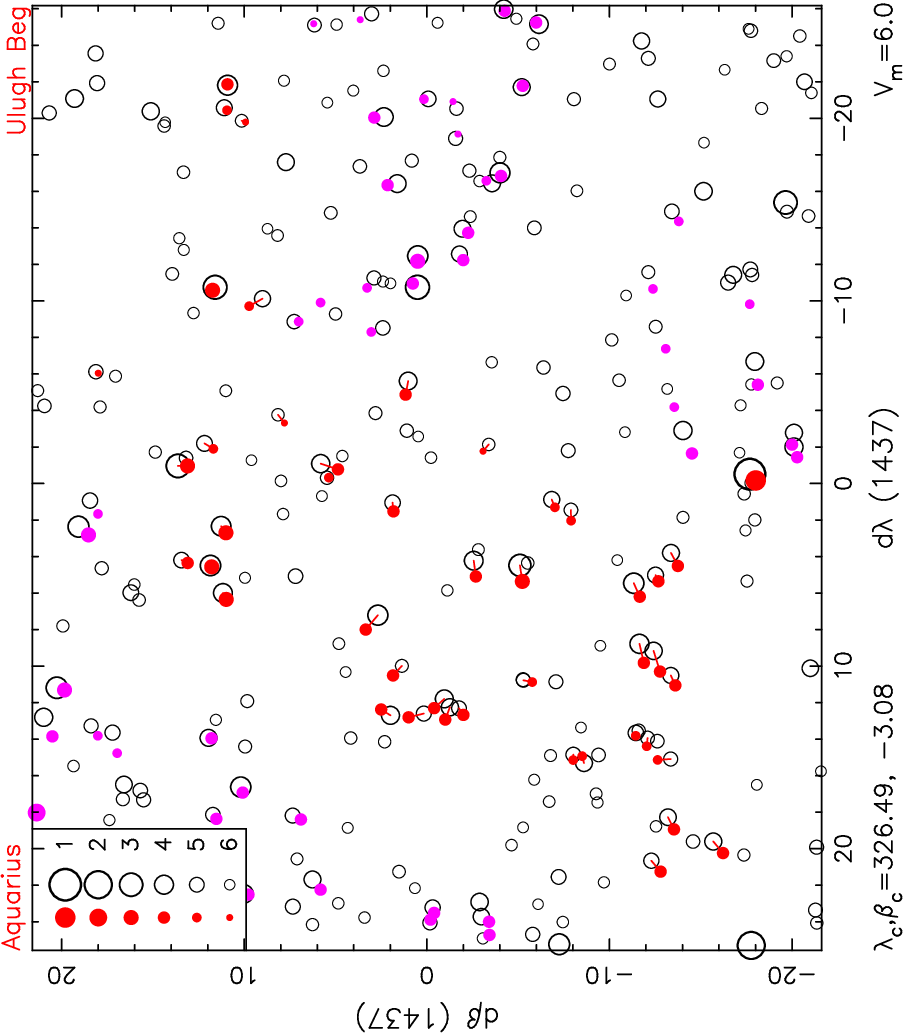}}
 \caption{Aquarius     \label{f:aquarius}}
 \end{figure*}

\noindent {\bf U\,653} and {\bf U\,654}, near 12.8,$-$1.5 in
Fig.\,\ref{f:aquarius}: these two stars correspond to a single entry
in \ptolemaios, P\,657. The entry P\,651 has no counterpart in
\ulughbeg.

\noindent {\bf U\,655}, near 10.9,$-$5.8 in Fig.\,\ref{f:aquarius},
corresponds to the close (2\farcm4) pair HIP\,115125/6.

\noindent {\bf U\,656, U\,657}, near 15,$-$8.3 in Fig.\,\ref{f:aquarius},
are identified by us with HIP\,116971 and HIP\,116758, respectively,
i.e.\ we preserve the north-south ordering. Knobel preserves the
ordering in longitude,  and thus identifies U\,656
with HIP\,116758 and U\,657 with HIP\,116971.
The corresponding stars in \ptolemaios, P\,659 and P\,660,
are identified with HIP\,116758 and  HIP\,116971, respectively.
Our identification of the pair therefore implies that their ordering
is switched in \ulughbeg\ with respect to \ptolemaios.

\noindent {\bf U\,660}, near 15.1,$-$12.6 in Fig.\,\ref{f:aquarius}, is
identified by Knobel with HIP\,117218, near 14.1,$-$12.6.

\begin{figure*}
 \centerline{\includegraphics[angle=270,width=\columnwidth]{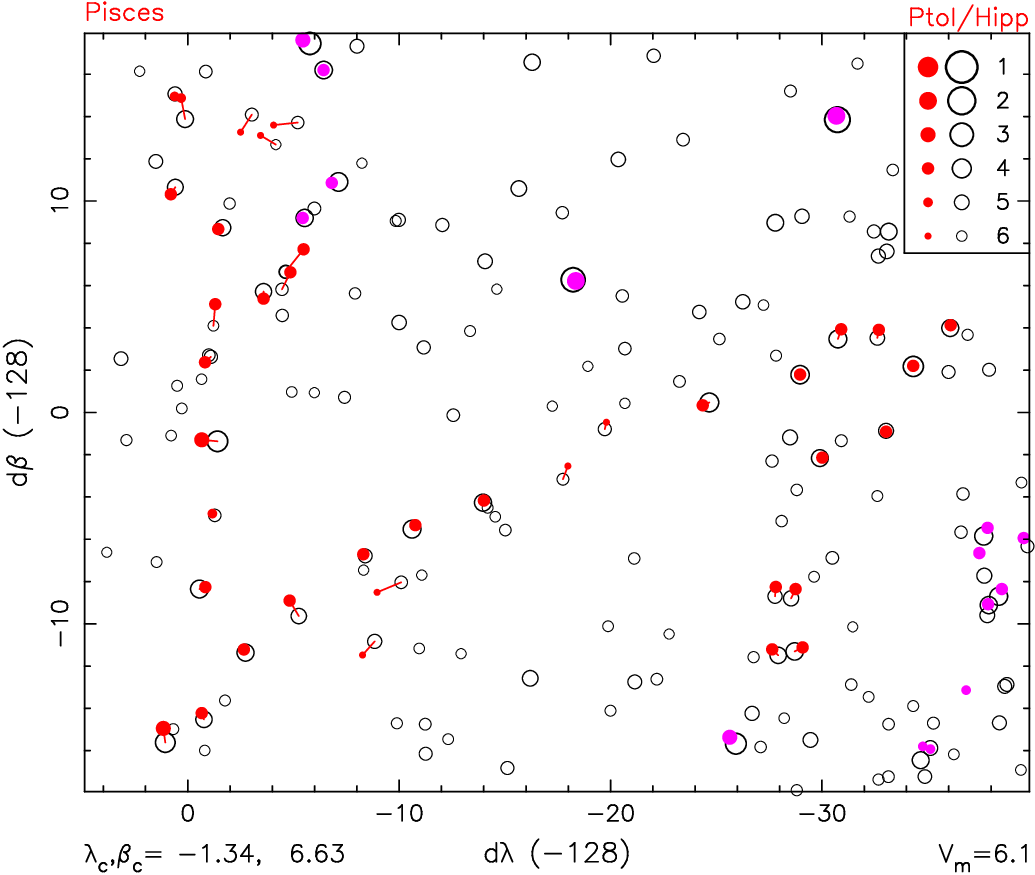}
\hfill\includegraphics[angle=270,width=\columnwidth]{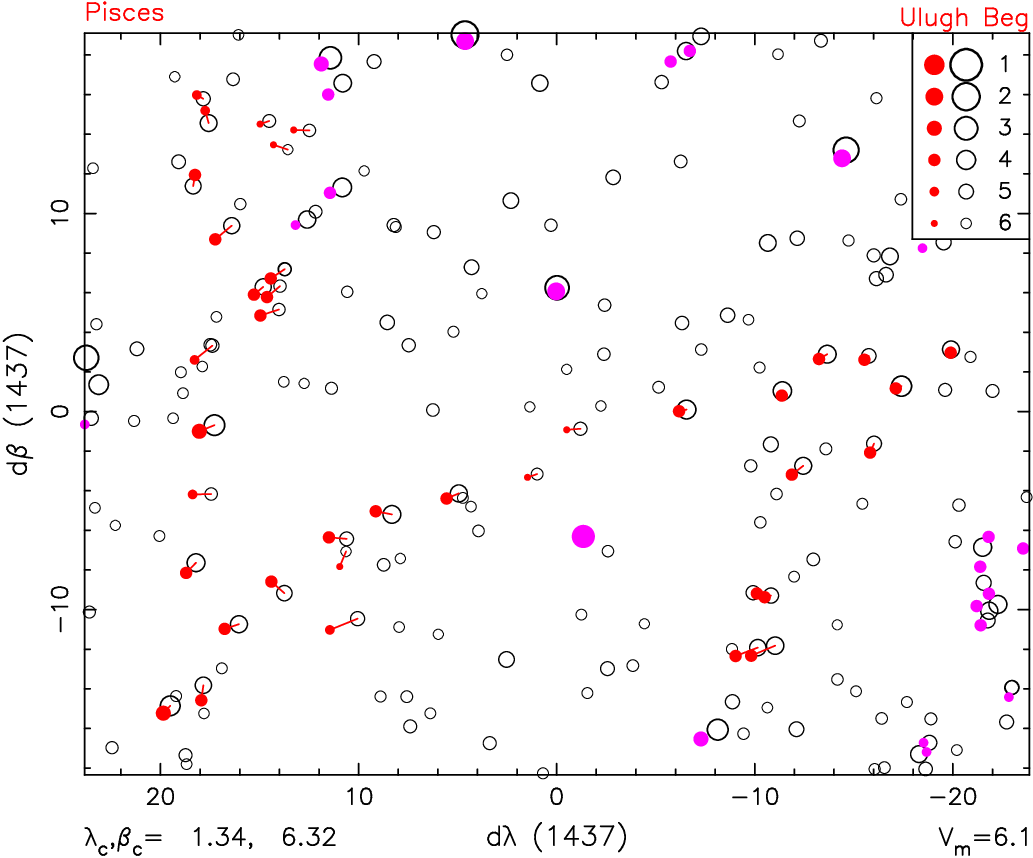}}
 \caption{Pisces       \label{f:pisces}}
 \end{figure*}

\noindent {\bf U\,684}, near 11.0,$-$7.8 in Fig.\,\ref{f:pisces}, is
identified by Knobel with HIP\,5346, the counterpart of the
corresponding star in \ptolemaios, P\,687.

\begin{figure*}
 \centerline{\includegraphics[angle=270,width=\columnwidth]{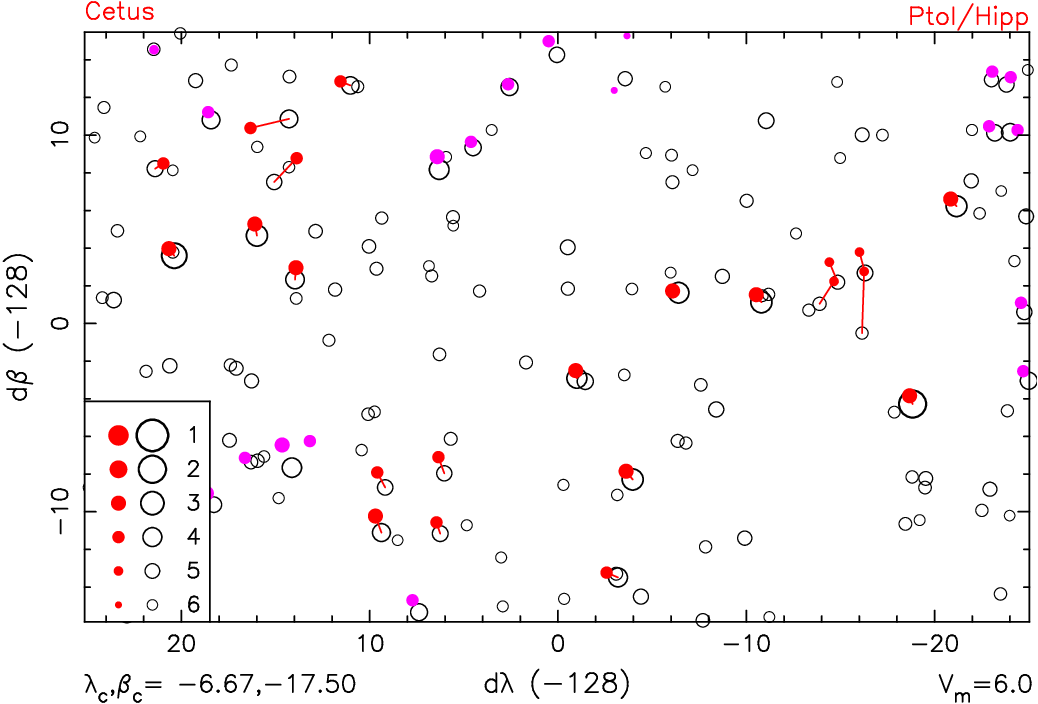}
\hfill\includegraphics[angle=270,width=\columnwidth]{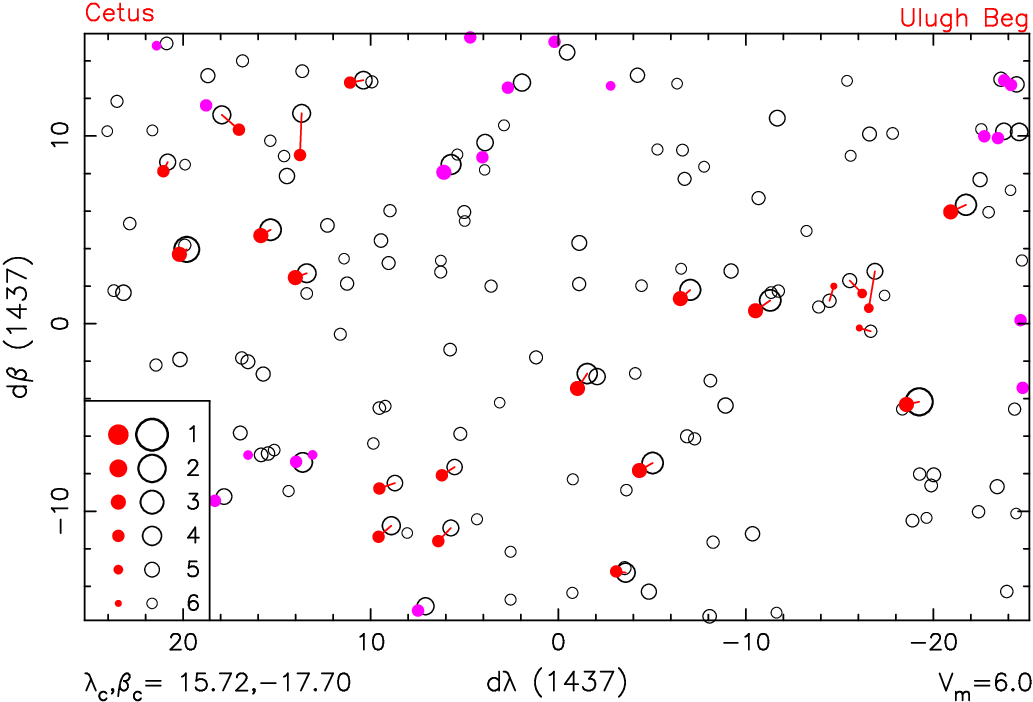}}
 \caption{Cetus        \label{f:cetus}}
 \end{figure*}

\noindent {\bf U\,713}, near 13.8,9.0 in Fig.\,\ref{f:cetus}, is identified
tentatively by Knobel with HIP\,12093 ($\nu$\,Cet), near 14.5,7.9,
closer but fainter ($d$=79\farcm1, $V$=4.9) than our preferred counterpart.

\begin{figure*}
 \centerline{\includegraphics[angle=270,width=\columnwidth]{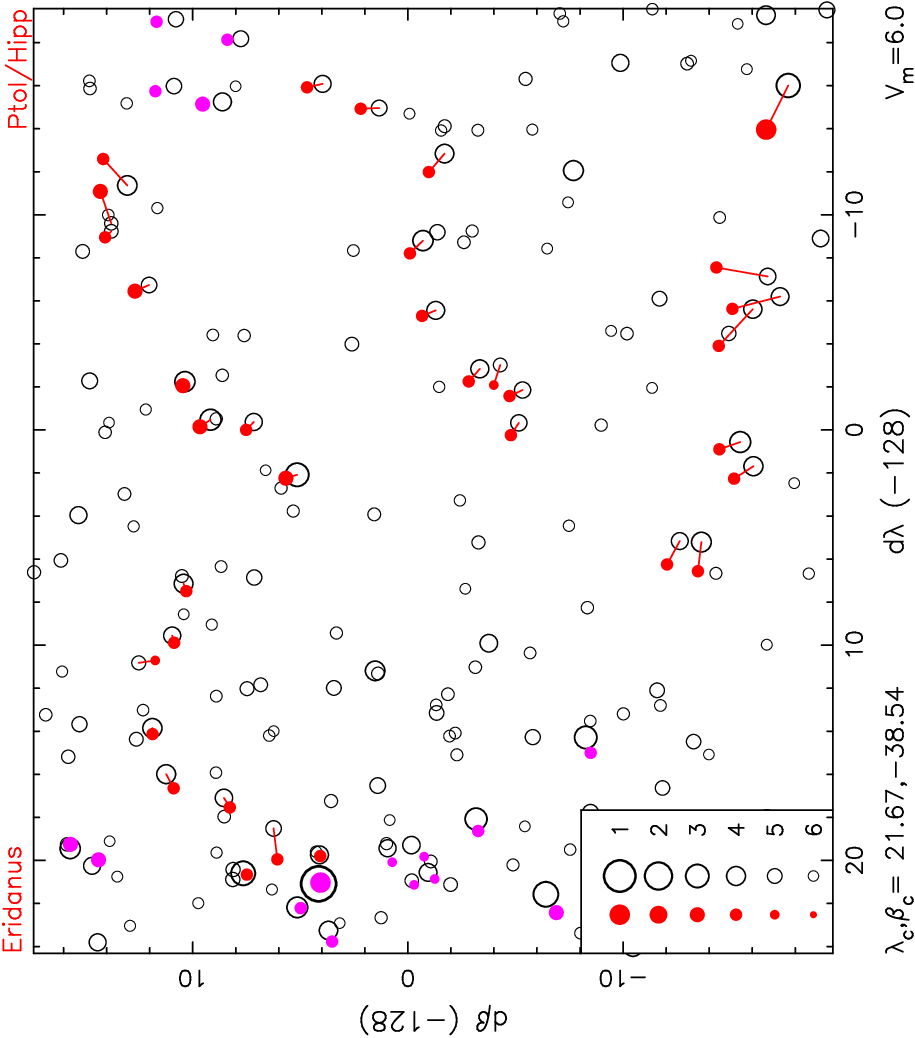}
\hfill\includegraphics[angle=270,width=\columnwidth]{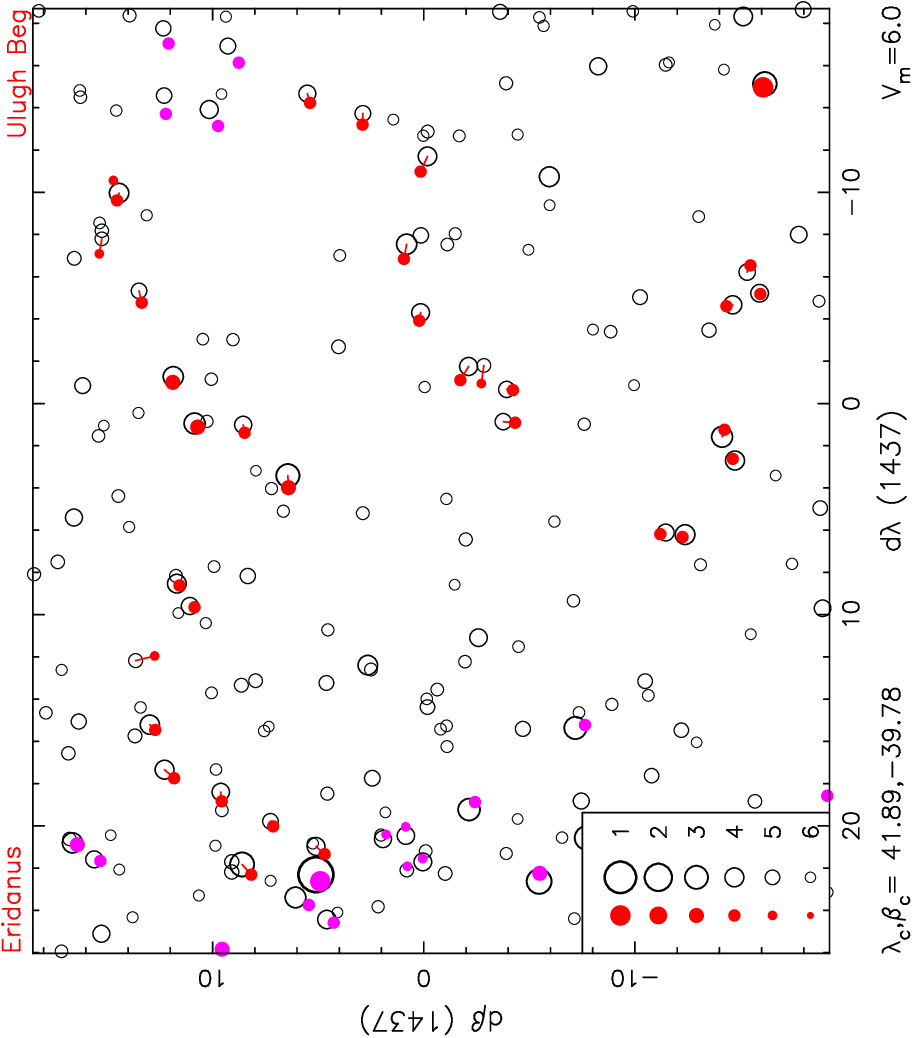}}
 \caption{Eridanus     \label{f:eridanus}}
 \end{figure*}

\noindent {\bf U\,785}, near $-$10.6,14.7 in Fig.\,\ref{f:eridanus}, is
identified by Knobel with HIP\,13421, which at $V$=6.3 is considered
by us too faint to be a probable counterpart.
We leave U\,785 unidentified.

\begin{figure*}
 \centerline{\includegraphics[angle=270,width=\columnwidth]{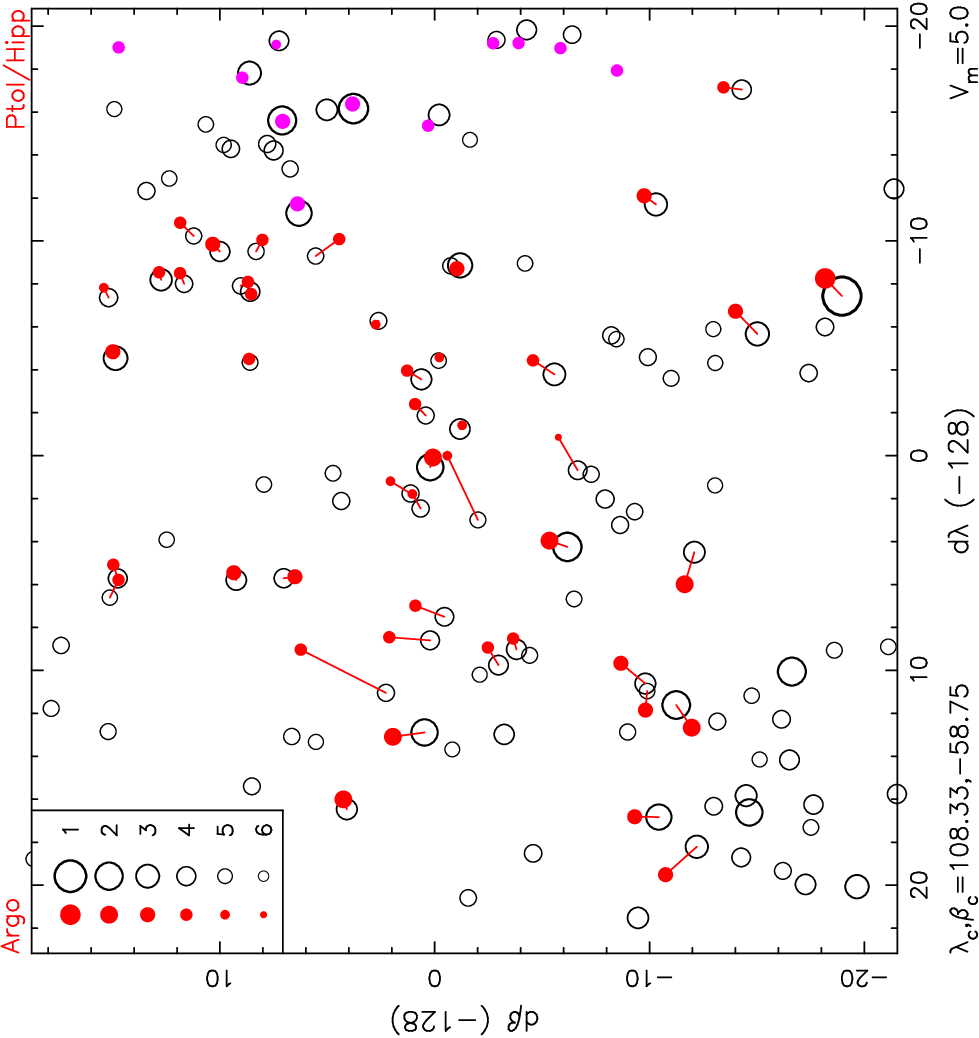}
\hfill\includegraphics[angle=270,width=\columnwidth]{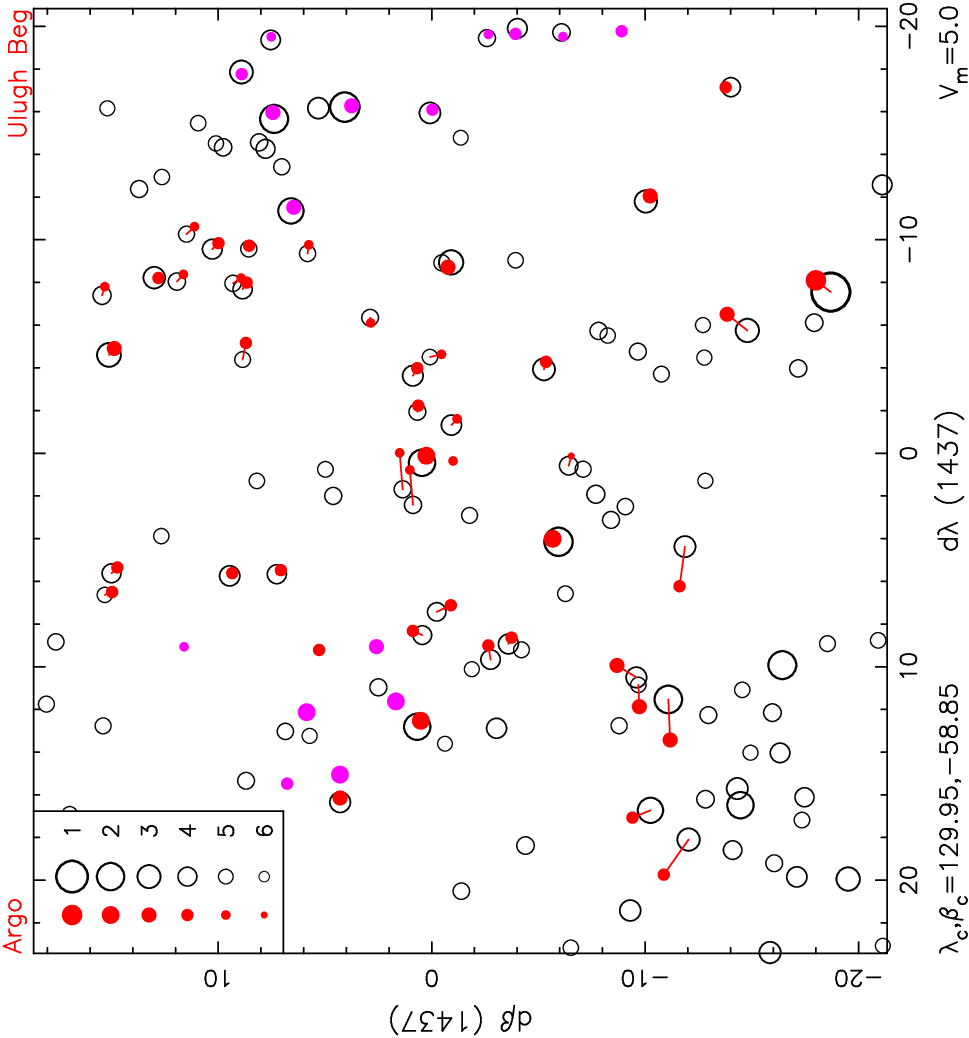}}
 \caption{Argo: bright Hipparcos stars only         \label{f:argbright}}
 \end{figure*}

\begin{figure*}
 \centerline{\includegraphics[angle=270,width=\columnwidth]{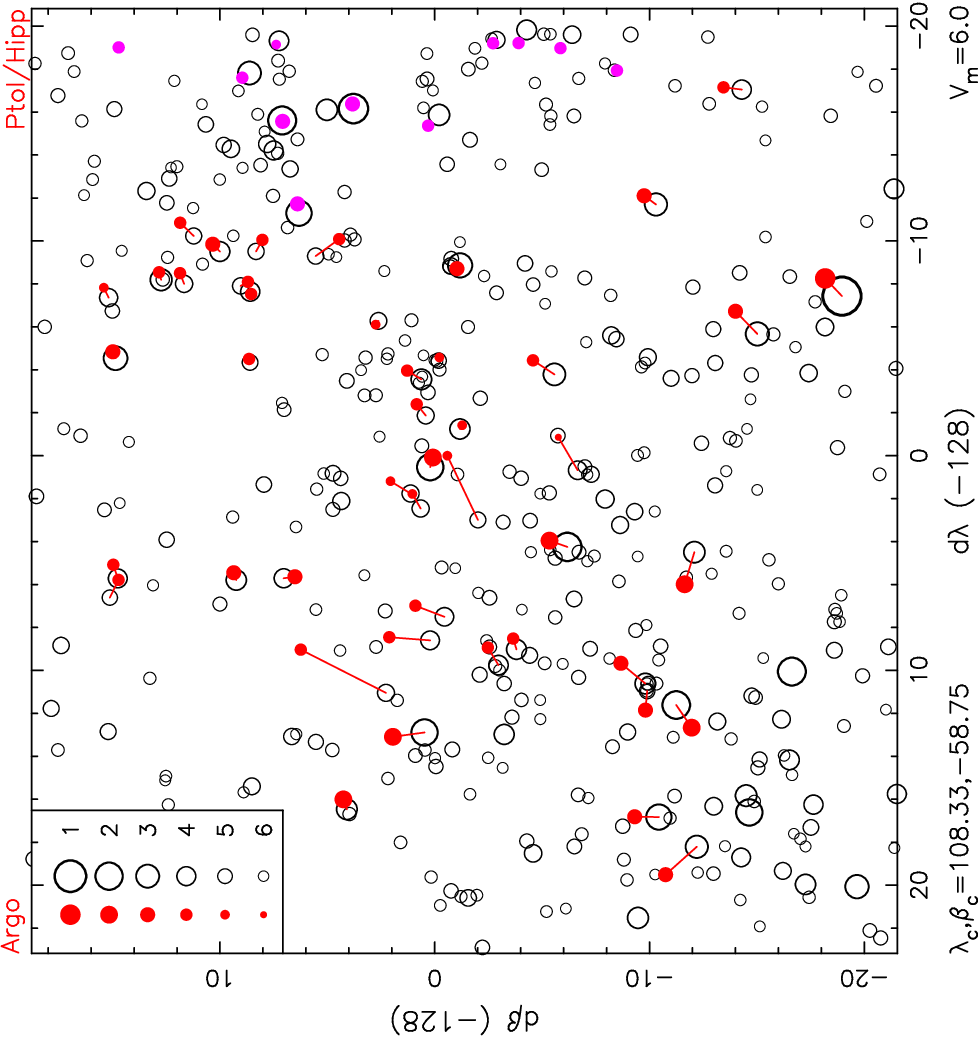}
\hfill\includegraphics[angle=270,width=\columnwidth]{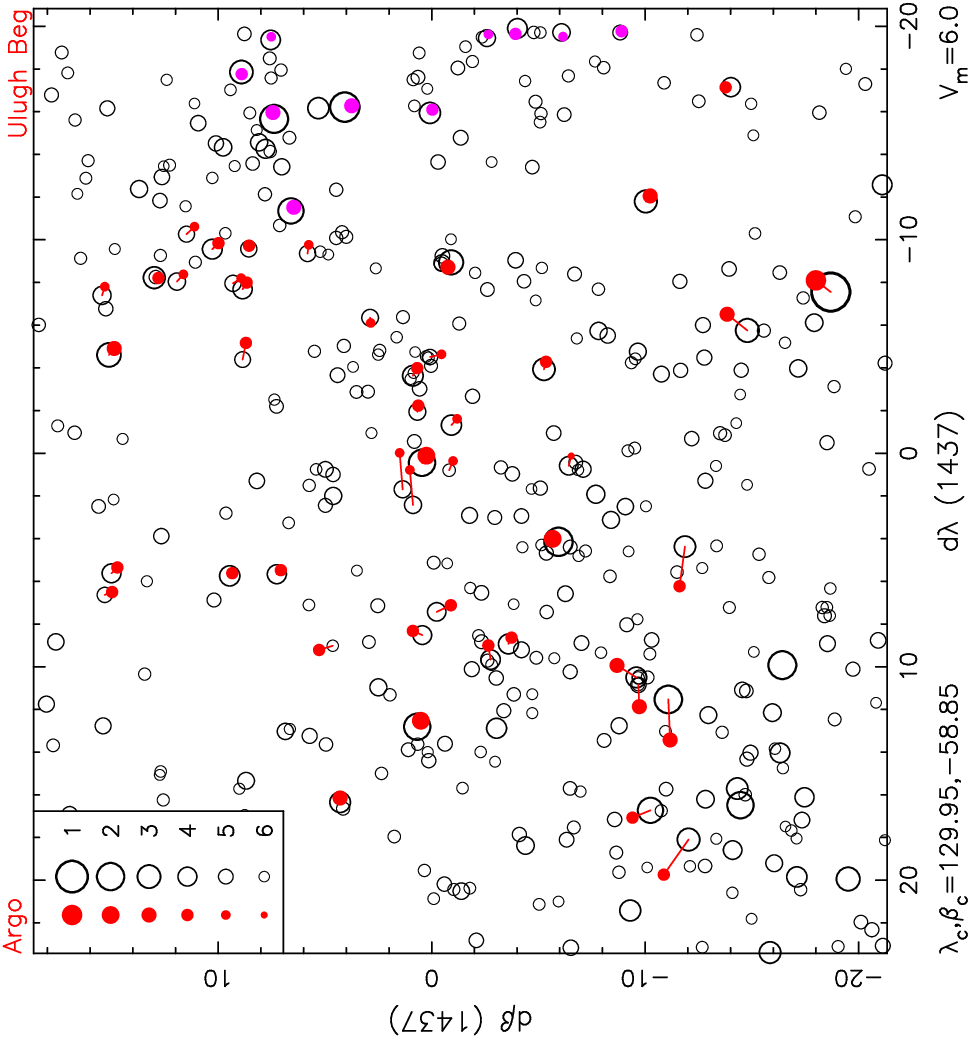}}
 \caption{Argo         \label{f:argo}}
 \end{figure*}

\noindent {\bf U\,882-884}, near 11,$-$10 in Figs.\,\ref{f:argbright} and \ref{f:argo}, are
identified by us with the three nearest stars. Knobel identifies
U\,884 with the fainter, almost equally distant HIP\,43105 ($V$=4.5,
$d$=128\farcm8), near 12.3,$-$13.0, and U\,883 with our
identification of U\,884.

\begin{figure*}
 \centerline{\includegraphics[angle=270,width=18.0cm]{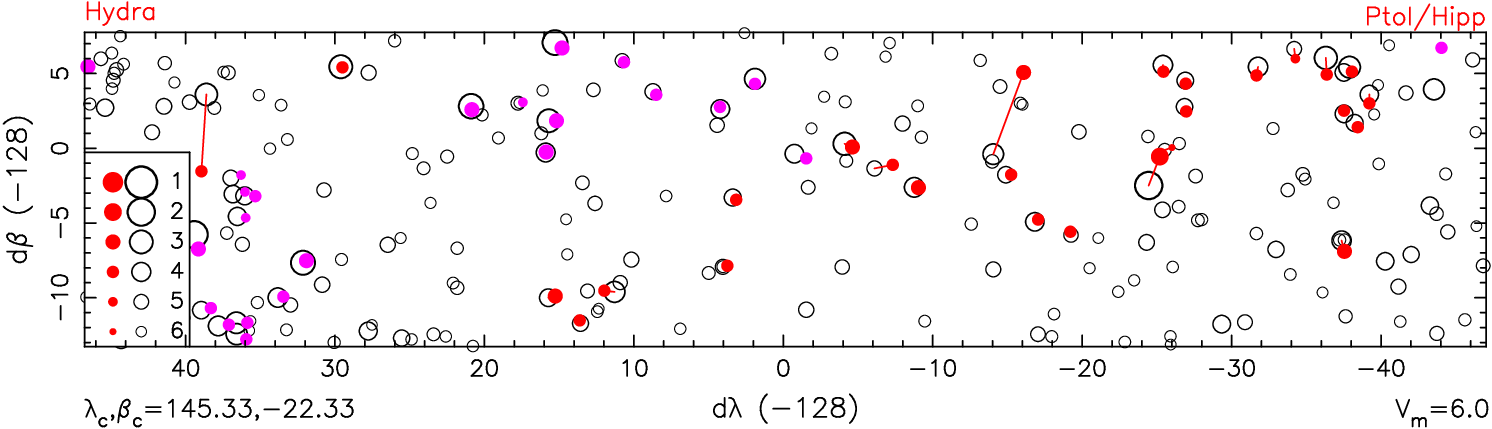}}

 \centerline{\includegraphics[angle=270,width=18.0cm]{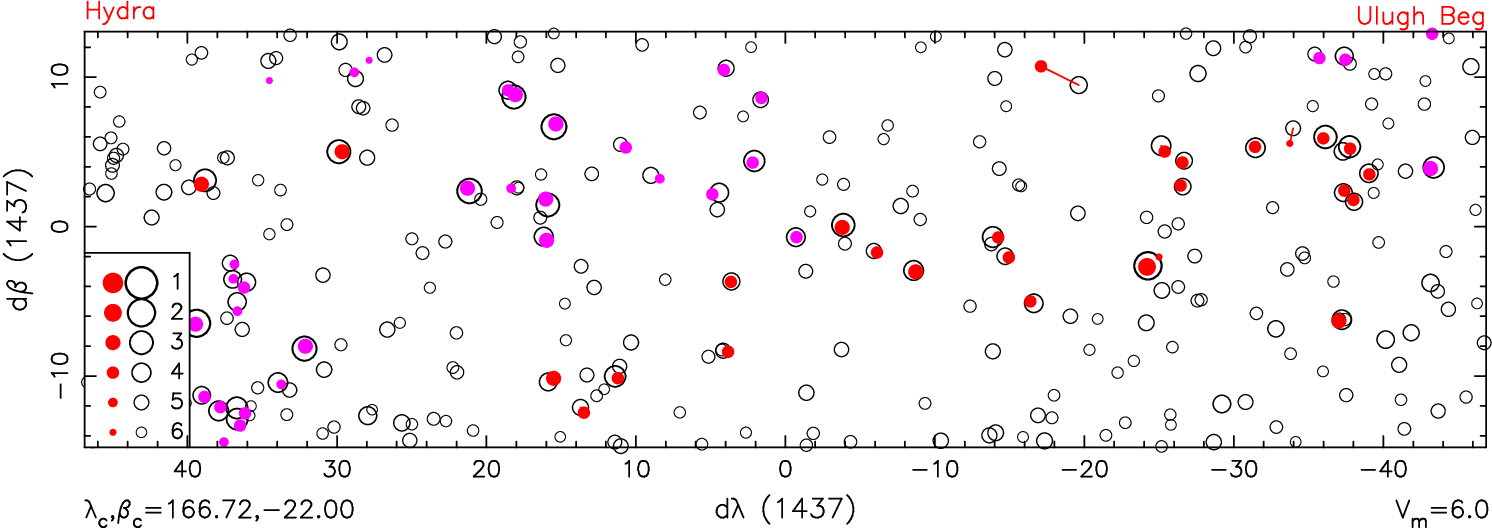}}
 \caption{Hydra        \label{f:hydra}}
 \end{figure*}

\noindent {\bf U\,901}, the faint star near $-$25.0,$-$2.0 in
Fig.\,\ref{f:hydra}, is identified by Knobel with HIP\,46404, near
$-$25.4,$-$0.3. We consider this star too far, and perhaps too
bright, to be a probable counterpart, and leave U\,901 unidentified.

\noindent {\bf U\,903-905}, near $-$16.4,$-$5.0, $-$15.0,$-$2.0, and
$-$14.2,$-$0.7, respectively in Fig.\,\ref{f:hydra}, correspond in
description to P\,906-908 in \ptolemaios, but their positions are
shifted, such that U\, 903 has the same Hipparcos counterpart as P\,907 and
U\,904 as P\,908. U\,905 has the Hipparcos counterpart which is 
our tentative identification for P\,920, the last star of Argo in
\ptolemaios. 

\noindent {\bf U\,952} corresponds to P\,955 and is the globular cluster
$\omega$\,Cen.

\noindent {\bf U\,961}, corresponding to P\,964, has no position from Ulugh
Beg, as he could not see it.

\noindent {\bf U\,979}, is tentatively identified by Knobel with Lacaille
5709; HIP\,67652 is the nearest Hipparcos star with $V<$7.5 (at
3\farcm3). At $V=7.2$ this is not a possible counterpart for
U\,979. We leave U\,979 unidentified.

\end{appendix}

\begin{appendix}
\section{Figures \label{s:figures}}

To illustrate and clarify our identifications we provide a Figure for
each constellation.  To facilitate comparison 
each figure shows the graph for  Ptolemaios
on the left and for Ulugh Beg on the right.
In these figures the stars listed with the
constellation in \ptolemaios\ / \ulughbeg\ are shown red,
and other stars from the same catalogue in purple.
Positions in the edition of the star catalogue of
Ptolemaios by Manitius (1913), when very different from
those in the Toomer (1998) edition, are indicated in brown,
connected by a dashed line to the position in the edition by Toomer (1998).

To minimize deformation of the constellations, we determine the center
of the constellation $\lambda_c,\beta_c$ from the extremes in
$\lambda$ and $\beta$, compute the rotation matrix which moves this
center to $(x,y,z)=(1,0,0)$, and then apply this rotation to each of
the stellar positions $\lambda_i,\beta_i$.  (For exact details see
Verbunt \&\ van Gent 2010a.)  The resulting $y,z$ values correspond roughly to
differences in longitude and latitude, exact at the center
$\lambda_c,\beta_c$ and increasingly deformed away from the center. We
plot the rotated positions of the stars in \ptolemaios\ / \ulughbeg\ as
$d\lambda\equiv y$ and $d\beta\equiv z$ with filled circles.  The same
rotation matrix is applied to all stars down to a magnitude limit
$V_\mathrm{m}$ (usually $V_m=6.0$) from the {\em Hipparcos Catalogue}
and those in the field of view are plotted as open circles.  The
symbol sizes are determined from the magnitudes as indicated in the
legenda. The used values for $\lambda_\mathrm{c}$, $\beta_\mathrm{c}$
and $V_\mathrm{m}$ are indicated with each Figure.

\begin{figure*}
 \centerline{\includegraphics[angle=270,width=0.9\columnwidth]{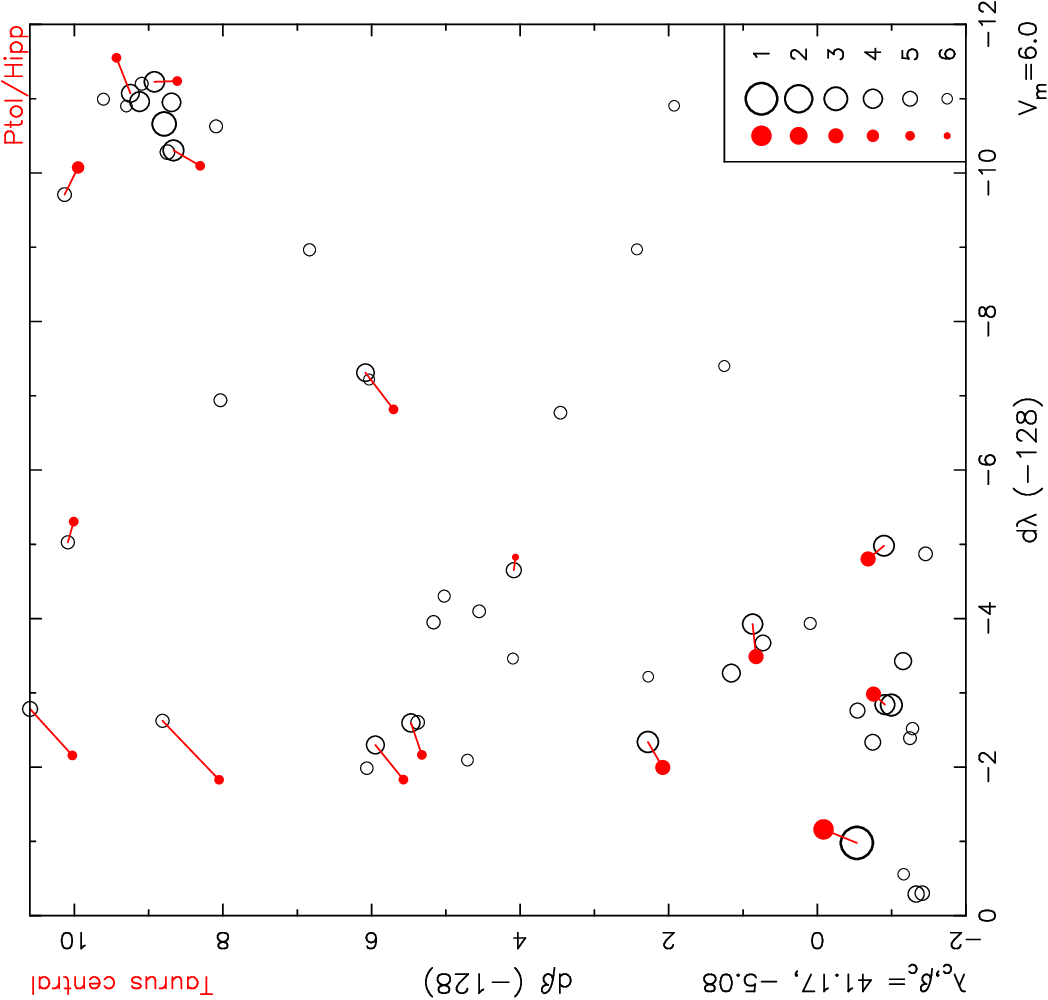}
\hfill\includegraphics[angle=270,width=0.9\columnwidth]{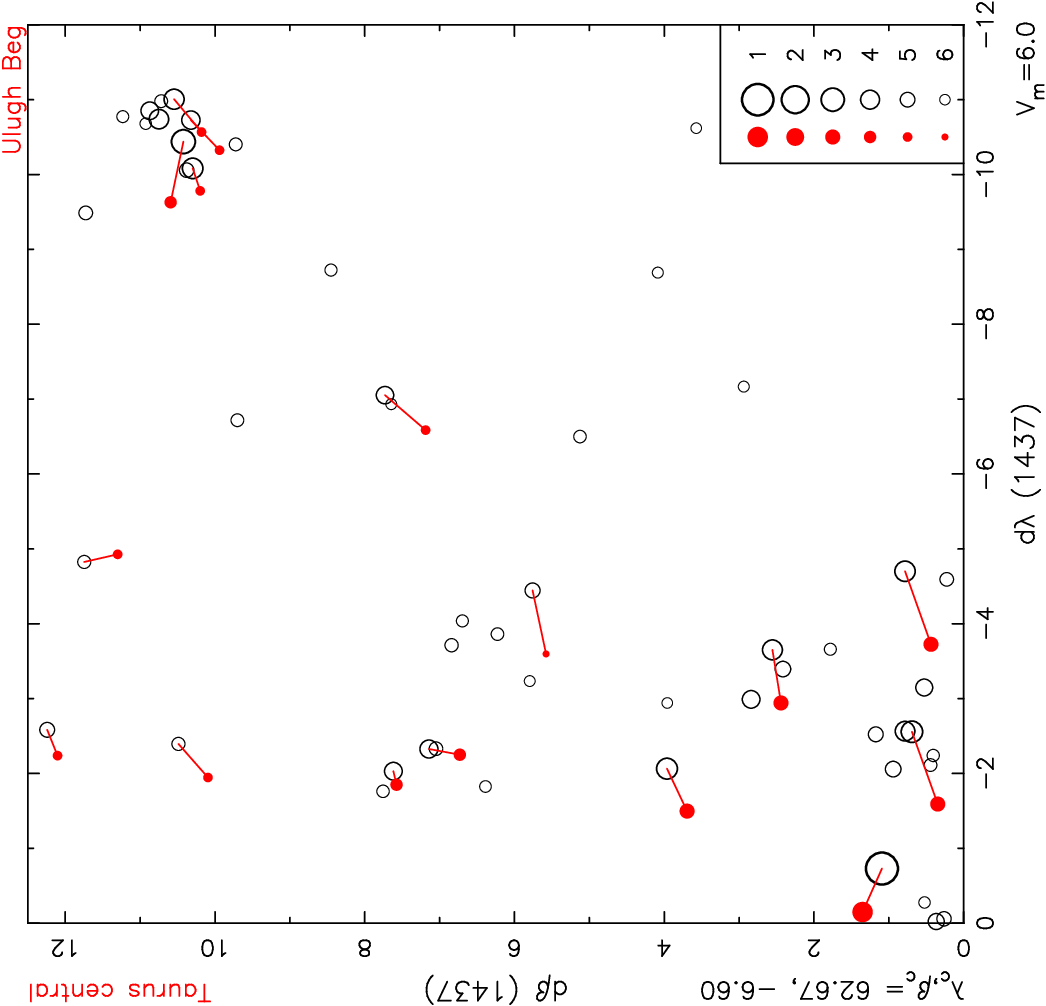}}
 \caption{Central area of Taurus       \label{f:taucent}}
 \end{figure*}

We show enlarged detail in Figures\,\ref{f:orisword},
\ref{f:bootesdet}, \ref{f:pleiades} and \ref{f:taucent}; for easy
comparison with the Figures showing the whole constellation, these
detail Figures use the same rotation center (and thus rotation
matrix).

 \begin{figure*}
 \centerline{
\includegraphics[angle=270,width=\columnwidth]{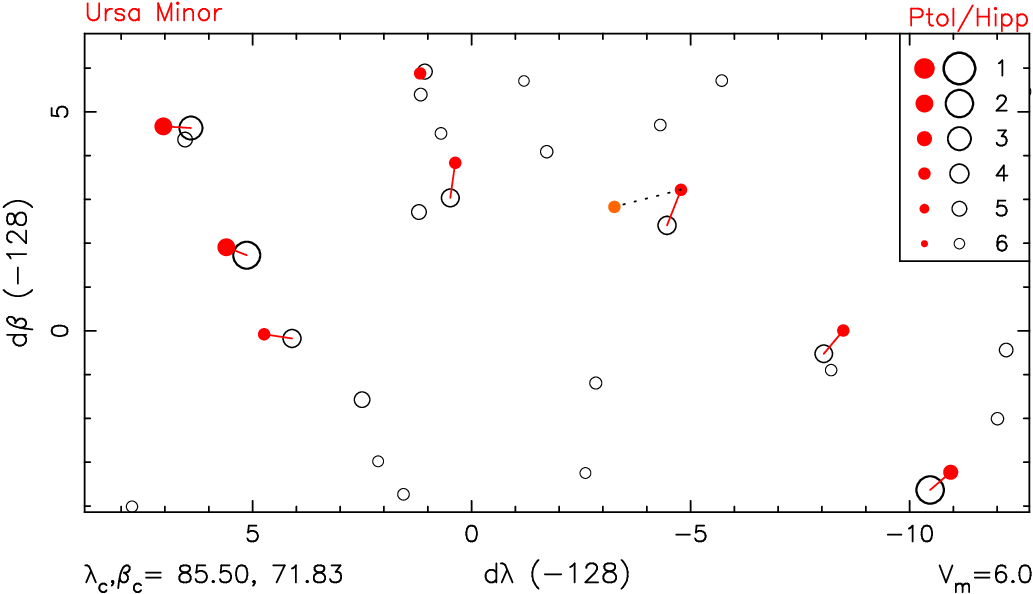} \hfill
\includegraphics[angle=270,width=\columnwidth]{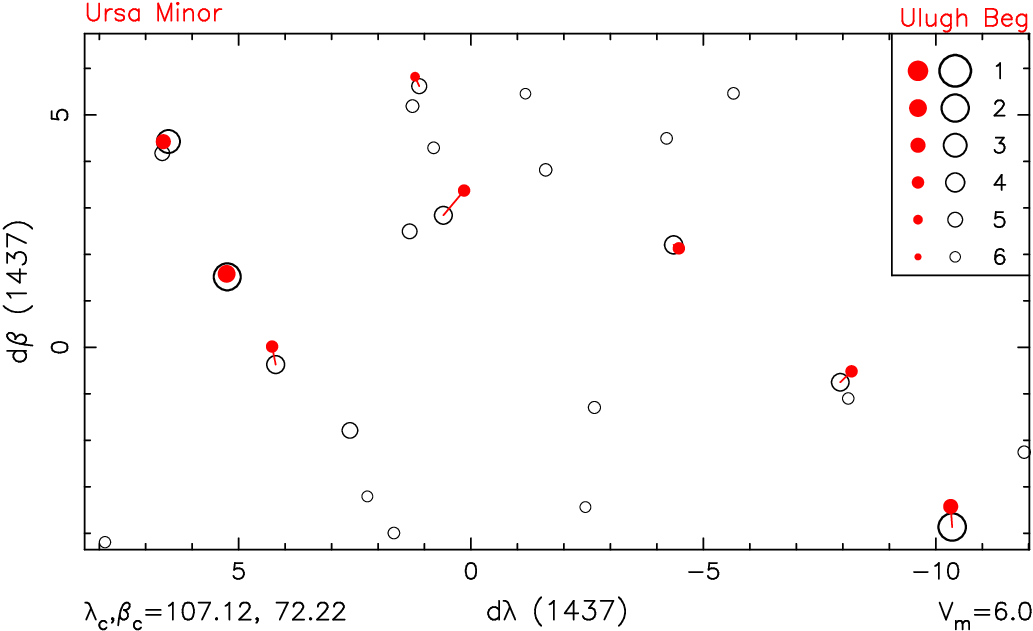}
}
 \caption{Ursa Minor   \label{f:ursaminor}}
 \end{figure*}
  
 \begin{figure*}
 \centerline{\includegraphics[angle=270,width=\columnwidth]{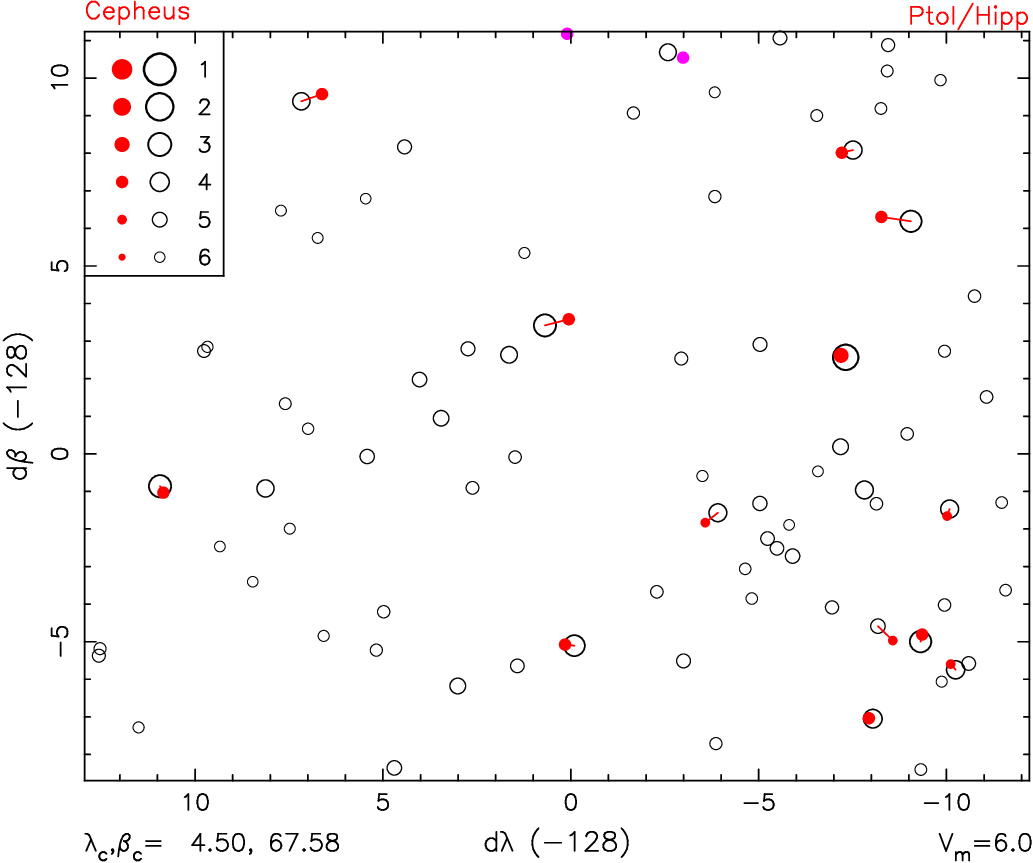}
\hfill\includegraphics[angle=270,width=\columnwidth]{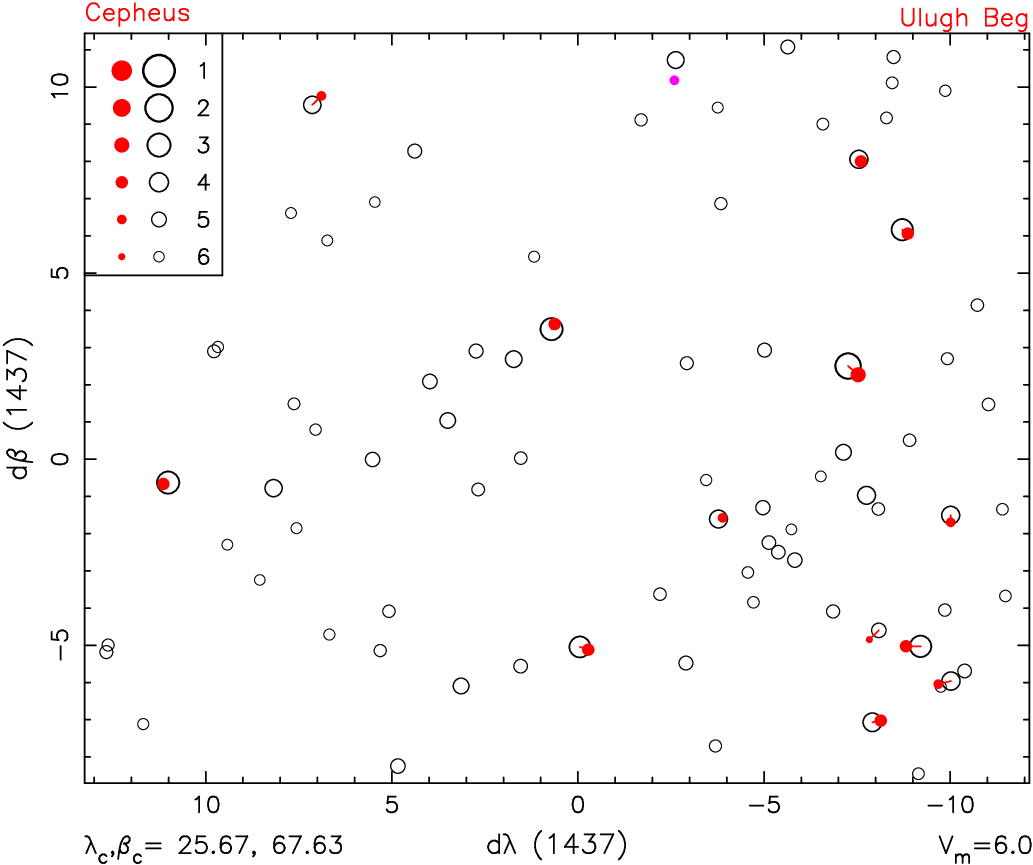}}
 \caption{Cepheus      \label{f:cepheus}}
 \end{figure*}

 \begin{figure*}
 \centerline{
\includegraphics[angle=270,width=\columnwidth]{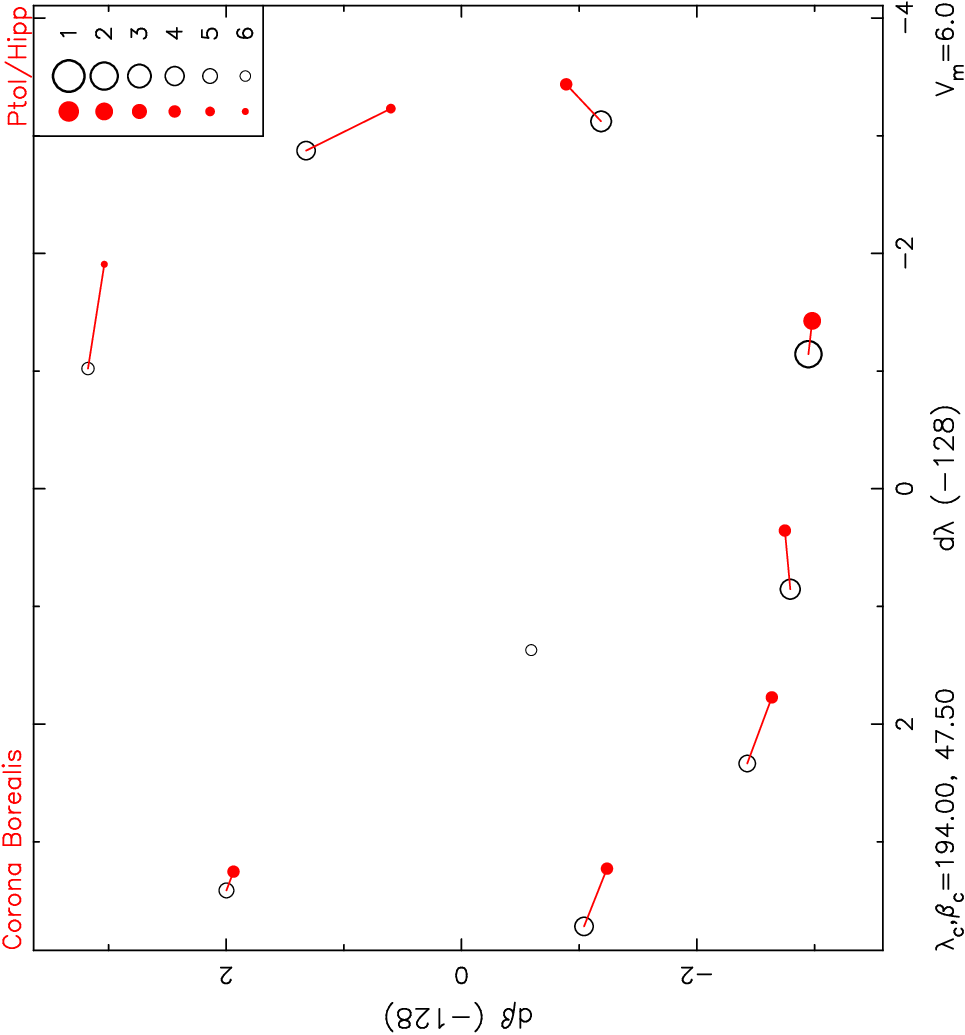}
\hfill\includegraphics[angle=270,width=\columnwidth]{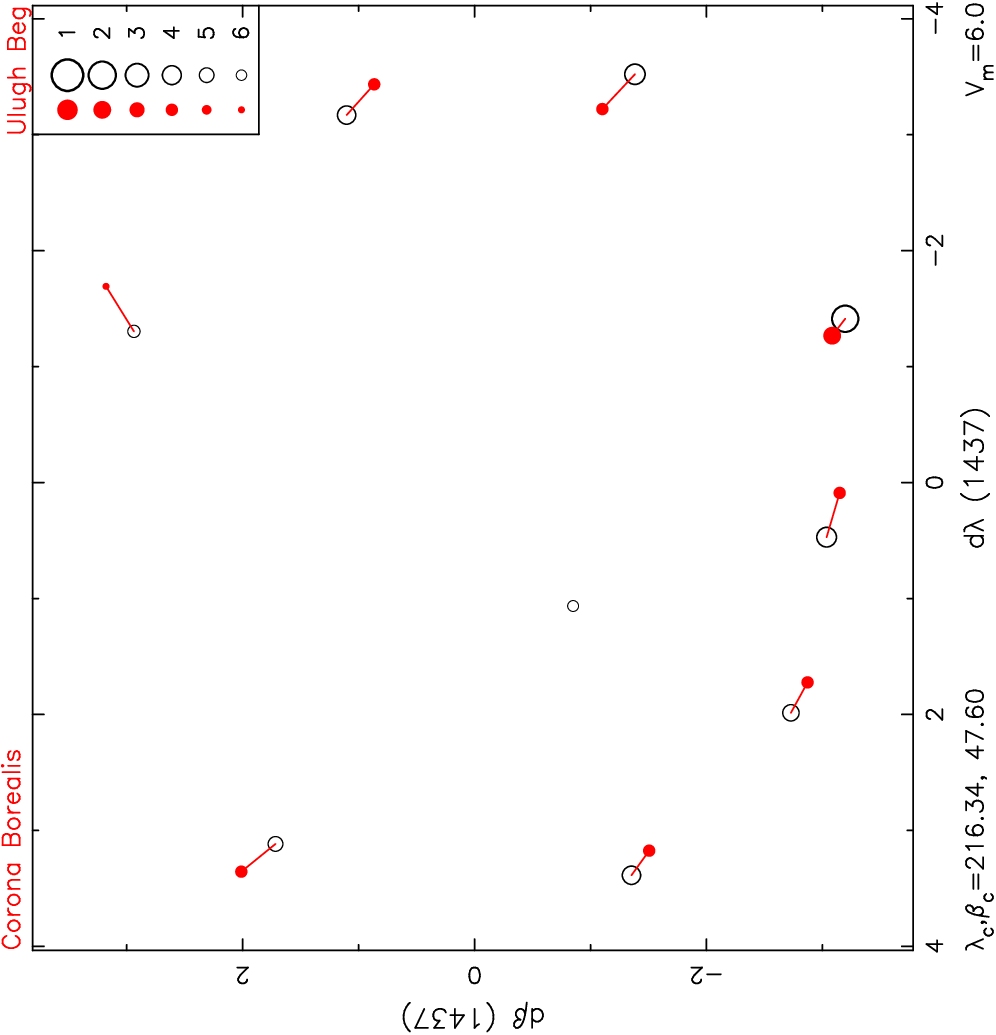}}
 \caption{Corona Borealis \label{f:coronaborea}}
 \end{figure*}
  
 \begin{figure*}
 \centerline{\includegraphics[angle=270,width=\columnwidth]{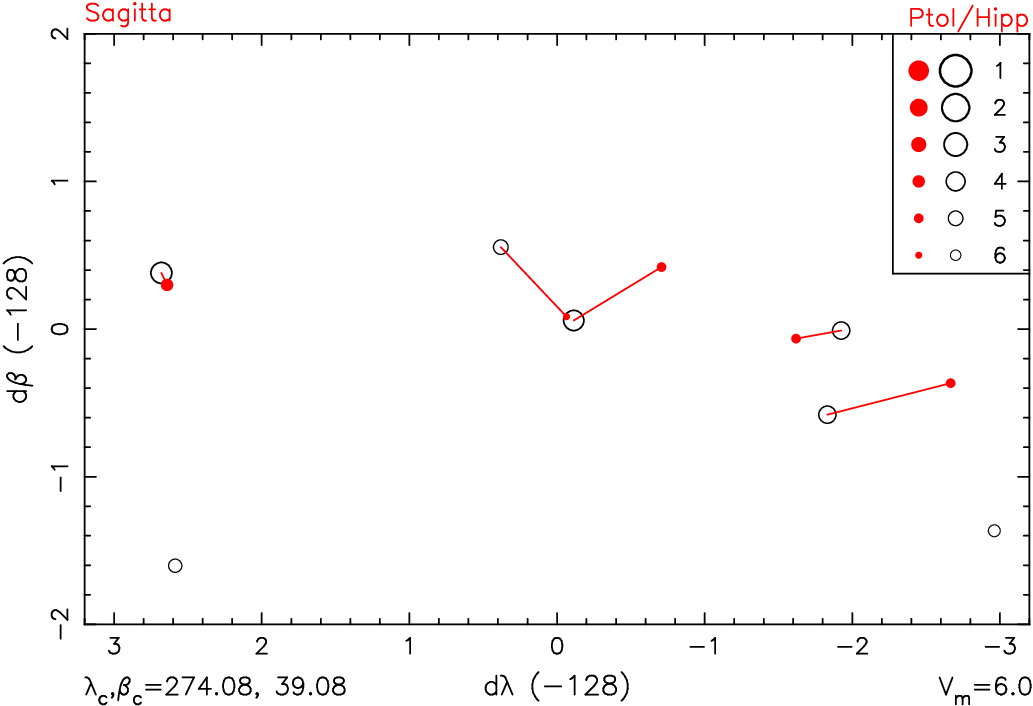}
\hfill\includegraphics[angle=270,width=\columnwidth]{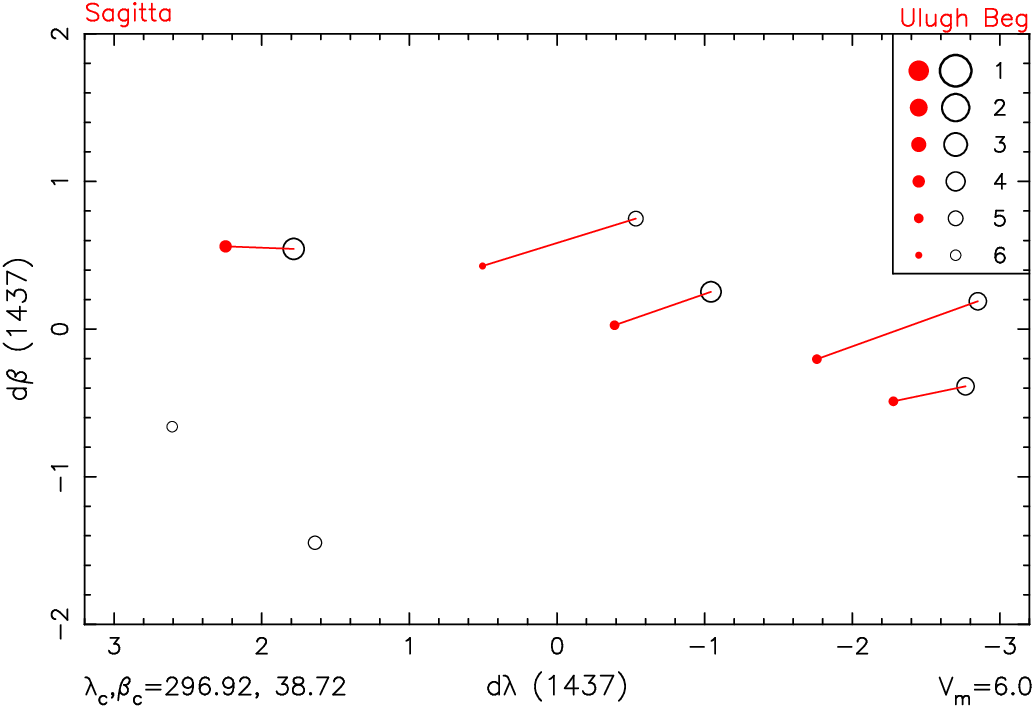}}
 \caption{Sagitta      \label{f:sagitta}}
 \end{figure*}
  
 \begin{figure*}
 \centerline{\includegraphics[angle=270,width=\columnwidth]{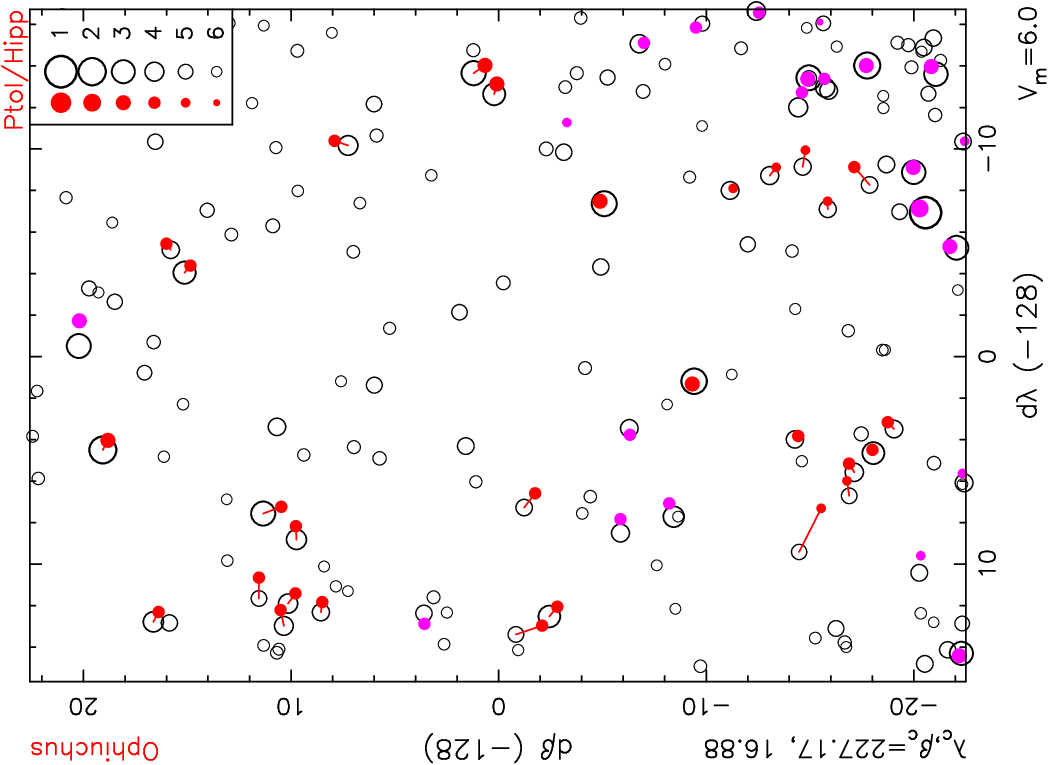}
\hfill\includegraphics[angle=270,width=\columnwidth]{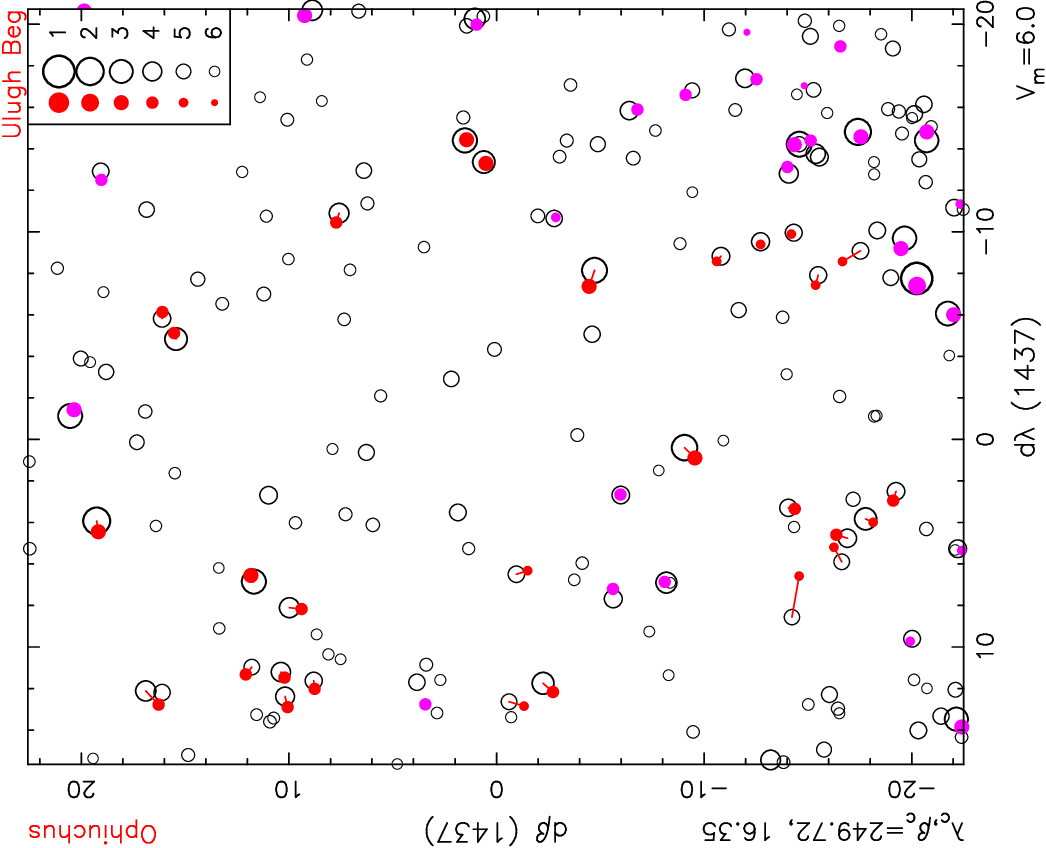}}
 \caption{Ophiuchus    \label{f:ophiuchus}}
 \end{figure*}
  
\clearpage
  
 \begin{figure*}
 \centerline{
\includegraphics[angle=270,width=\columnwidth]{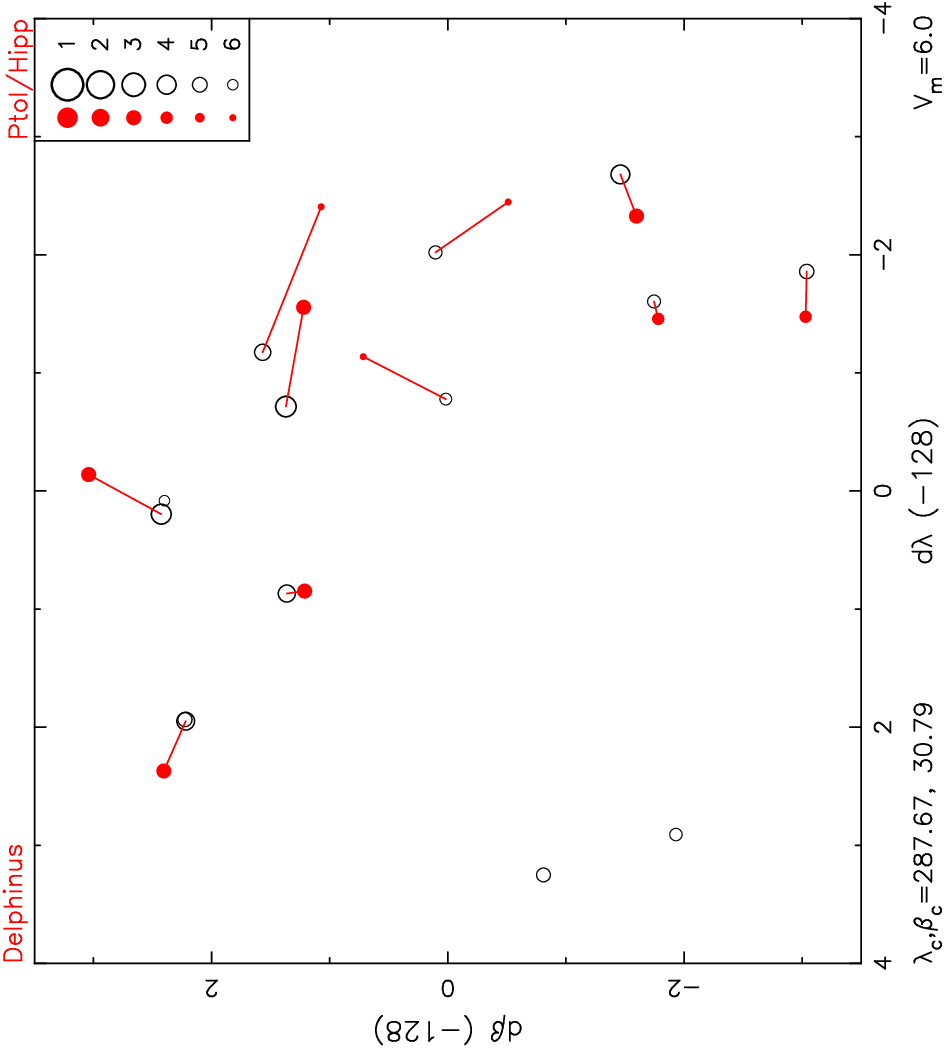}
\hfill\includegraphics[angle=270,width=\columnwidth]{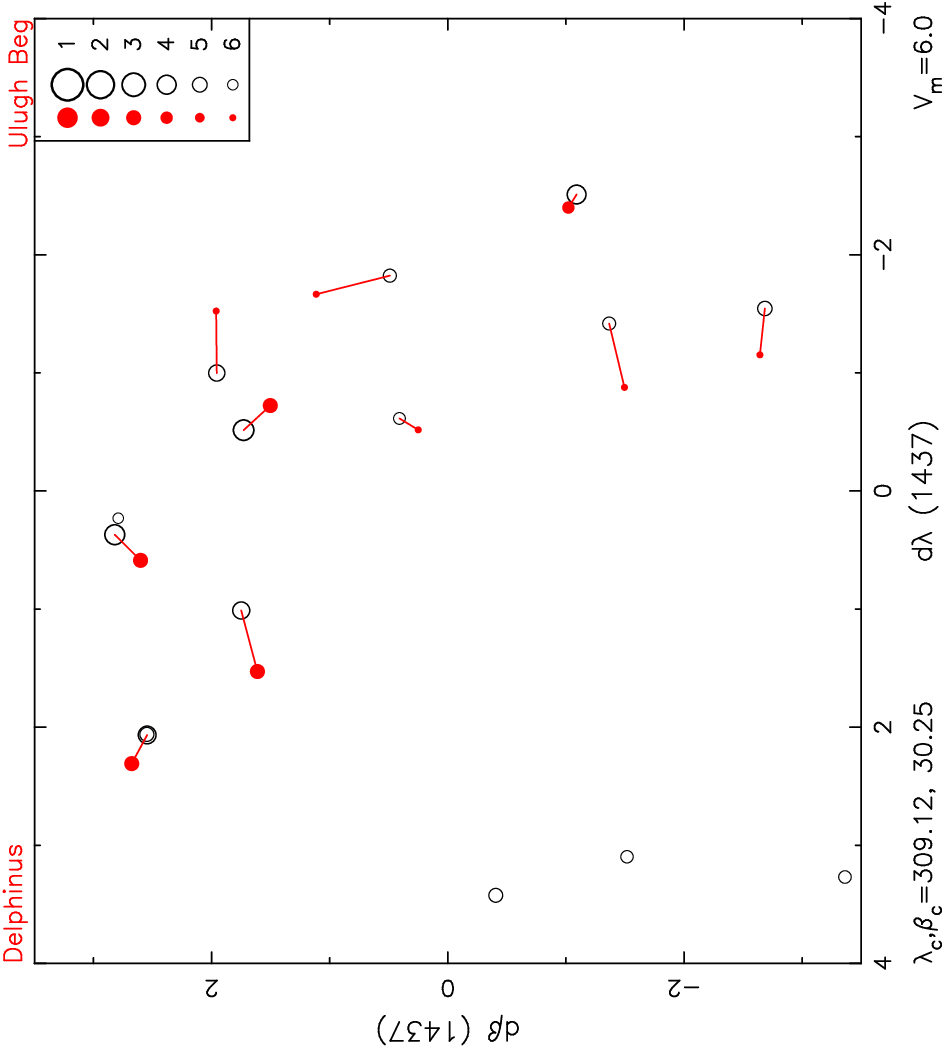}}
 \caption{Delphinus    \label{f:delphinus}}
 \end{figure*}

 \begin{figure*}
 \centerline{\includegraphics[angle=270,width=\columnwidth]{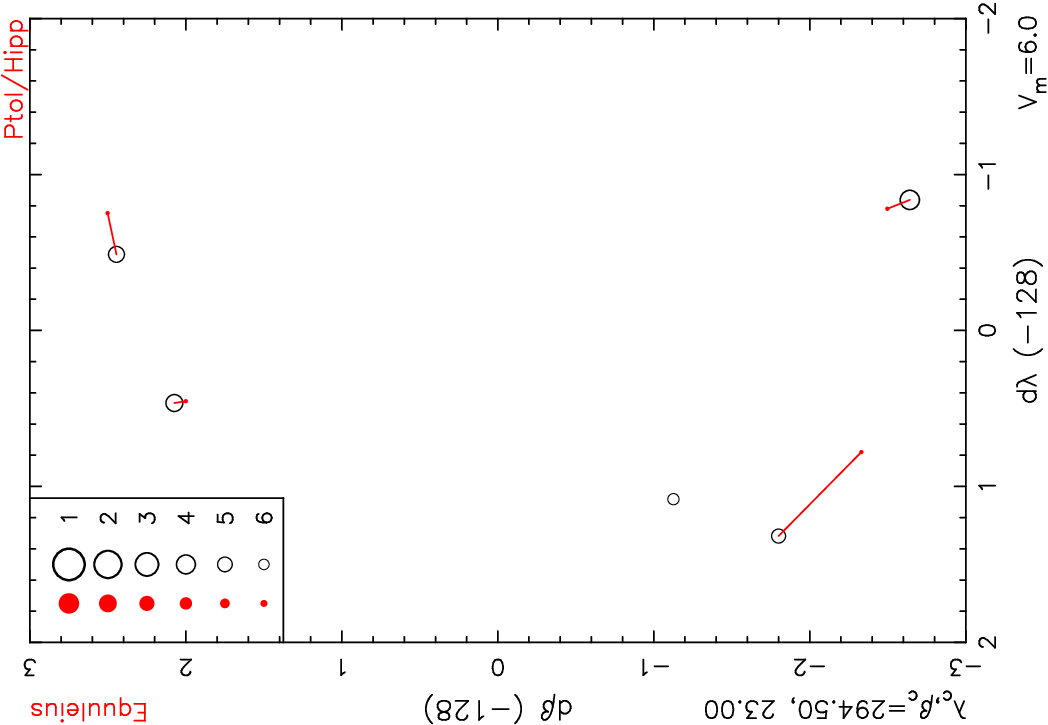}
\hfill\includegraphics[angle=270,width=\columnwidth]{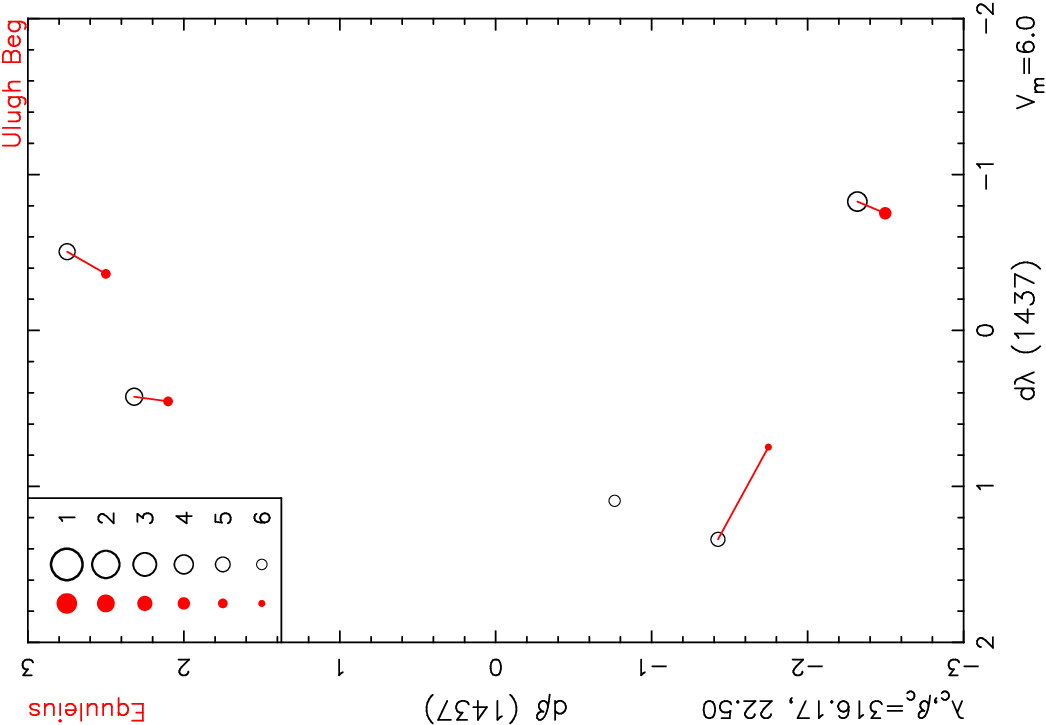}}
 \caption{Equuleius    \label{f:equuleius}}
 \end{figure*}
  
 \begin{figure*}
 \centerline{\includegraphics[angle=270,width=\columnwidth]{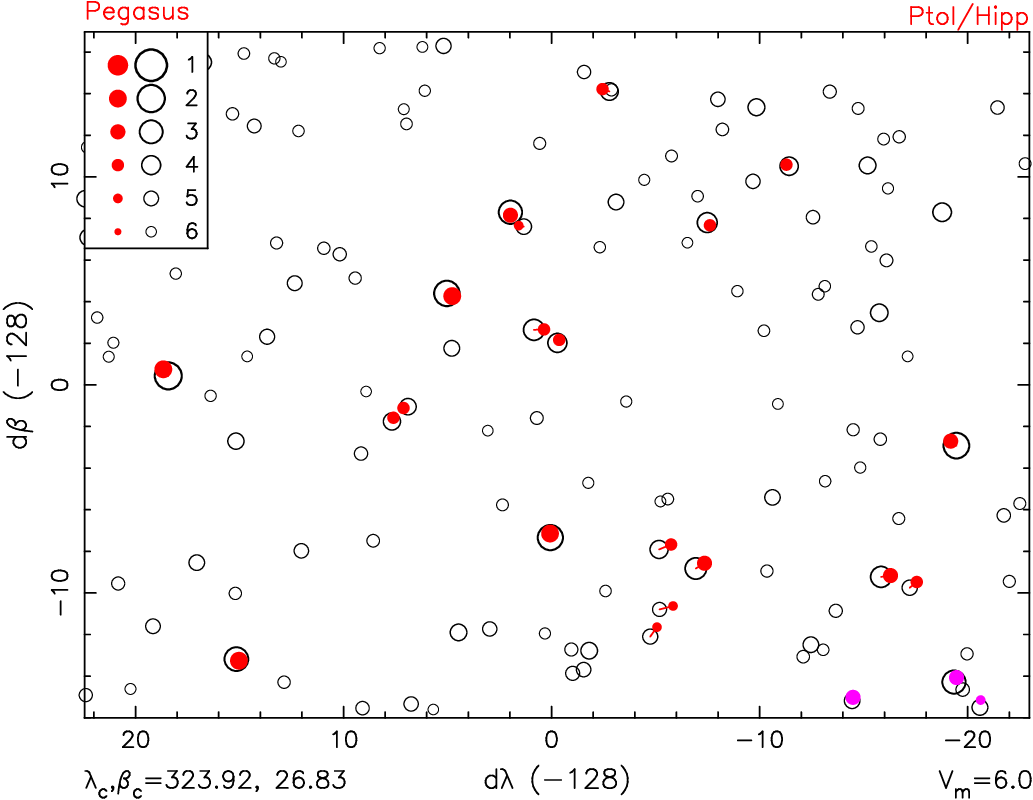}
\hfill\includegraphics[angle=270,width=\columnwidth]{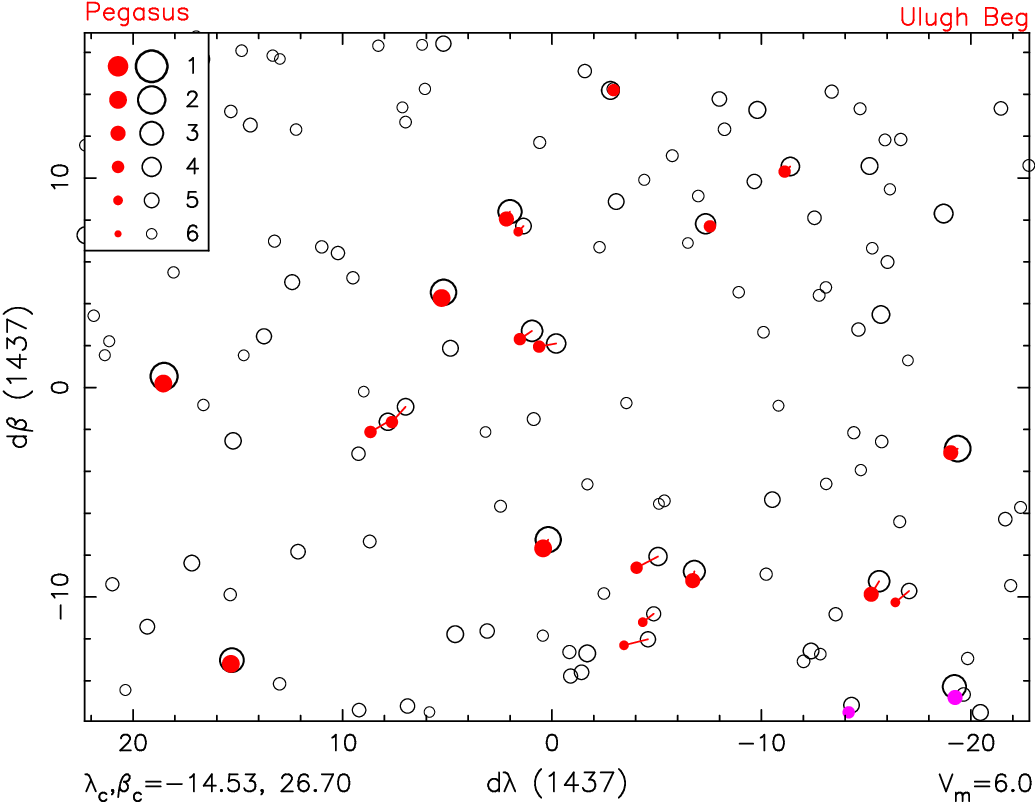}}
 \caption{Pegasus      \label{f:pegasus}}
 \end{figure*}
  
 \begin{figure*}
 \centerline{\includegraphics[angle=270,width=0.9\columnwidth]{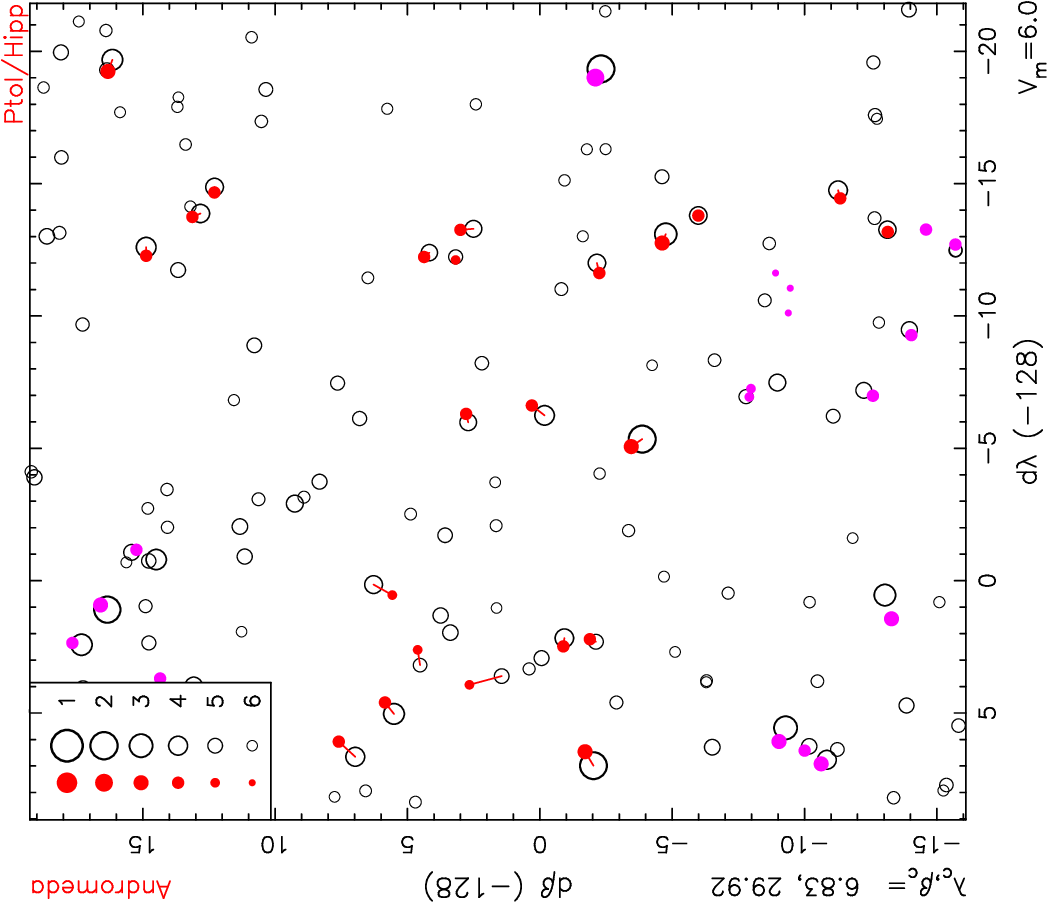}
\hfill\includegraphics[angle=270,width=\columnwidth]{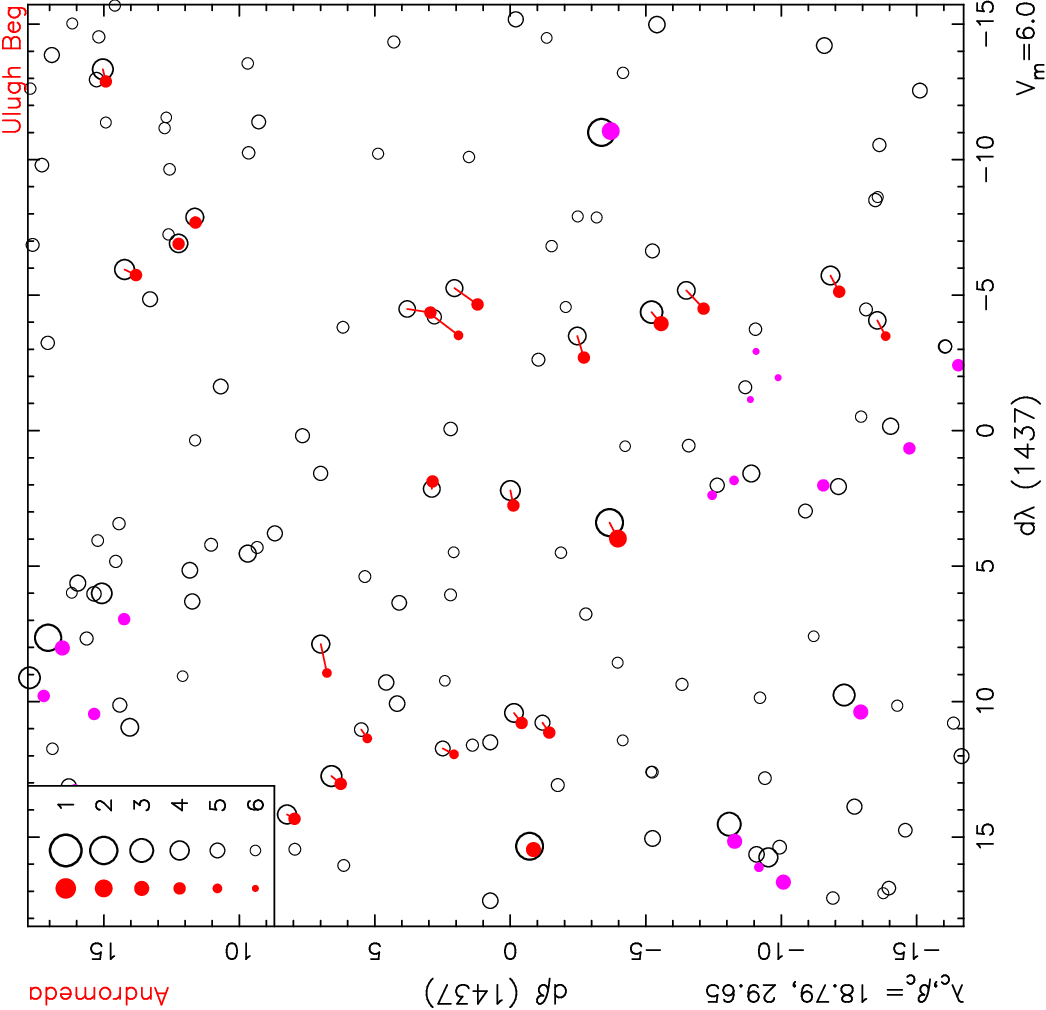}}
 \caption{Andromeda    \label{f:andromeda}}
 \end{figure*}
  
 \begin{figure*}
 \centerline{\includegraphics[angle=270,width=0.8\columnwidth]{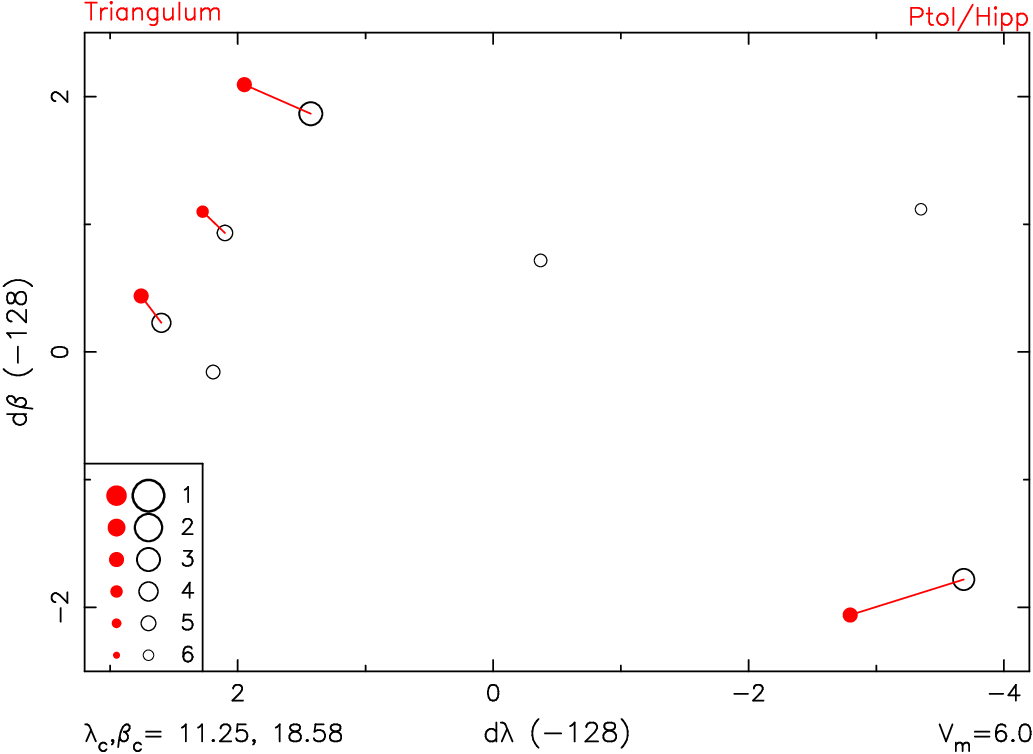}
\hfill\includegraphics[angle=270,width=0.9\columnwidth]{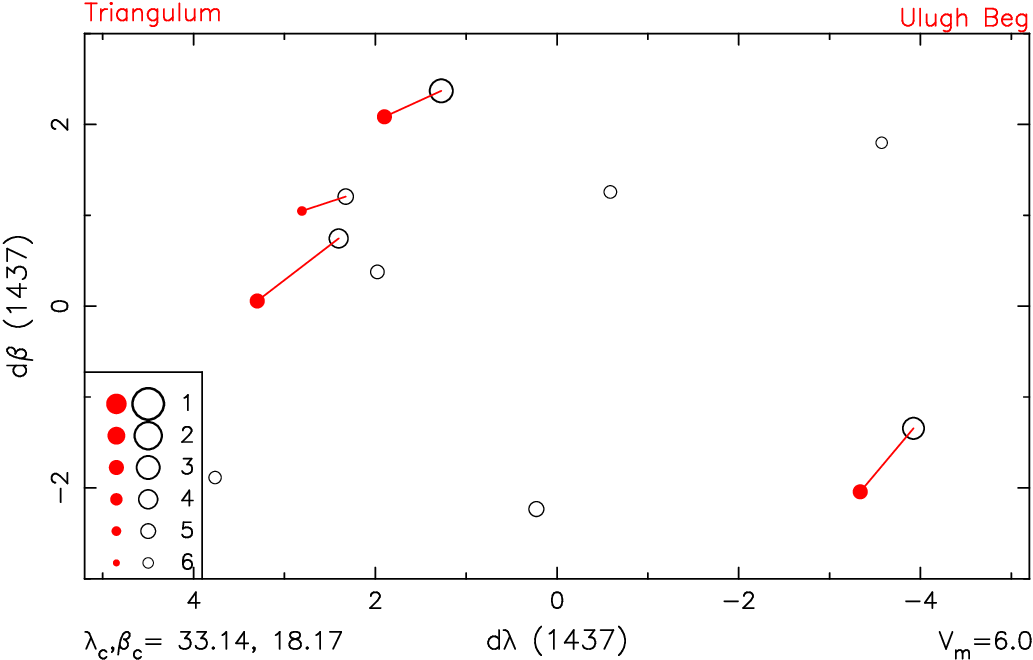}}
 \caption{Triangulum   \label{f:triangulum}}
 \end{figure*}
  
\clearpage
 
 \begin{figure*}
 \centerline{\includegraphics[angle=270,width=\columnwidth]{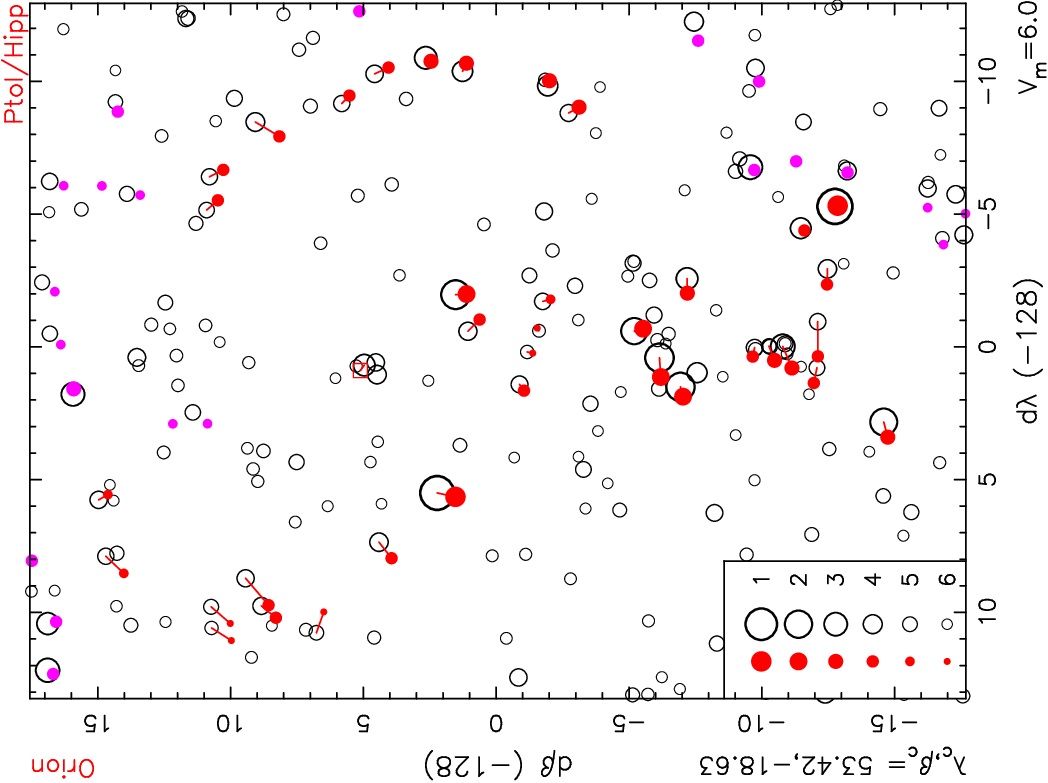}
\hfill\includegraphics[angle=270,width=\columnwidth]{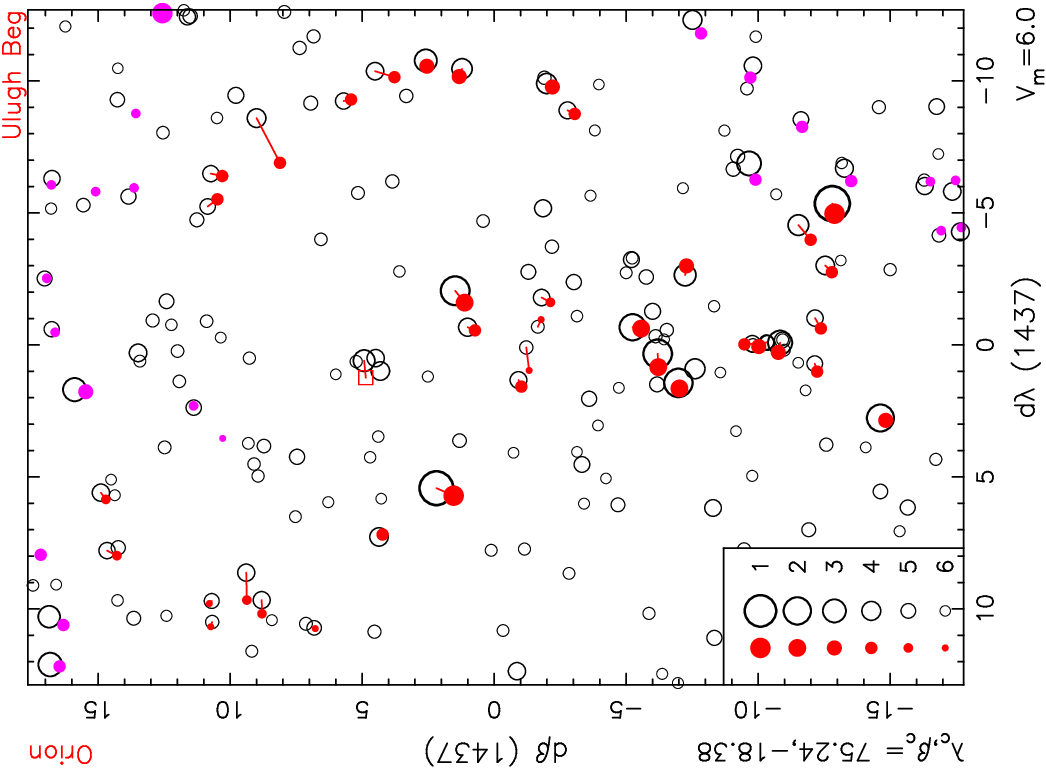}}
 \caption{Orion        \label{f:orion}}
 \end{figure*}
  
 \begin{figure*}
 \centerline{\includegraphics[angle=270,width=\columnwidth]{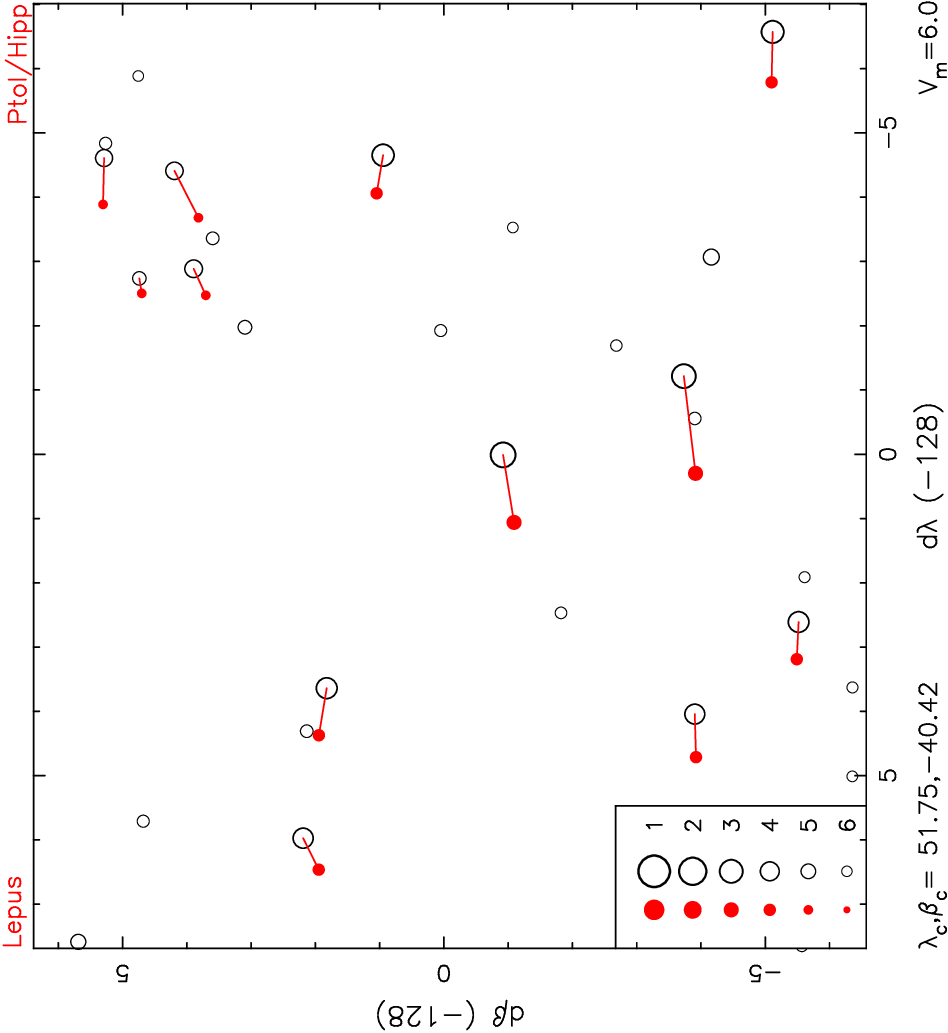}
\hfill\includegraphics[angle=270,width=\columnwidth]{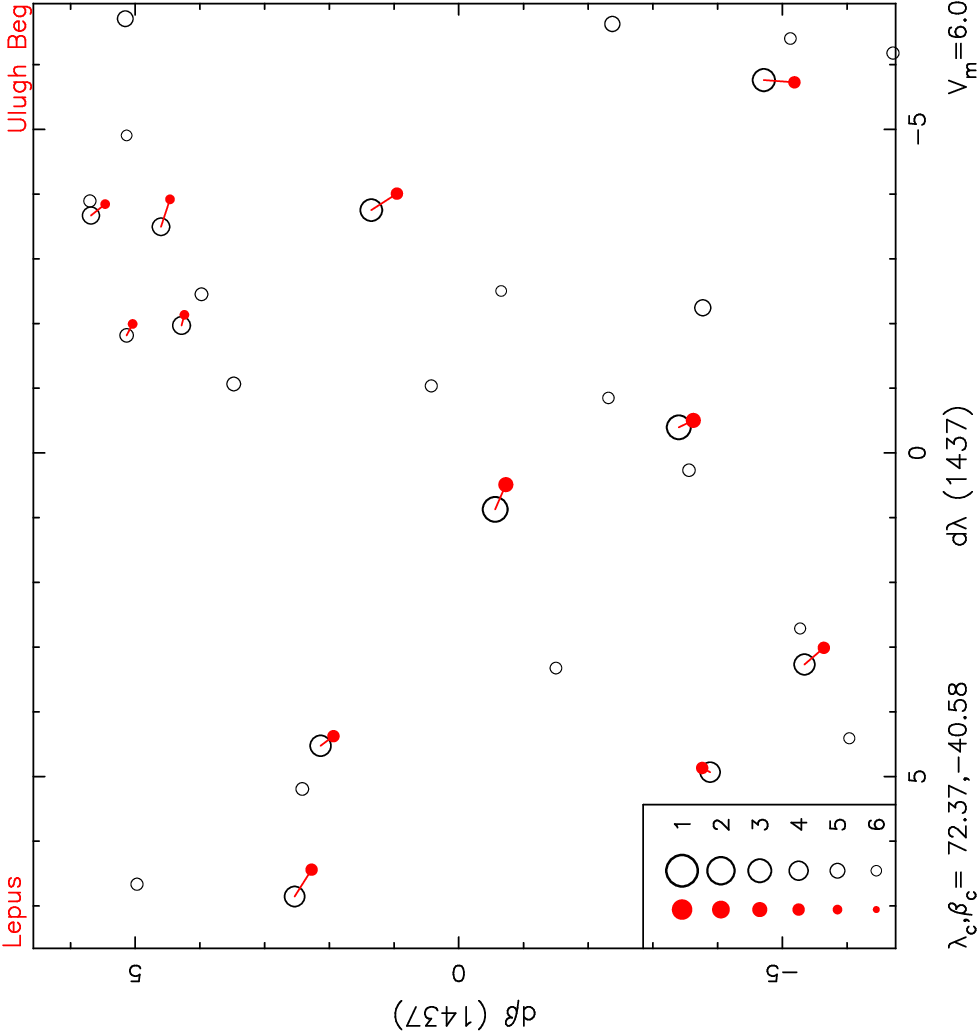}}
 \caption{Lepus        \label{f:lepus}}
 \end{figure*}
  
 \begin{figure*}
 \centerline{\includegraphics[angle=270,width=\columnwidth]{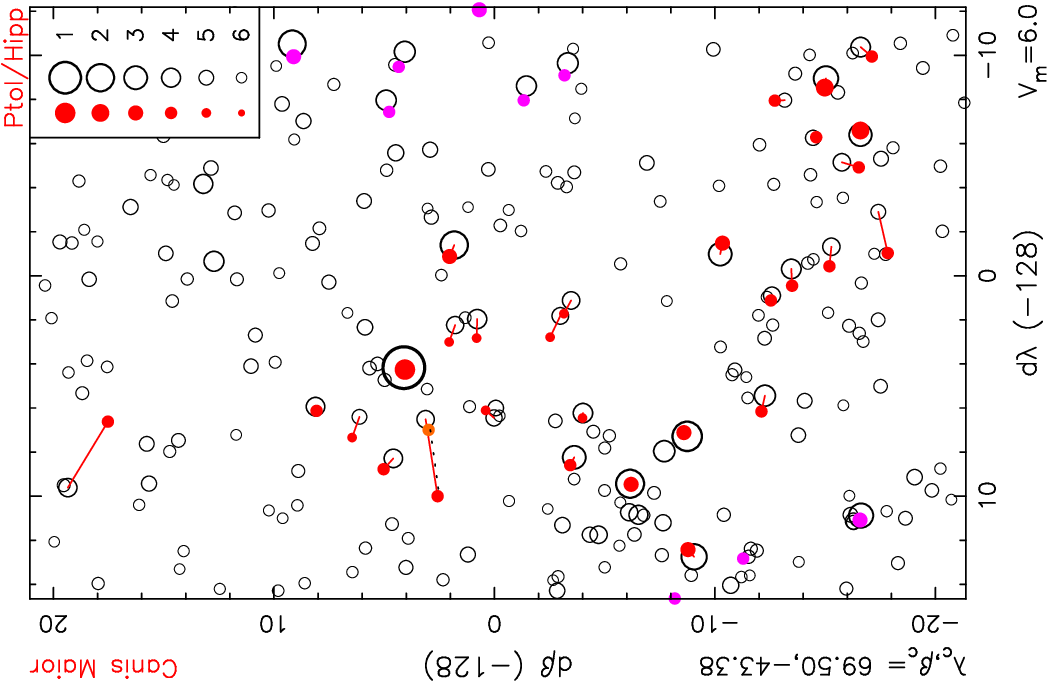}
\hfill\includegraphics[angle=270,width=\columnwidth]{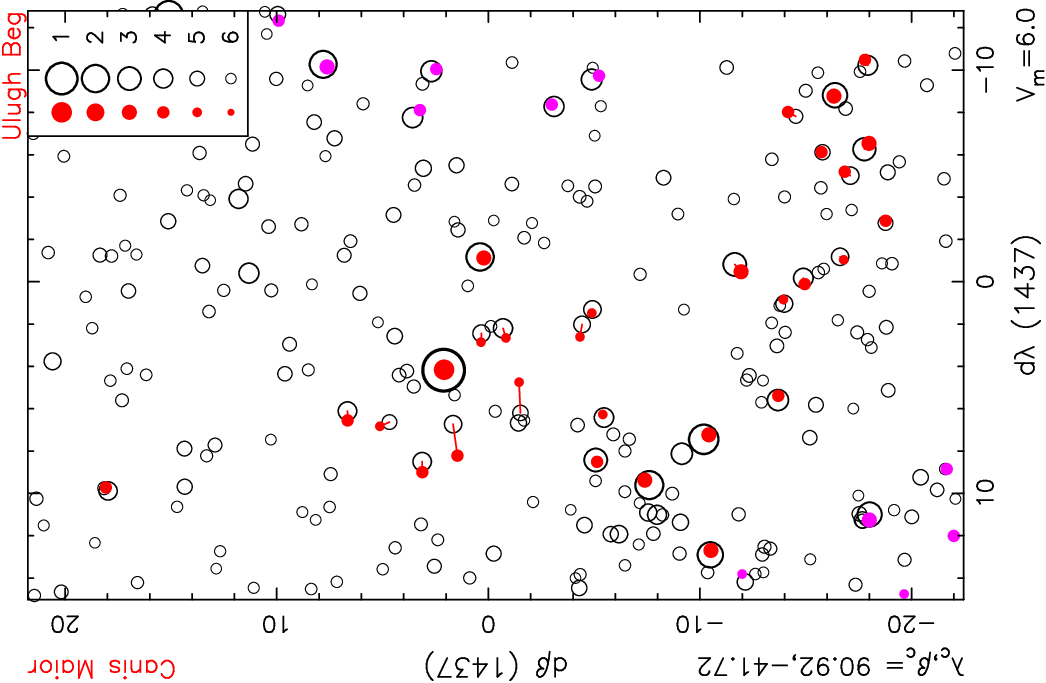}}
 \caption{Canis Maior  \label{f:canismaior}}
 \end{figure*}
  
 \begin{figure*}
 \centerline{\includegraphics[angle=270,width=0.9\columnwidth]{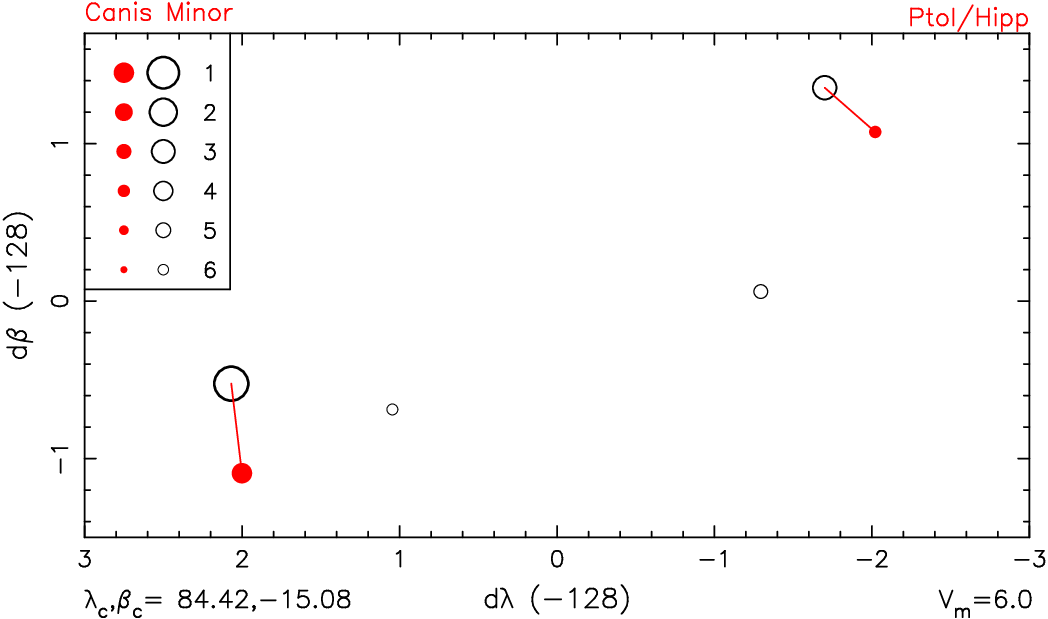}
\hfill\includegraphics[angle=270,width=0.9\columnwidth]{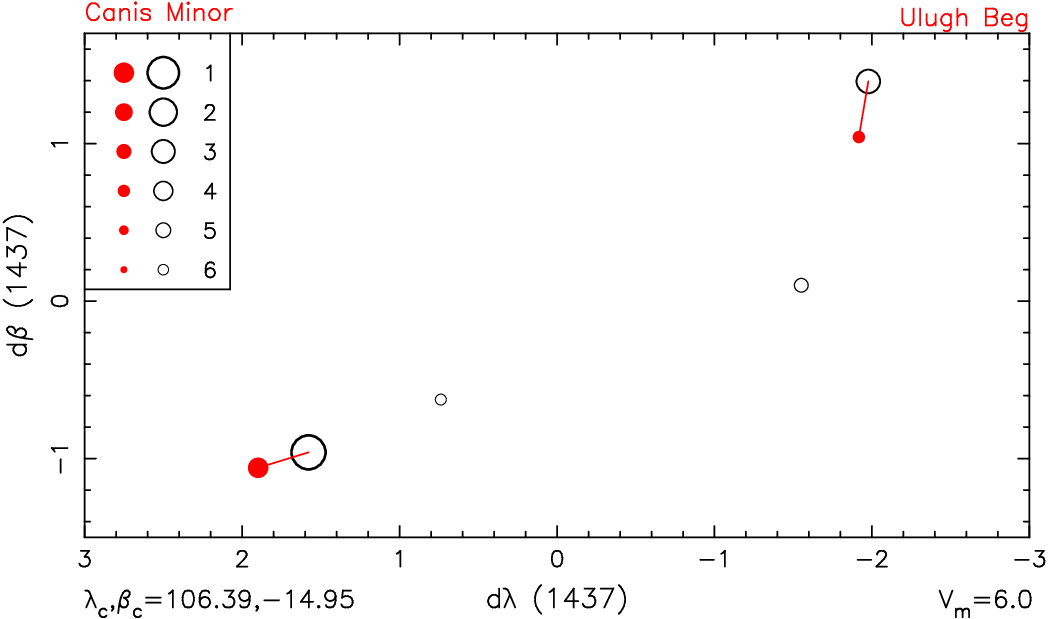}}
 \caption{Canis Minor  \label{f:canisminor}}
 \end{figure*}
  
 \begin{figure*}
 \centerline{\includegraphics[angle=270,width=\columnwidth]{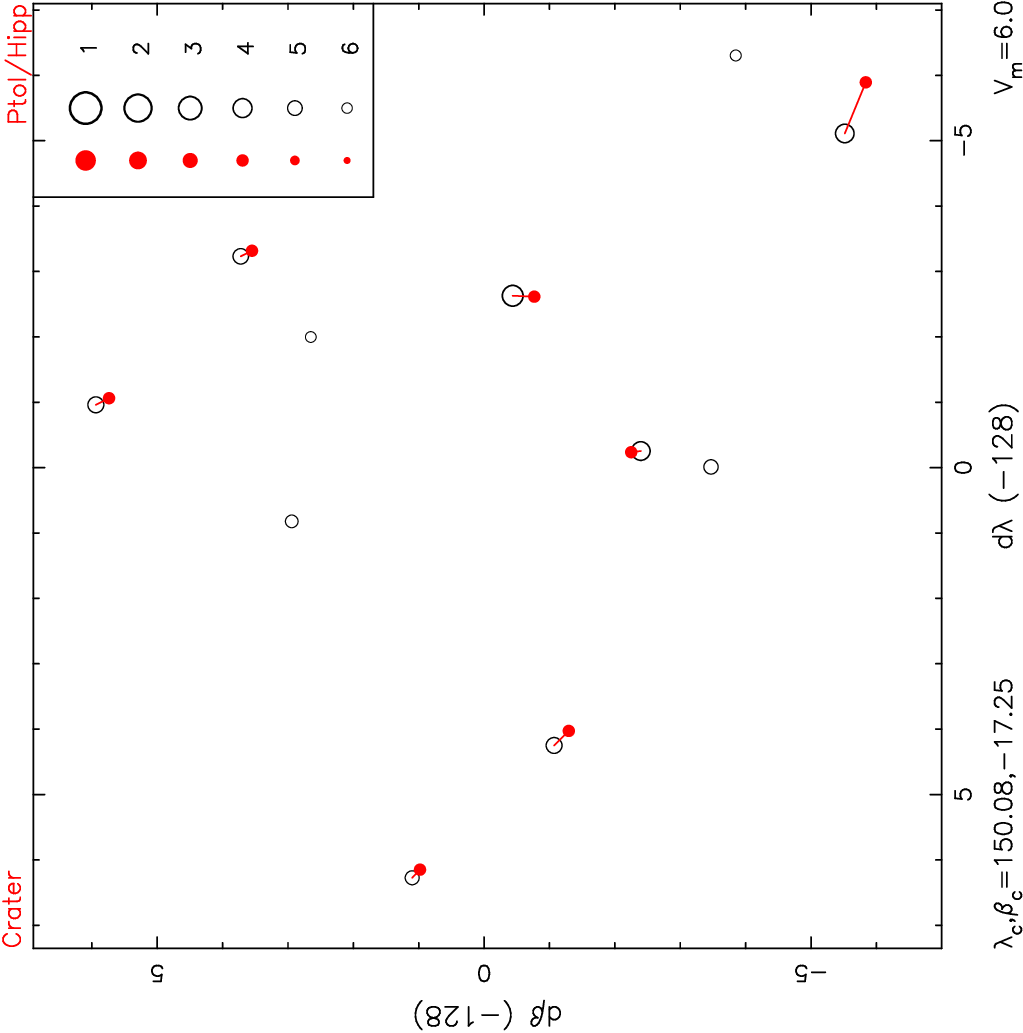}
\hfill\includegraphics[angle=270,width=\columnwidth]{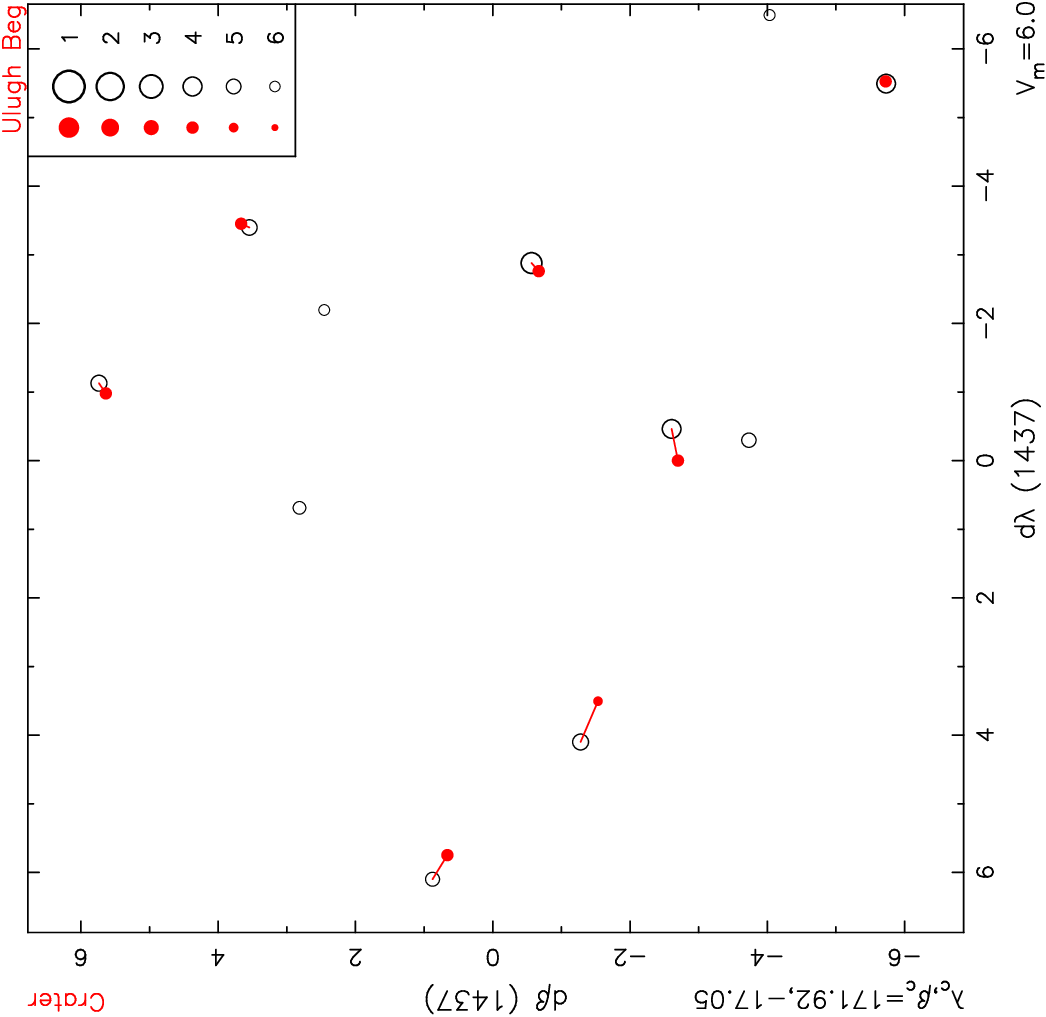}}
 \caption{Crater       \label{f:crater}}
 \end{figure*}
  
 \begin{figure*}
 \centerline{\includegraphics[angle=270,width=\columnwidth]{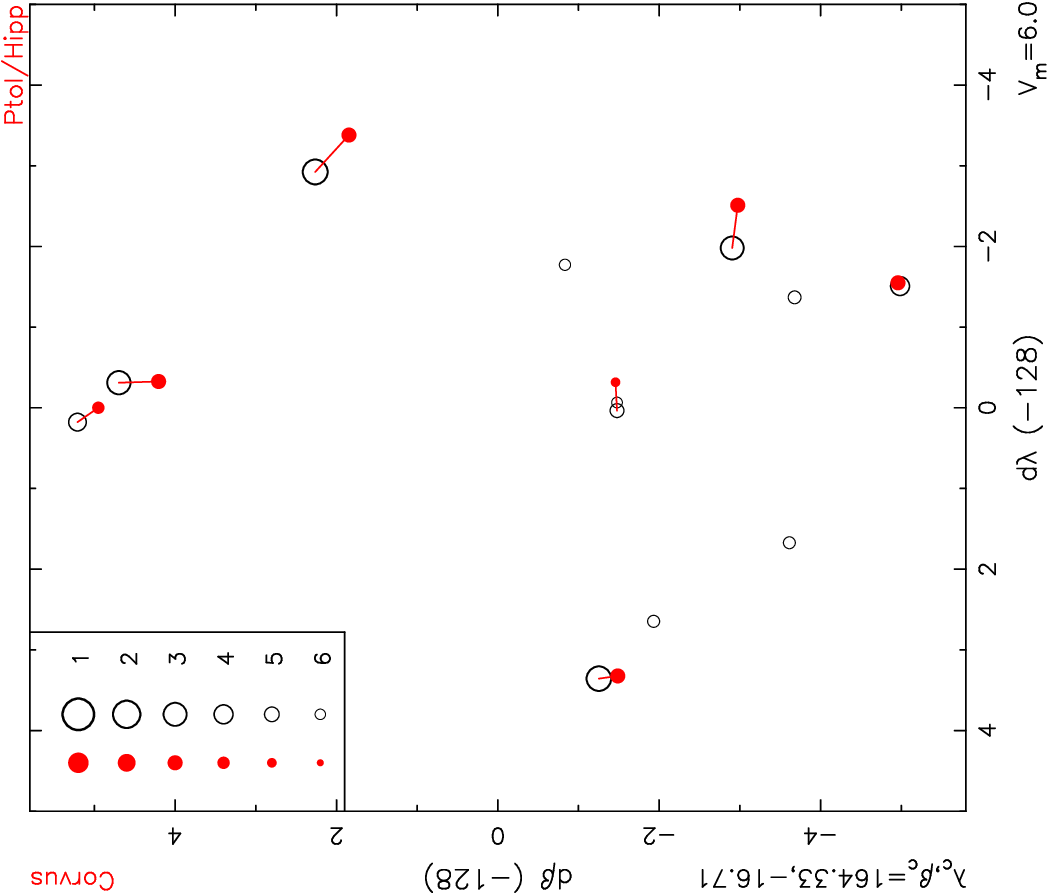}
\hfill\includegraphics[angle=270,width=\columnwidth]{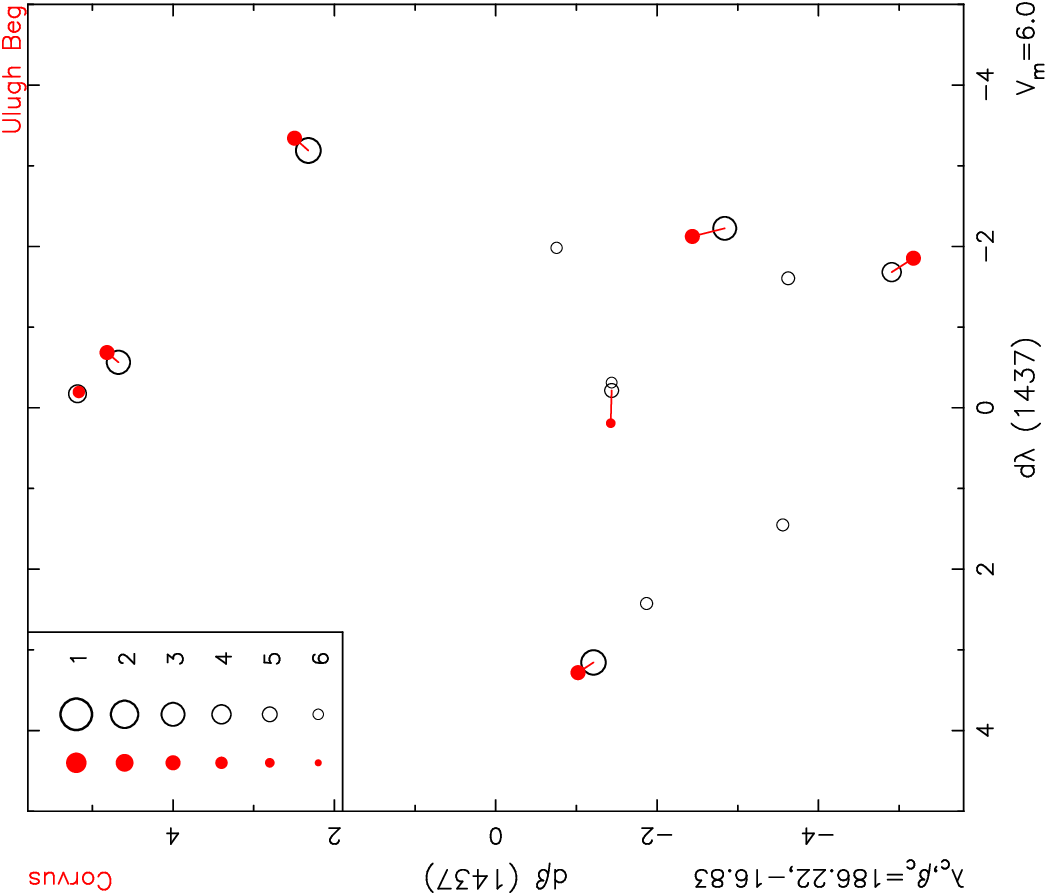}}
 \caption{Corvus       \label{f:corvus}}
 \end{figure*}
  
 \begin{figure*}
 \centerline{\includegraphics[angle=270,width=\columnwidth]{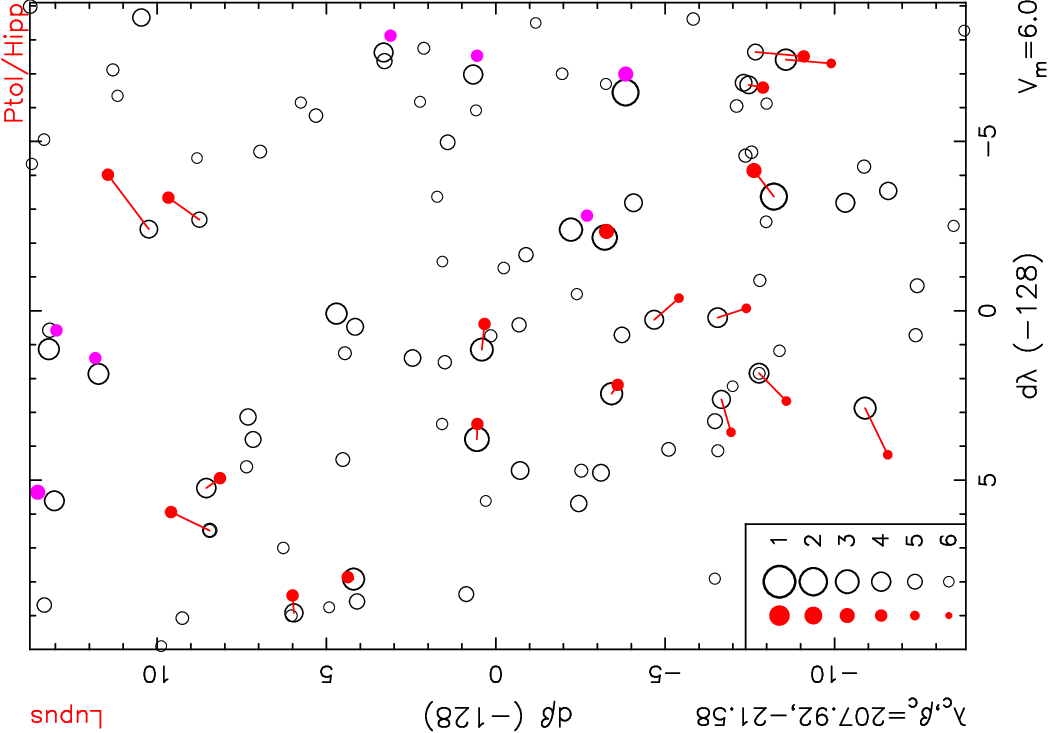}
\hfill\includegraphics[angle=270,width=\columnwidth]{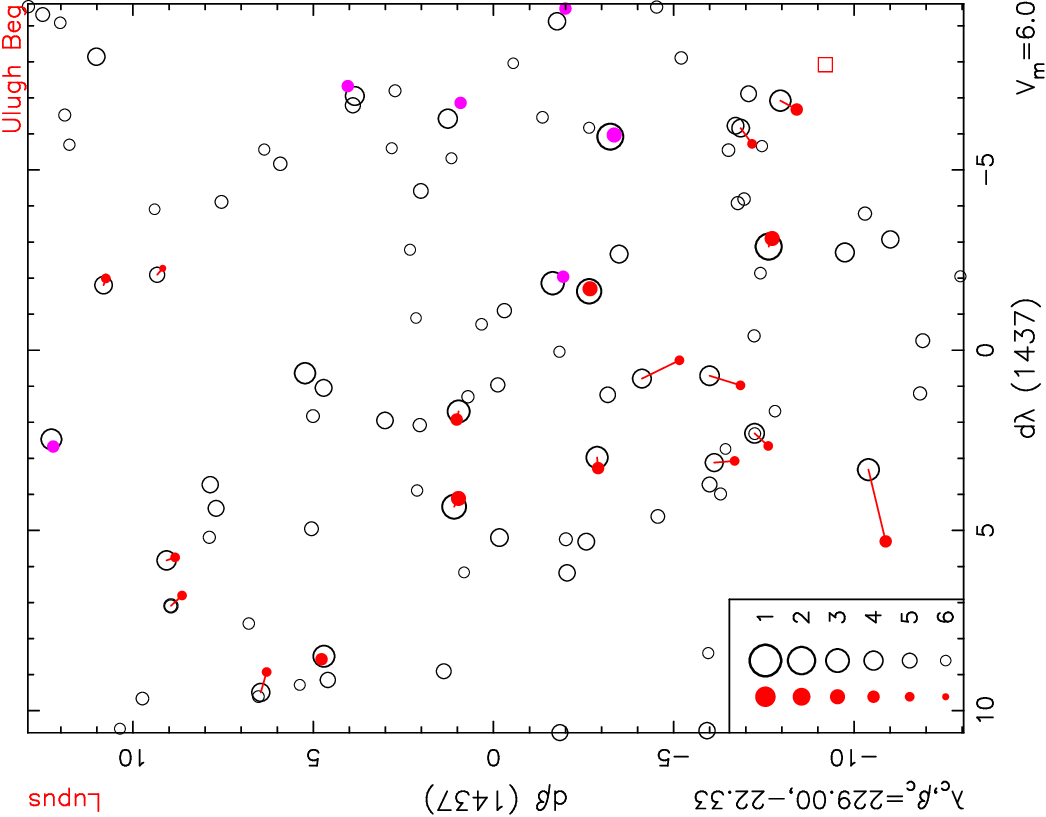}}
 \caption{Lupus        \label{f:lupus}}
 \end{figure*}
  
 \begin{figure*}
 \centerline{\includegraphics[angle=270,width=\columnwidth]{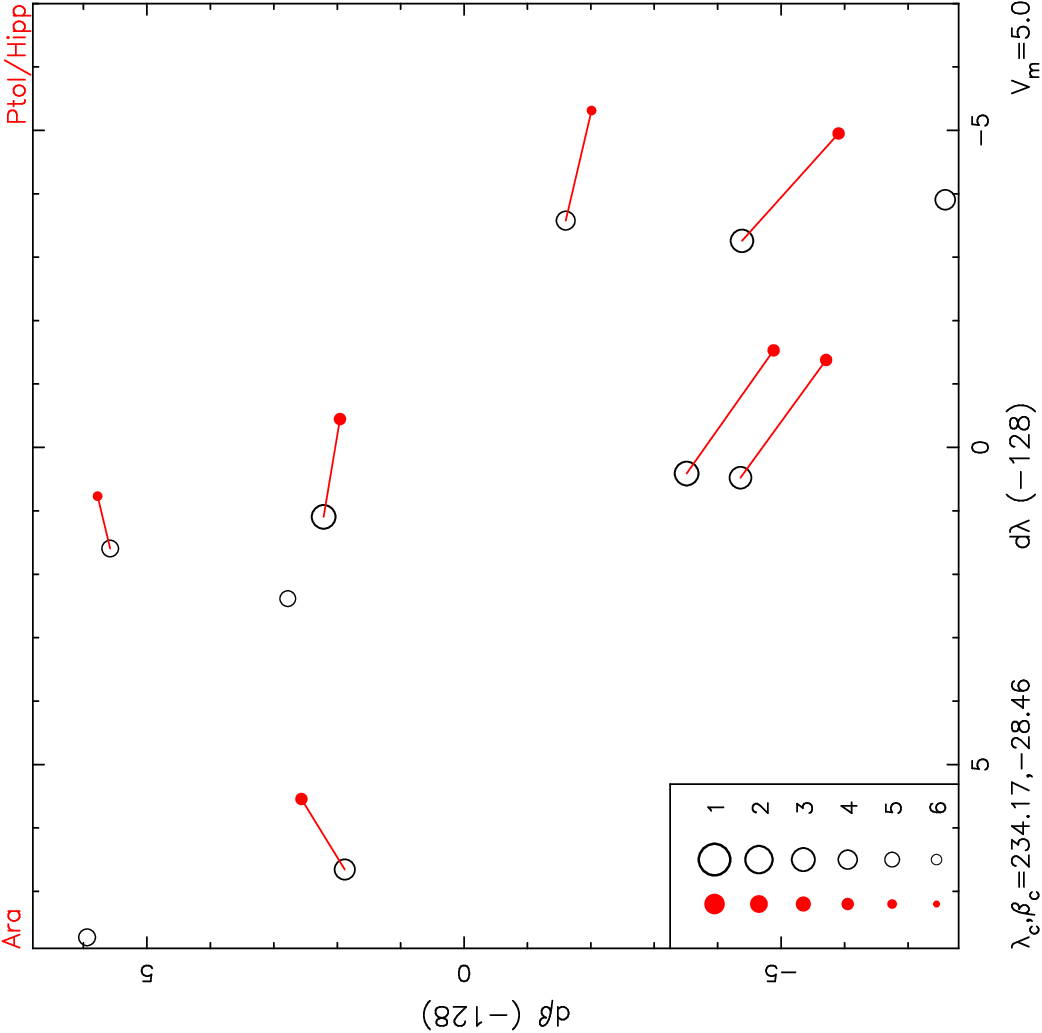}
\hfill\includegraphics[angle=270,width=\columnwidth]{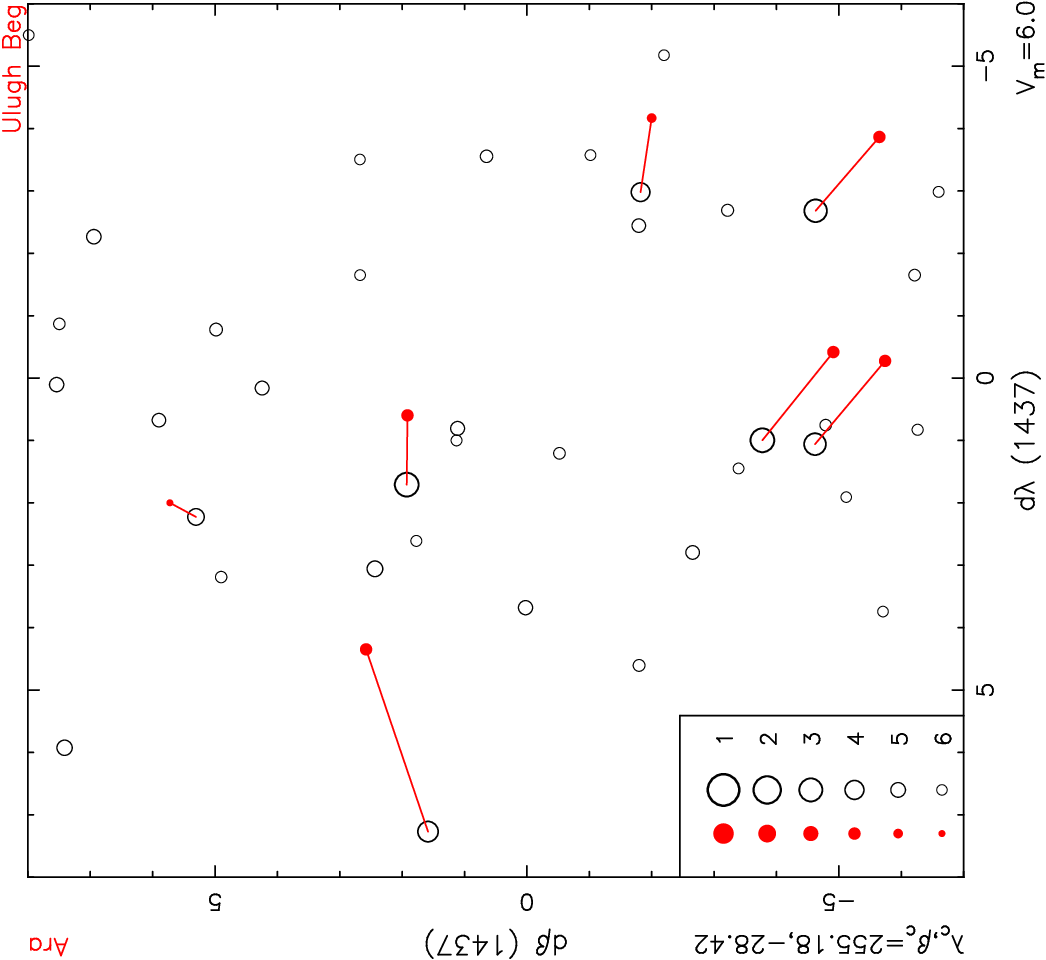}}
 \caption{Ara          \label{f:ara}}
 \end{figure*}

\end{appendix}
\end{document}